\documentclass[universe,review,accept,moreauthors,pdftex]{mdpi_modified}

\firstpage{1} 
\pubvolume{1}
\issuenum{1}
\articlenumber{0}
\pubyear{2022}
\copyrightyear{2022}
\datereceived{4 June 2022} 
\dateaccepted{20 June 2022} 
\datepublished{} 
\hreflink{https://doi.org/} 

\usepackage{natbib}

\Title{The Past and Future of Mid-Infrared Studies of AGN}

\TitleCitation{The Past and Future of Mid-Infrared Studies of AGN}

\Author{Anna Sajina$^{1,}$*\orcidA{}, Mark Lacy $^{2}$ and Alexandra Pope $^{3}$}

\AuthorNames{Anna Sajina, Mark Lacy and Alexandra Pope}

\AuthorCitation{Sajina, A.; Lacy, M.; Pope, A.}

\address{%
$^{1}$ \quad Department of Physics \& Astronomy, Tufts University, Medford, MA 02155, USA 
\\
$^{2}$ \quad National Radio Astronomy Observatory, 520 Edgemont Road, Charlottesville, VA 22903, USA;  mlacy@nrao.edu \\
$^{3}$ \quad Department of Astronomy, University of Massachusetts, Amherst, MA 01003, USA; pope@astro.umass.edu}
\corres{Correspondence: Anna.Sajina@tufts.edu} 

\abstract{Observational studies of AGN in the mid-infrared regime are crucial to our understanding of AGN and their role in the evolution of galaxies. Mid-IR-based selection of AGN is complementary to more traditional techniques allowing for a more complete census of AGN activity across cosmic time. Mid-IR observations including time variability and spatially resolved imaging have given us unique insights into the nature of the obscuring structures around AGN. The wealth of fine structure, molecular, and dust features in the mid-IR allow us to simultaneously probe multiple components of the ISM allowing us to explore in detail the impact on the host galaxy by the presence of an AGN---a crucial step toward understanding galaxy-SMBH co-evolution. This review gives a broad overview of this wide range of studies. It also aims to show the evolution of this field starting with its nascency in the 1960s, through major advances thanks to several generations of space-based and ground-based facilities, as well as the promise of upcoming facilities such as the {\sl James Webb Space Telescope (JWST)}. 
}


\begin{document}

\section{Introduction}

Observed trends, such as the close relationship between black hole mass and velocity dispertion (the M-$\sigma$ relation)~\citep{Magorrian1998,Gultekin2009} and the similarity of the histories of  cosmic star-formation and black hole accretion~\citep{MadauDickinson2014} have led to the conclusion that likely all massive galaxies host supermassive black holes and, moreover, that the growth of these black holes and their host galaxies are closely coupled~(see \citep{HeckmanBest2014} for a recent review). Although only about 1\% of galaxies are at any given time observed as AGN (Active Galactic Nuclei), it is in this phase, when the black holes are actively accreting material, that we are most likely to witness the processes by which galaxies and their central black holes co-evolve.

A recent review of the large range of phenomenology of AGN across the electromagnetic spectrum is presented in~\citet{Padovani2017}. Much of this observed diversity can be understood via the orientation-based unification model (see~\citep{Netzer2015} for a recent review). An alternative (or at least complementary) interpretation for the observed diversity in AGN properties, however, exists in the form of merger-driven evolution, where the time around coalescence corresponds to both the most intense stellar build-up and strongest AGN activity.  At this peak, both stellar mass build up and black hole growth take place in heavily dust-obscured systems, with obscuration decreasing post coalescence due to feedback processes~\citep{Hopkins2008}. This leads to IR-luminous (dust-obscured) galaxies commonly showing at least some level of AGN activity, including a significant fraction appearing as composite objects, where both star-formation and AGN contribute to the dust heating~\citep{Sajina2012,Kirkpatrick2012,Kirkpatrick2015}.  
Such a merger-driven model is also supported by the observation of increased fractions of obscured AGN among galaxies showing evidence of recent/ongoing mergers or vice versa (e.g.,~\citep{Zamojski2011,Kocevski2015}), a trend that is, however, largely confined to the most luminous AGN~\citep{Treister2012}. For a recent review of obscured AGN see~\citet{Hickox&Alexander2018_review}. By contrast to the merger-driven model, secular AGN triggering, through dynamical processes internal to the host galaxy, dominates among lower luminosity, but much more numerically common, AGN~\citep{Treister2012,KormendyHo2013,Lambrides2021}.  

Overall, much of the cosmic black hole accretion history is happening in obscured AGN~\citep{Gilli2007,lacy&sajina2020,Runburg2022} just like the bulk of the cosmic star-formation is dust obscured~\citep{MadauDickinson2014}. Infrared AGN selections have therefore been crucial in building up a more complete census of AGN as they can find obscured AGN missed by other selection methods such as X-rays and optical spectroscopy~\citep{lacy2004,Stern2005,Donley2007,Dey2008,Assef2015}. 

Beyond merely identifying AGN, detailed spectrally, spatially and temporarily resolved infrared observations of AGN have already given us great insights into the obscuring structures around the AGN. These have allowed us to probe both small scales, from tens to hundreds of parsecs with 8--10 m class \linebreak telescopes~\citep{Packham2005,Roche2006,Honig2010,Alonso-Herrero2016} associated directly with the central AGN, as well as galactic ($\stackrel{>}{_{\sim}}$1\,kpc) scales~\citep{Lyu2019}.  Indeed, the AGN driving-away or at least exciting the host galaxy's ISM is a crucial component in our understanding of galaxy-black hole co-evolution by providing a mechanism by which the AGN can serve to regulate star-formation within its host galaxy. For example, recent MUSE IFU data gives us direct evidence of how AGN fueling is influenced by galactic scale structures and in turn affects the wider galaxy through AGN-driven feedback~\citep{Juneau2022}.  

This review focuses in particular on the mid-IR regime, roughly 3--30\, microns in the rest-frame. Thanks to a rich array of diagnostic broad spectral features and emission lines, this decade in wavelength is ideally suited for the study of star formation in dusty galaxies and AGN, and also for probing key ISM characteristics that may influence or be influenced by the presence of an AGN. The study of AGN in the mid-IR dates back to the 1960s but took large leaps forward thanks to the {\sl Infrared Space Observatory (ISO)}~\citep{Kessler1996}; the {\sl Spitzer} Space Telescope~\citep{Werner2004} \endnote{See also~\citet{lacy&sajina2020} for a recent review of AGN studies with {\sl Spitzer}.}; the {\sl Wide-field Infrared Survey Explorer (WISE)}~\citet{Wright2010}; and ground-based instruments such as MIDI~\citep{Leinert2003} and MATISSE~\citep{Lopez2022}. This study is now on the cusp of another huge leap forward with the advent of the James Webb Space Telescope (JWST) with its MIRI instrument,~\citet{Rieke2015}. For a more complete discussion of AGN properties in the infrared, we also refer the reader to two other reviews in this Special Issue. These are: ``Infrared Spectral Energy Distribution and Variability of Active Galactic Nuclei: Clues to the Structure of Circumnuclear Material'' by Jianwei Lyu and George Rieke \citep{LyuRieke_review2022}; and ``The Role of AGN in Infrared Galaxies from a Multiwavelength Perspective'' by Vivian U \citep{U_review2022}. 

Finally, this review is aimed as an accessible entry point for students and researchers who are new to the field, providing a broad overview of the questions addressed via mid-IR spectroscopic or imaging observations of AGN. It is organized as follows. In Section\,\ref{sec:history}, we present a brief history of mid-IR studies of AGN. In Section\,\ref{sec:obs_properties}, we present the observed characteristics of AGN in the mid-IR, as well as discuss mid-IR AGN selection and its implications for AGN demographics, including the AGN luminosity function and the cosmic black hole accretion rate. In Section\,\ref{sec:physical_properties}, we discuss a variety of constraints on the  physical properties of AGN achieved through mid-IR observations including spectroscopy, spatially resolved observations and variability studies. In Section\,\ref{sec:agn_host}, we discuss the gas and dust properties around AGN (i.e., how the AGN affects its host galaxy's ISM), as well as disentangling how much of the dust heating of a given galaxy is due to star-formation vs. AGN activity---both topics are critical to the broader study of black hole-galaxy co-evolution. Section\,\ref{sec:future} presents an overview and comparison of past, present and upcoming mid-IR spectrographs and in particular addresses the anticipated AGN science expected from the recently launched {\sl JWST}. Finally, in Section\,\ref{sec:summary}, we present a summary and key open~questions.


\section{A Brief History}
\label{sec:history}

The history of the study of AGN in the infrared goes hand-in-hand with the history of infrared astronomy itself. Underlying it of course is the history of the development of infrared detectors---originally largely motivated by military applications (for a comprehensive review see 
~\citep{Rogalski2012}). It is interesting to note, that the first observations of astronomical objects in the infrared were performed by W. Coblentz more than 100 years ago, in 1915~\citep{Rogalski2012}. The development of increasingly sensitive detectors out to the mid-IR regime allowed for ground-based observations that revealed several basic characteristics of the mid-IR spectra of normal galaxies and AGN. The earliest observations obtained ground-based constraints at 2, 10 and 22\,$\upmu$m of bright sources such as 3C273 and others~\citep{Low1965,Low1968}. These showed an upturn in the continua of galaxies toward the infrared that, in nearly all cases, could not be explained as simple extrapolation of the radio synchrotron emission. Rather, it was already interpreted as a distinct component, specifically thermal dust emission due to dust grains absorbing optical/UV light from the nuclear source~\citep{Rees1969}. Such thermal dust emission was limited to the infrared by the dust sublimation temperature of $\approx$1500 K~\citep{Rees1969}. 
A key early study was~\citet{RiekeLow1972} which showed ubiquitous mid-IR (10 $\upmu$m) emission both among regular spirals and Seyferts, as well as an early hint of the IR-radio correlation. Mid-IR variability on the scale of months was detected in several active galaxies including NGC1068 and 3C273~\citep{Rieke1972b}. A 8--13 $\upmu$m spectrometer with $\Delta\lambda/\lambda\sim0.02$ at the Kitt Peak National Observatory 2m telescope allowed for the first mid-IR spectra of just a handful of bright galaxies including the Seyfert NGC1068 and the starburst galaxy M82~\citep{Gillett1975,Kleinmann1976}. These revealed several now familiar features such the 9.7 $\upmu$m silicate absorption, the [NeII]\,12.8~$\upmu$m line, and a 11.3 $\upmu$m feature of then unknown origin, but later attributed to PAH. Overcoming the very challenging spectrometry (only available for a handful of galaxies),~\citep{LebofskyRieke1979} used an ingenious technique of nested broad and narrow filters to detect and study the 9.7 $\upmu$m silicate absorption feature in a larger sample. Shorter wavelength low-resolution spectrometry also showed the 3.3 $\upmu$m PAH emission feature~\citep{Russell1977a}.

Ground-based studies out to 3\, $\upmu$m or at $\approx$10\,$\upmu$m are challenging but still possible thanks to relatively lower atmospheric absorption there. A spectrometer mounted on the balloon-born Kuiper Astronomial Observatory (KAO) allowed for the first 4--8 $\upmu$m spectra to be taken, revealing the 6.2 and 7.7 $\upmu$m PAH (then unidentified) features, as well as hints of multiple fine structure features of Ar, S, Ne, and Mg~\citep{Russell1977b}. These early studies already revealed the richness of the mid-IR spectra of both galaxies and AGN but left many unanswered questions and were limited to the brightest sources with photometric samples on the order of a hundred and spectroscopic samples on the order of five.  Many parts of the spectrum were virtually unexplored due to strong atmospheric absorption. This motivated the development of space-based facilities which included first the InfraRed Astronomical Satellite~(IRAS~\citep{Neugebauer1984}) (launched 1983); then the Infrared Space Observatory ({\sl ISO};~\citep{Kessler1996}, launched 1995); the {\sl Spitzer} Space Telescope~\citep{Gehrz2007} (launched 2003), and the Wide Field Infrared Survey Explorer ({\sl WISE};~\citep{Wright2010}, launched 2009).  {\sl WISE} mapped out the whole sky at 3.4, 4.6, 12 and 22\,$\upmu$m. 

\begin{figure}[H]
\includegraphics[width=\textwidth]{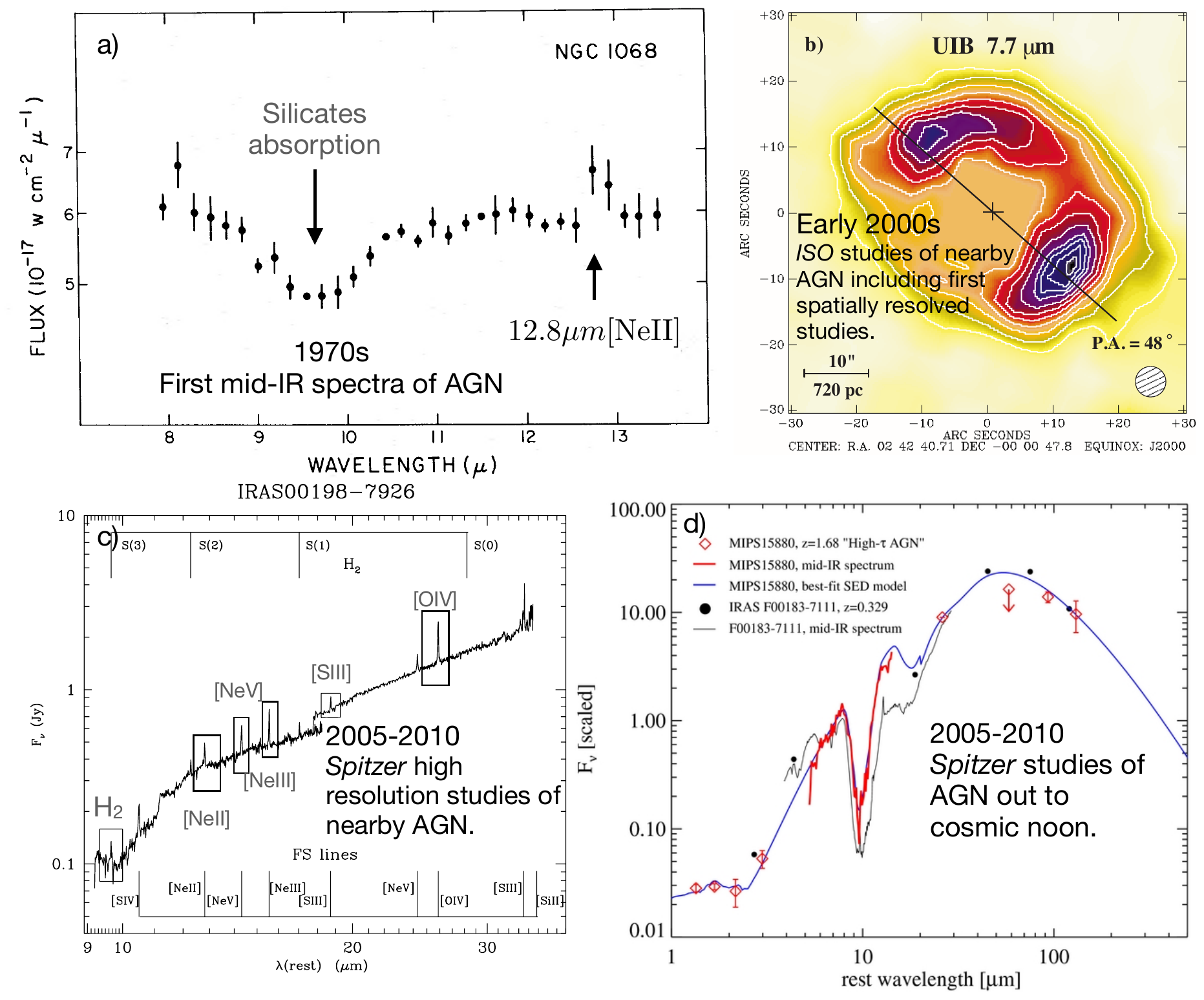}
\caption{A 
 broad view of the history of mid-IR spectroscopic studies of AGN. ({\bf a}) Shows the 1976 spectrum of the Seyfert 2 NGC1068 ($\approx 5".3$ beam), already identifying the presence of silicate absorption at 9.7\,$\upmu$m as well as the fine structure [NeII]\,12.8\,$\upmu$m line, adapted from 
 \linebreak \citet{Kleinmann1976}. ({\bf b}) Shows an {\sl ISO} image highlighting the spatial distribution of the PAH 7.7\,$\upmu$m feature (here the continuum is subtracted) in NGC1068, adapted from~\citet{LeFloch2001}. The dearth of PAH emission at the center is consistent with the idea that the AGN radiation destroys the PAH molecules in its immediate vicinity, but not on galaxy-wide scales. ({\bf c}) Shows the {\sl Spitzer} IRS high resolution spectrum of a local Seyfert 2 source IRAS00198-7926, highlighting a few key lines, adapted from~\citet{Tommasin2008}. We see a continuum-dominated spectrum superimposed with a wide range of fine structure lines including from high-ionization species such as [NeV], as well as features associated with shocked H$_2$. ({\bf d}) Shows an example of a heavily obscured AGN at cosmic noon, characterized by weak PAH but extremely deep silicate absorption, adapted from \linebreak \citet{Sajina2012}. {\sl Spitzer}/IRS was the first instrument to allow for mid-IR spectroscopic studies beyond the local Universe. ({\bf a},{\bf c}) and d are AAS\textsuperscript{©}, panel b is ESO\textsuperscript{©} . Reproduced with permission. 
\label{fig:intro}}
\end{figure}

Figure\,\ref{fig:intro} shows only a small subset of the history of mid-IR spectroscopic studies of AGN across the last several decades. Figure\,\ref{fig:intro}a is a 1970s example~\citep{Kleinmann1976} of the earliest ground-based mid-IR spectra that were available for a handful of bright nearby AGN such as NGC1068. Figure\,\ref{fig:intro}b shows that {\sl ISO} not only greatly expanded mid-IR spectroscopic studies of nearby galaxies and AGN but also provided some spatially resolved results such as this map of the 7.7\,$\upmu$m PAH feature \endnote{At the time, these features were designated UIB or Unidentified Infrared Bands.} in NGC1068~\citep{LeFloch2001}---a precursor to what we expect to see from JWST MIRI beyond the local Universe. Figure\,\ref{fig:intro}c shows the much more detailed and high-resolution spectra available for nearby AGN through {\sl Spitzer}~\citep{Tommasin2008}. Figure\,\ref{fig:intro}d highlights that {\sl Spitzer} allowed for the first time mid-IR spectroscopic studies beyond the local Universe for both galaxies and AGN~\citep{Yan2007,Sajina2007a,Pope2008,Dasyra2008,MartinezSansigre2008,Seymour2008,Fadda2010}. These studies not only confirmed features and trends already seen by earlier studies, especially with {\sl ISO}~\citep{Genzel1998,Sturm2000}, but also highlighted the increasing importance of previously largely unknown/underappreciated classes of objects such as AGN with extremely deep silicate absorption features (Figure\,\ref{fig:intro}d).  These sources are likely to represent an interesting stage in the co-evolution of galaxies and their central black holes.

\section{Observed Characteristics and Mid-IR AGN Demographics}
\label{sec:obs_properties}
  
This section summarizes the key observational characteristics of AGN in the mid-IR, especially as contrasted to the mid-IR properties of pure star-forming galaxies. These characteristics are used to select AGN in a way that is complementary to other traditional selection methods such as in the X-rays. Therefore, here, we also summarize the lessons learned from using mid-IR observations to select AGN with the aim of a more complete census of AGN activity throughout cosmic time.

\subsection{The Observed AGN Mid-IR Emission}
\label{sec:observed_properties_lowres}
 
\begin{figure}[H]
\includegraphics[width=0.75\textwidth]{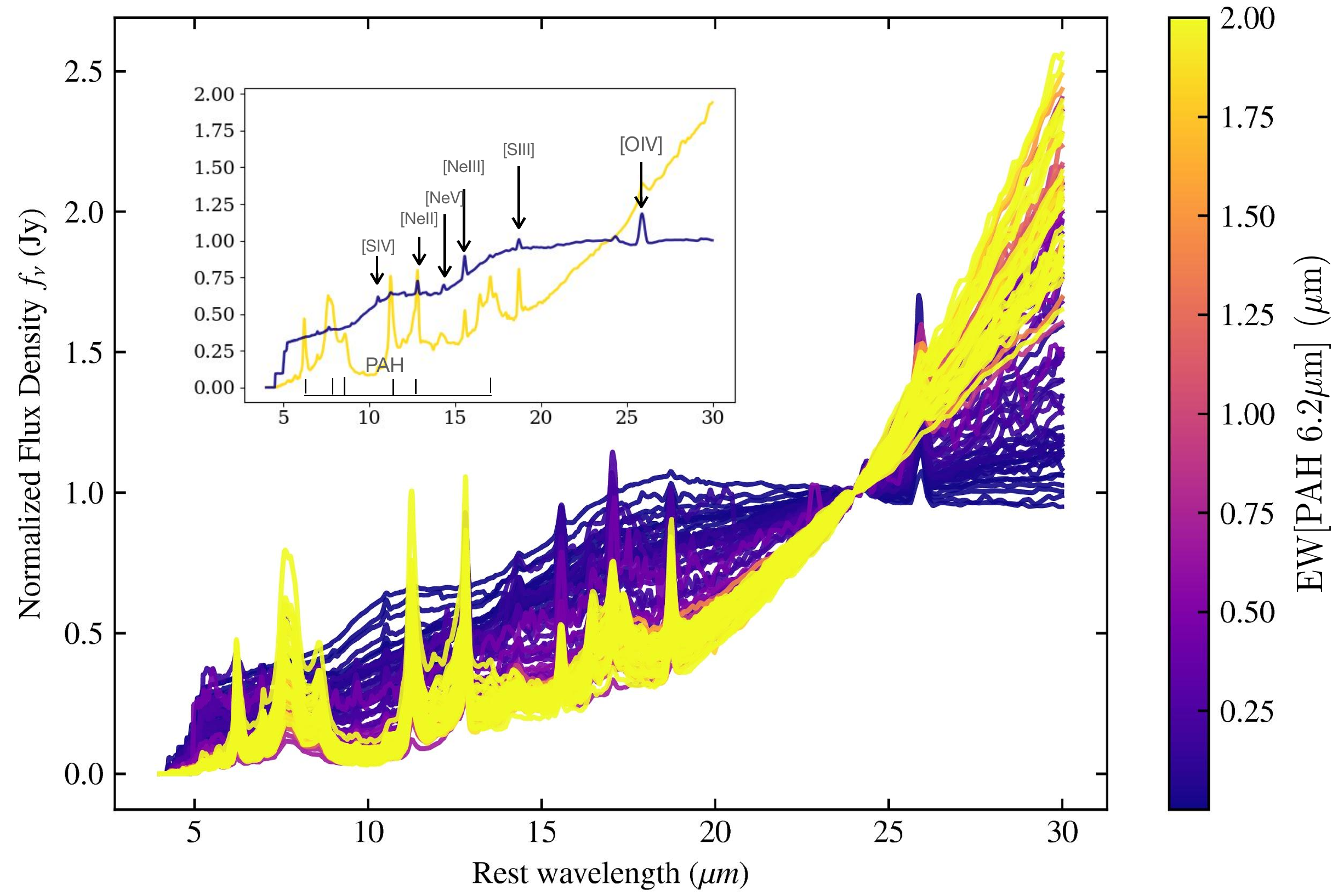}
\caption{Stacked SEDs in bins of EW6.2\,$\upmu$m where high EW galaxies are star-formation dominated, whereas low EW galaxies are AGN-dominated, at least in the mid-IR. The inset shows just two representative templates, one high EW6.2\,$\upmu$m (this looks much like the stacked SED of local pure starburst galaxies
~\citep{Brandl2006}) and one low EW6.2\,$\upmu$m (this appears very similar to the SED of an unobscured/Type 1 AGN (e.g.,~\citep{Tommasin2008,lacy&sajina2020}), see also Figure\,\ref{fig:intro}c. Among IR galaxies, we see a continuum of equivalent widths corresponding to a continuum of AGN fractions. Besides the equivalent width, it is clear that star-forming galaxies have steeper 30\,$\upmu$m/15\,$\upmu$m slopes.  Adapted from
 \linebreak \citet{Lambrides2019}. 
\label{fig:mirseds}}
\end{figure}

\textls[-10]{We already saw in Figure\,\ref{fig:intro}c that the mid-IR spectra of AGN are characterized by a strong, roughly power--law, continuum with superimposed fine structure lines. By contrast, the mid-IR spectrum of pure star-forming galaxies is dominated by the prominent PAH features at 6.2, 7.7, 8.6, 11.3, 12.6 and 17.0\,$\upmu$m~\citep{Brandl2006}. Indeed, the luminosities of these lines have been shown to correlate with traditional SFR indicators such as $L_{H_{\alpha}}$~\citep{Shipley2016}.  Because of this, a commonly used measure to determine the fraction of the mid-IR emission that is due to an AGN as opposed to star-formation, is the equivalent width of the 6.2\,$\upmu$m PAH feature (EW6.2), where values of EW6.2 < 0.5 $\upmu$m are typically considered AGN-dominated systems. 
Figure\,\ref{fig:mirseds} shows stacked SEDs in bins of EW6.2\,$\upmu$m. We see a gradual transition from PAH-dominated to continuum-dominated sources with weak or absent PAH features. These features of mid-IR AGN spectra vs. star-formation dominated spectra are commonly seen in spatially un-resolved space-based spectra (e.g.,~\citep{Sajina2007a,Tommasin2008,Hernan-Caballero2009,Hernan-Caballero2011,Sajina2012,Kirkpatrick2012}), and have been further explored in spatially-resolved studies with ground-based data~\citep{Honig2010,Esquej2014,Alonso-Herrero2014,Alonso-Herrero2016}. The mid-IR spectra of both are also rich in fine structure lines which allow us to probe the physical conditions of the gas,~\citet{Inami2013}.} These also serve as AGN diagnostics, especially using line ratios involving high and low ionization potential species such as [NeV]~14.3\,$\upmu$m/[NeII]~12.8\,$\upmu$m or [NeIII] 15.5\,$\upmu$m/[NeII] 12.8\,$\upmu$m, where such ratios are found to correlate with the PAH equivalent widths~\citep{Genzel1998,Armus2007} further supporting the use of the latter as AGN diagnostics in lower resolution mid-IR spectra. 
We discuss in more detail the diagnostics provided by the fine structure lines in Section\,\ref{sec:observed_properties_highres}.  
\vspace{-6pt}

\begin{figure}[H]
\includegraphics[width=0.75\textwidth]{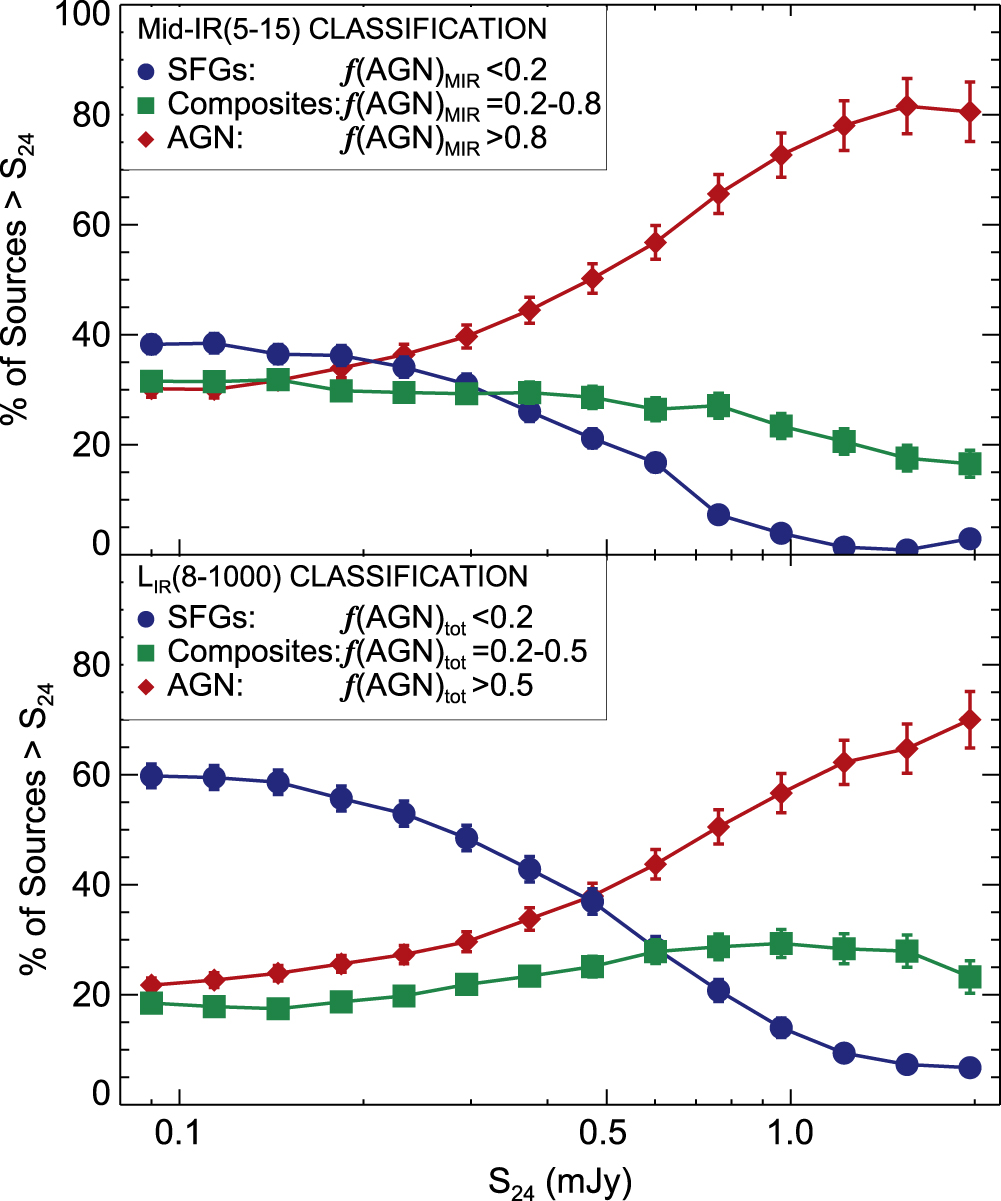}
\caption{The relative contribution of SF-dominated, AGN-dominated and composite sources as a function of 24\,$\upmu$m flux. These classifications are based on low resolution {\sl Spitzer}-IRS spectra which are fitted with a model including PAH emission and continuum emission, the relative strength of which provides the fraction of AGN contribution to the mid-IR ($f(AGN)_{MIR}$). Adapted from~\mbox{\citet{Kirkpatrick2015}}, © AAS. Reproduced with permission. 
\label{fig:role_composites}}
\end{figure}

It is critical to note that in the mid-IR, we do not see a bimodal star-forming/AGN distribution but rather a continuum that gradually transitions from the star-formation dominated (strong-PAH) to the AGN-dominated (weak-PAH, predominantly power law continuum) spectrum. Typically, therefore, researchers in the field have modeled the mid-IR spectrum as a sum of a star-forming and AGN component with the relative amplitudes as free parameters and defined a mid-IR AGN fraction ($f_{AGN,MIR}$) on the basis of integrating (over, e.g., 5--15 $\upmu$m) the best-fit AGN-only component divided by the integral of the best-fit total model, (e.g.,~\citep{Sajina2007a,Sajina2012,Kirkpatrick2012,Kirkpatrick2015}). This approach has also led to our considering sources as SF-dominated if the SF component accounts for $\geq$80\% of the total mid-IR emission or AGN-dominated, if the AGN component accounts for $\geq$20\% of the total mid-IR emission (over the 5-15\,$\upmu$m range) or a composite if the source lies in between. Figure\,\ref{fig:role_composites} shows the relative distribution of SF-dominated, AGN-dominated and composite objects among the 24\,$\upmu$m population. Note that while, unsurprisingly, AGN-dominated sources dominate among the $S_{24}>1$\,mJy sources, composites remain an important component (20-25\%) of the overall mid-IR source population even at the faintest fluxes.  

In addition to the characteristics of the PAH features, mid-IR spectra of both galaxies and AGN are also characterized by silicate features at $\approx$10 and 18\,$\upmu$m. These features are commonly observed in absorption, as shown in Figure\,\ref{fig:intro}a,d. Indeed, accounting for silicate absorption is critical for obtaining accurate PAH equivalent widths, especially for the 11.3\,$\upmu$m feature. For face-on orientations (Type 1 AGN), they are expected to be in emission~\citep{PierKrolik1992,Nenkova2000}, which was only first observed with {\sl Spitzer}~\citep{Siebenmorgen2005,Hao2005}. We discuss further the silicate features and their implications for the dust around AGN in Section\,\ref{sec:dust}. 

\begin{figure}[H]
\begin{adjustwidth}{-\extralength}{0cm}
\centering 
\includegraphics[width=18cm]{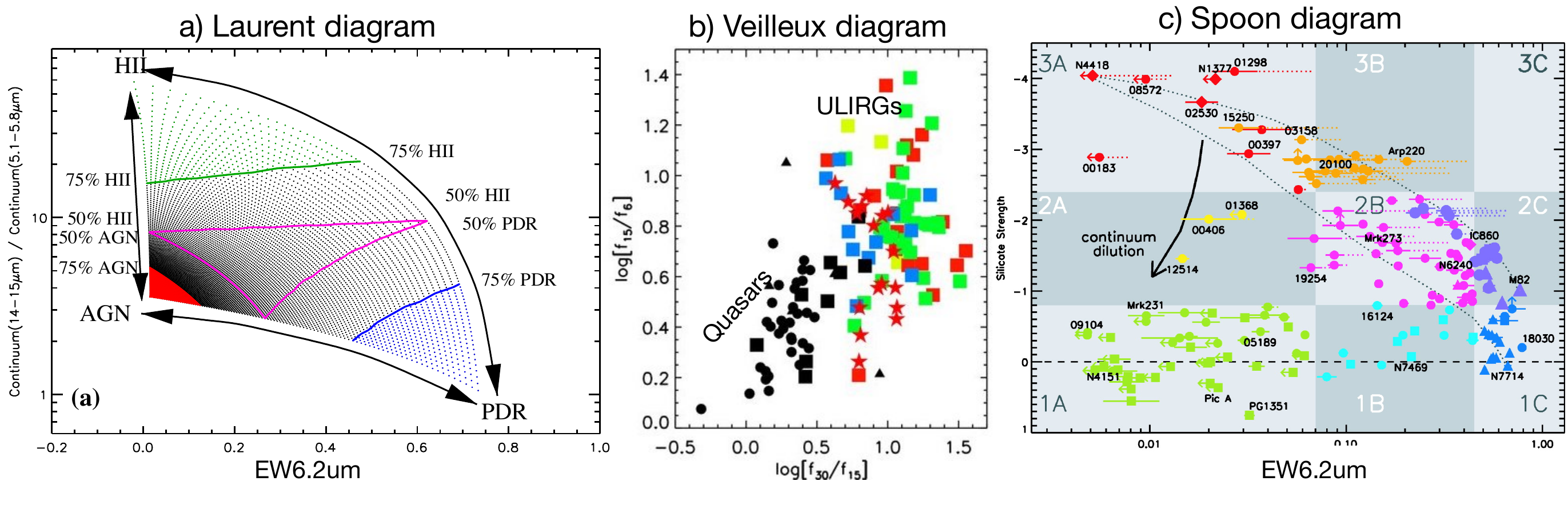}
\end{adjustwidth}
\caption{This 
 shows three important mid-IR diagnostic diagrams where we see the continuous transition from star-formation dominated to AGN-dominated spectra. ({\bf a}) The ``Laurent diagram''~\citep{Laurent2000}, indicates a mixing diagram between AGN-like, HII region-like and PDR-like spectra involving the PAH equivalent width and the mid-IR 15 to 5\,$\upmu$m continuum color. ({\bf b}) The ``Veilleux diagram''~\citep{Veilleux2009} shows the same 15 to 5\,$\upmu$m continuum color on the y-axis, but now compares it with the 30 to 15\,$\upmu$m color on the x-axis. All black symbols represent Seyfert 1 optical spectra, the blue squares are Seyfert 2 optical spectra. All other colors and the star symbols indicate sources with pure star-forming optical spectra. ({\bf c}) The ``Spoon diagram''~\citep{Spoon2007} has the same x-axis as the Laurent diagram but has the depth of the silicate feature along the y-axis, highlighting a potential embedded nucleus evolutionary phase. ({\bf a}) is ESO\textsuperscript{©}, ({\bf b},{\bf c}) are AAS\textsuperscript{©}. Reproduced with permission.   
\label{fig:diagnostic_diagrams}}
\end{figure}

Figure\,\ref{fig:mirseds} shows that besides the relative strength of the PAH features, the mid-IR spectra of sources also show different slopes/colors of the 15--5
\,$\upmu$m continuum, as well \linebreak as 30-to-15\,$\upmu$m continuum depending on whether the source is star-formation or AGN dominated. This behavior gives us additional diagnostic power and can serve to identify AGN in photometric data alone. Figure\,\ref{fig:diagnostic_diagrams} shows a series of three diagnostic diagrams that include both the principle emission (PAH) and absorption (silicates) features in mid-IR spectra, as well as these continuum colors. These plots also highlight that we have a continuum from SF-dominated to AGN-dominated sources, but it also highlights that the PAH equivalent width alone does not capture the full range of spectra either for the non-AGN or the AGN components. The Laurent diagram represents a mixing diagram between AGN-like, PDR-like and HII region-like mid-IR spectra. It shows again that the PAH equivalent width can be used to determine the level of AGN contribution relative to star-formation in a galaxy (since typical star-forming galaxies have PDR-like spectra with strong PAH, see also~\citep{Brandl2006}). However, it also shows that the spectra of pure HII regions (which are also consistent with the spectra of some dwarf starburst galaxies) have weak PAH features same as AGN spectra. The two can be distinguished by the much steeper 15--5\,$\upmu$m spectra of HII regions. 
The Veilleux diagram (Figure\,\ref{fig:diagnostic_diagrams}b) shows that both the 15--5\,$\upmu$m and the 30--15\,$\upmu$m colors are much steeper in star-formation dominated systems than in AGN-dominated systems (such as the classical quasars shown in black filled circles). Some sources with pure star-forming spectra in the optical have AGN-like 15--5\,$\upmu$m colors (consistent with their being more PDR-like mid-IR spectra). They can however be distinguished from AGN on the basis of much redder 30--15\,$\upmu$m colors.  However, systems with Seyfert 2, such as optical spectra (the blue squares in the plot), are hard to distinguish on this plot from star-forming systems on the basis of their 5--30\,$\upmu$m mid-IR colors alone. Indeed,~\citet{Veilleux2009} used multiple mid-IR spectral diagnostics (including fine structure line ratios, PAH equivalent width and colors) to determine mid-IR AGN fractions since, as we point out here as well, no single diagnostic works for all systems. However, the issue of why these ``Seyfert 2'' systems have similar 30--15\,$\upmu$m colors to star-forming galaxies has a lot to do with the question of whether or not the 30\,$\upmu$m ``warm dust'' continuum can be powered by AGN. If AGN powered, it would need to be further out from the region of the classical dust torus, most likely due to dust in the NLR region. In a detailed study of a large sample of sources with both mid-IR spectra and photometric coverage across the full IR SED, \linebreak \citet{Kirkpatrick2015} argued that the warm dust can indeed be AGN powered, unlike the cold dust which dominates the far-IR/sub-mm regime and is associated with star-formation, see Section\,\ref{sec:agn_host} for further discussion. 

Finally, the Spoon diagram shows that AGN-dominated systems in the mid-IR split into two distinct tracks---one with weak silicate absorption (such as classical quasars) and one with deep silicate features which are often referred to as ``buried AGN'', see, e.g., Figure\,\ref{fig:intro}d. They argue that there is an evolutionary sequence from the latter to the former. It is important to emphasize though that these observed mid-IR properties are indicative of ``buried nuclei'' and said nuclei can, in principle, be powered by either an AGN or a compact nuclear starburst~\citep{Spoon2022}. This uncertainty can be addressed by introducing, for example, radio data which is insensitive to dust obscuration. For example,~\citet{Sajina2007b} find that nearly 40\% of such deep silicate absorption sources are radio-loud, supporting their AGN nature, and indeed the source shown in Figure\,\ref{fig:intro}d presents a classical double-lobed radio morphology.  The review article by Eric Murphy in this Special Issue discusses further such synergies between infrared and radio surveys.  

Another approach is to combine multiple diagnostics. For example, building on the diagnostic plots in Figure\,\ref{fig:diagnostic_diagrams},~\citet{Marshall2018} presented a unified model for deeply obscured systems where a 3D diagram including PAH equivalent width, silicate feature depth and the 15-5\,$\upmu$m color can be used to distinguish whether a dusty nucleus including only an AGN+clumpy torus, an AGN and clumpy torus plus nuclear starburst or only a nuclear starburst. They note that in the case of the most deeply obscured AGN, $>$75\% of the total bolometric emission is powered by the AGN. They also show that such buried AGN spectra require covering fractions of near 100\%, even 10\%  unobscured lines of sight toward the nucleus would make the source appear unobscured in the mid-IR.  The basic trends in the observed diagrams shown in Figure\,\ref{fig:diagnostic_diagrams} have also been shown to hold in GADGET+radiative transfer simulations of isolated galaxies and mergers~\citep{Snyder2013}.

Lastly, we note the recently published work by~\citet{Spoon2022} on the Infrared Database of Extragalactic Observables from {\sl Spitzer} (IDEOS 
 \endnote{\url{http://ideos.astro.cornell.edu/}}) which presents a public database of homogeneously derived set of 77 different observable properties based on low-resolution 5--36\,$\upmu$m {\sl Spitzer} IRS spectra for $>$3000 galaxies. 

\subsection{Fine Structure Mid-IR Lines}
\label{sec:observed_properties_highres}

\begin{figure}[H]
\includegraphics[width=0.75\textwidth]{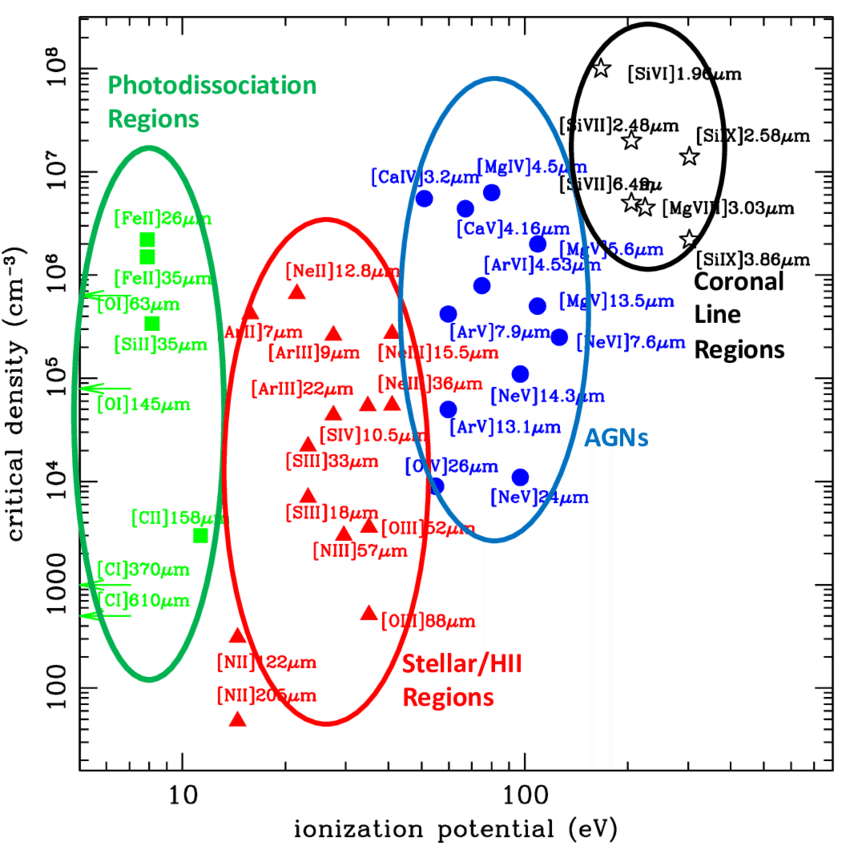}
\caption{The wealth of mid-IR fine structure diagnostic lines. The ratio of lines from say the AGN group vs. from the HII group (e.g., $\rm{[N_V]/[N_{II}]}$) serve as diagnostics of the relative strength of AGN in a galaxy. Adapted from~\citet{Spinoglio2017}. \label{fig:fine_structure}}
\end{figure}

Above, we focus on the broad emission/absorption and continuum features such as seen in the lower resolution {\sl Spitzer} IRS spectra---the only mid-IR spectra available beyond the local Universe so far. However, Figure\,\ref{fig:fine_structure}, based on the original work of~\citet{SpinoglioMalkan1992}, shows the wealth of mid-IR fine structure lines displayed in the ionization potential vs. critical density plane. The presence of any of the lines within the AGN or coronal line regions can be considered an indication of the presence of an AGN in the system. Ratios of AGN to HII-like lines (e.g., [NeV]/[NeII]) can be used as diagnostics of the relative strength in the mid-IR of the AGN to the star-formation. Note that the AGN-like spectrum in the insert in Figure\,\ref{fig:mirseds} shows high ionization potential (also known as AGN) lines including [NeV] 14.3\,$\upmu$m, and [OIV] 26\,$\upmu$m. We also see the star-formation line [NeII] 12.8\,$\upmu$m and the borderline [NeIII]\,15.5\,$\upmu$m. Indeed, the [NeIII]/[NeII] ratio itself can be used as a diagnostic of the degree of AGN activity in a galaxy. Indeed, there is a strong correlation between the NLR lines [NeV] and [NeIII] spanning several orders of magnitude in luminosity~\citep{Gorjian2007}, which places constraints on allowed photoionization models and explains why these two ratios both work as AGN diagnostics. As mentioned already in Section\,\ref{sec:observed_properties_lowres}, the [NeIII]/[NeII] ratio has been shown to correlate with the PAH equivalent width~\citep{Armus2007} supporting the latter's use as an AGN diagnostic. Of course, besides serving as indications of the presence or strength of an AGN, these mid-IR fine structure lines can serve as probes of the physical conditions (e.g., metallicity, temperature, density) of the gas in the NLR, as well as more broadly in the AGN host galaxy~\citep{Fernandez-Ontiveros2021}. For a recent review of emission line diagnostics including in the mid-IR see~\citet{kewley2019}. We return to this point in more detail in Section\,\ref{sec:ism}. Lastly we note that here we focus on the mid-IR diagnostic lines only, but of course ALMA has been observing the far-IR emission lines, especially the PDR [CII] 158\,$\upmu$m line, out to $z \sim$6. As an example, we refer the reader to the paper of A. Faisst on the results of the ALPINE survey, presented in this Special Issue. 

\subsection{Mid-IR AGN Selection: Implications for AGN Demographics}
\label{sec:miragn}
 
The previous section discussed how we select and characterize AGN using low- and high-resolution mid-IR spectra. This section summarizes how we select AGN in mid-IR photometric surveys, and what such mid-IR selected AGN tell us about the AGN luminosity function and the cosmic history of black hole mass build-up.   

\subsubsection{Mid-IR AGN Selection Techniques}
\label{sec:agn_color_selection}

The characteristic red, often featureless mid-infrared SED of AGN (see Figure\,\ref{fig:mirseds}) enables effective selection techniques based solely on broad band colors, provided the AGN emission dominates the observed-frame mid-infrared (see~\citet{lacy&sajina2020} for a recent review). Bright mid-IR flux limited samples are also effective at selecting AGN (see Figure\,\ref{fig:role_composites}) as was made use of in {\sl IRAS} data for their 12\,$\upmu$m-limited sample,~\citep{1993ApJS...89....1R}. Some of the AGN diagnostics already discussed in Section\,\ref{sec:observed_properties_lowres}, such as those in~\citet{Laurent2000} became the basis for photometric selection of AGN with {\em ISO} surveys~\citep{2004A&A...419L..49H}. Photometric broad-band selections were later developed for {\em Spitzer} IRAC surveys (e.g.,~\citep{Sajina2005}) and became very valuable tools for finding obscured AGN in the {\em Spitzer} surveys~\citep{lacy2004,Alonso-Herrero2006,Stern2005,2012ApJ...748..142D} and subsequently for {\em WISE}~\citep{Mateos2012,2012ApJ...753...30S, 2018ApJS..234...23A}. Techniques have also been proposed for {\em JWST}~\citep{Kirkpatrick2017}.

Selection techniques can be based on colors, e.g., with IRAC~\citep{lacy2004,Stern2005}, or by requiring a power--law spectrum~\citep{Alonso-Herrero2006,Donley2007}. These techniques are all discussed extensively in the review by Lyu and Rieke in this Special Issue to which we refer the reader for a more complete coverage. Here, we only re-iterate some key points. First, all mid-IR AGN selection techniques only work for sources where the AGN dominates the mid-IR emission (see discussion on role of composite objects in Section\,\ref{sec:observed_properties_lowres}, a point also made in~\citet{Hickox&Alexander2018_review}). Second, while such techniques in general pick out both unobscured (Type 1) and obscured (Type~2) AGN, they do fail for the most extremely obscured AGN where the short wavelength IRAC emission from the AGN might be largely absorbed by dust in the nuclear regions or extended galaxy (see, e.g.,~\citet{Snyder2013},~\citet{Marshall2018}). One could make a rough argument that if for a typical attenuation curve the ratio of the attenuation at 3\,$\upmu$m to that in the V-band is about 10, then AGN become optically-thick at 3\,$\upmu$m if subject to dust attenuation of $A_V>10$. Much higher levels of attenuation are seen in high spatial resolution studies of the nuclear regions around buried AGN (e.g.,~\citep{Scoville2017}). That being said, there are several examples of Compton-thick AGN (even ones undetected in 2Ms X-ray images) that are nonetheless dominated by AGN-like emission in the mid-IR~\citep{Alexander2008,Bauer2010}. 


\subsubsection{Mid-IR Luminosity Functions}
\label{sec:agn_lf}


The luminosity function of AGN selected in the mid-infrared has been studied since the era of $IRAS$, when~\citet{1993ApJS...89....1R} made a local luminosity function for AGN selected at 12~$\upmu$m. In the near-infrared, some progress was made after the {\em IRAS} era with {\em 2MASS} when it was realized that a significant population of dust-reddened quasars exists~\citep{2002ASPC..284..127C,2004ApJ...607...60G}. The advent of $ISO$, however,  allowed space-based infrared surveys to attain sufficient depth to detect large numbers of highly-reddened AGN at $z>>0.1$.~\citet{2006A&A...451..443M} used AGN identified in $ISO$ surveys and $IRAS$ to make the first study of the cosmic evolution of the luminosity function for AGN selected in the mid-infrared. {\em Spitzer}, with its much faster mapping speed, was able to vastly increase the numbers of mid-infrared AGN and allow a determination of the evolution of the luminosity function up to $z \sim$4
\citep{Lacy2015}). Finally, the addition of {\em WISE} data has allowed even better constraints on the bright end of the AGN luminosity function~\citep{2018ApJ...861...37G}. 

\begin{figure}[H]
\includegraphics[width=\textwidth]{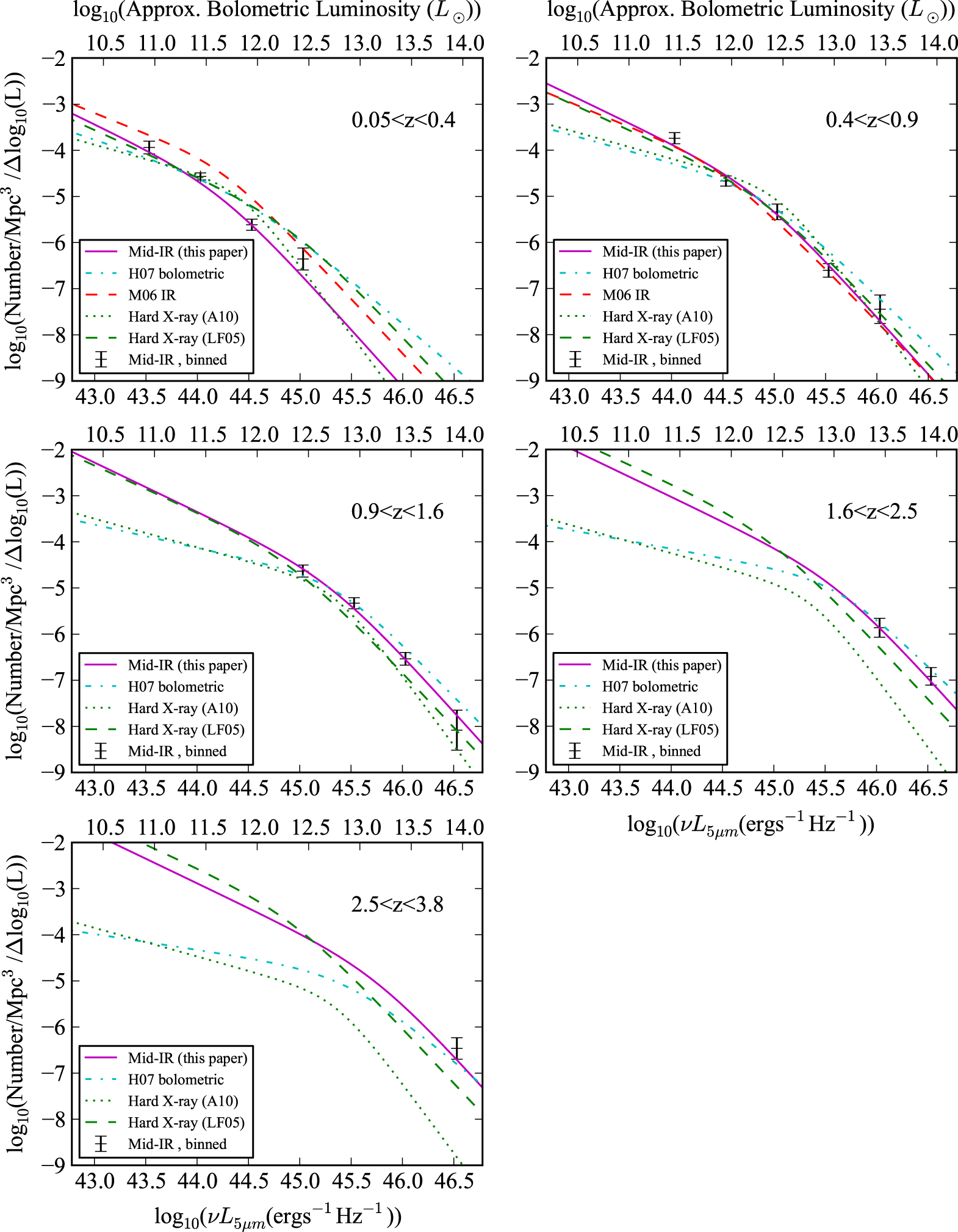}
\caption{Luminosity functions of mid-IR selected (and spectroscopically confirmed) AGN compared with estimated bolometric luminosity functions (H07;~\citep{Hopkins2007}) and luminosity functions based hard X-ray selected AGN (LF05;~\citep{LaFranca2005}) and (A10;~\citep{Aird2010}). More recent results on the mid-IR AGN LF are consistent with these at the high-L end, but show somewhat less steep slopes in the low-L regime~\citep{Runburg2022}.  Adapted from~\citet{Lacy2015}, AAS\textsuperscript{©}. Reproduced with permission. 
\label{fig:agn_lf}}
\end{figure}

Figure\,\ref{fig:agn_lf} shows the mid-IR AGN luminosity functions (LFs) from~\citet{Lacy2015} compared with the bolometric and hard X-ray LFs. It is clear that the mid-IR selected AGN come closer to the bolometric LF than those based on hard X-ray selected samples. It is worth noting that in that work the low-L end of the LF was not constrained. More recent determinations show somewhat shallower slopes below the knee at lower-z~\citep{Runburg2022}.  In the near future, {\em JWST} data will allow much better constraints to be placed on the mid-IR luminosity function at high redshifts and low luminosities. 

Indeed, such comparison to luminosity functions derived from AGN surveys at different wavelengths also allows the selection biases of AGN surveys in each waveband to be better understood.~\citet{2012MNRAS.423..464H} and~\citet{Lacy2015} compared the mid-infrared AGN luminosity functions with X-ray and optical luminosity functions from the literature.~\citet{Runburg2022} directly compared the mid-IR and X-ray-selected AGN populations in the 4.5 deg$^{2}$ XMM-LSS field.~\citet{Assef2015} compared the luminosity function of infrared-selected ``Hot DOG'' obscured AGNs in {\em WISE} with the optical quasar luminosity function. All these studies find broadly consistent results, namely that the AGN luminosity functions derived from mid-IR selected samples are higher than that for samples selected in the X-ray or optical at a given bolometric luminosity.

\subsubsection{Cosmic Black Hole Accretion Rate Density}
\label{sec:cosmic_bhard}

\begin{figure}[H]
\includegraphics[width=12.5 cm]{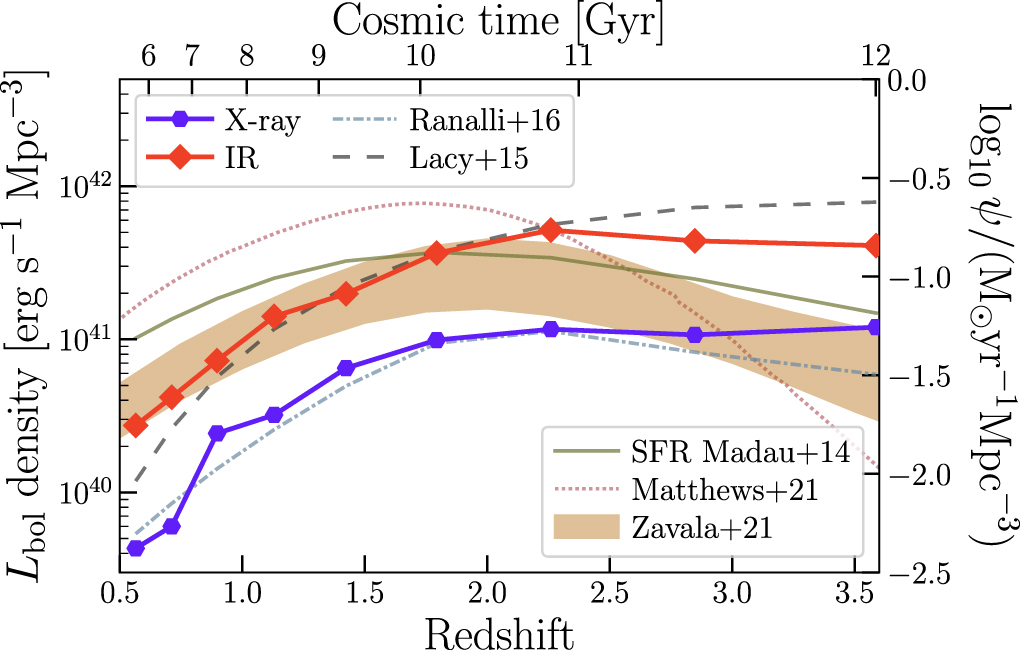}
\caption{The AGN luminosity density (proxy for the black hole accretion rate density) as derived from the IR vs. X-ray AGN luminosity functions. Note that at all redshifts, the IR selected AGN dominate the luminosity density. Morever, the shape of this function (for the IR), is similar to the more recent determination of the SFRD 
~\citep{Zavala2021} including the greater contribution of dusty galaxies at $z>2.5$. Adapted from~\citet{Runburg2022}. \label{fig:runburg_lum_density}}
\end{figure}

Integrating the luminosity functions allows us to obtain a luminosity density (e.g., Figure\,\ref{fig:runburg_lum_density}). The bolometric luminosity density is a proxy for the black hole accretion rate density through:
\begin{equation}
    \dot{M}=L_{bol}/\eta c^2
    \label{eq:accretionrate}
\end{equation}
where the efficiency, $\eta$, is estimated at $\approx0.18^{+0.12}_{-0.07}$~\citep{Lacy2015}.

Doing the above using AGN luminosity functions based on different selection methods can be used to assess what fraction of the growth of supermassive black holes occurs whilst the AGN is heavily obscured by gas and dust. Figure\,\ref{fig:runburg_lum_density} shows the AGN luminosity density derived, respectively, from the IR vs. X-ray (un-corrected for obscuration) AGN luminosity functions. It is clear that, at all redshifts, the mid-IR selected AGN dominate the luminosity density, although that is contingent on extrapolations to the uncertain faint-end of the luminosity functions. This result together with the other AGN luminosity function comparisons discussed in the previous section, are consistent with $\approx$50--70\% of accretion occurring in an obscured phase. 

Figure\,\ref{fig:runburg_lum_density} also compares the AGN luminosity density with the SFRD both as given in~\citep{MadauDickinson2014} and the more recent~\citep{Zavala2021}. It finds better agreement between the cosmic BHARD and SFRD using the latter estimate which has a relatively stronger contribution from dusty galaxies at $z>2.5$. We also refer the reader to the review of high redshift dust obscured in this Special Issue by J. Zavala and Caitlin Casey.  

\subsubsection{Mid-IR Selection of AGN in Low-Mass Systems}
\label{sec:agn_lowmass}

As discussed in Section\,\ref{sec:agn_color_selection}, the traditional mid-IR methods for finding AGN are biased against AGN that experience heavy nuclear obscuration or lower luminosity AGN that are overpowered by star-formation in the host galaxy (both issues arise for example in late stage mergers when simulations suggest black holes grow most rapidly yet obscuration peaks and the AGN activity is accompanied by a strong starburst). The mid-IR luminosity function of AGN (see Section\,\ref{sec:agn_lf}) is poorly constrained in the low-L end due to both such systematic uncertainties as well as limitations in sensitivity. The sensitivity issue will be significantly alleviated soon with the successful launch of JWST, as discussed in~\mbox{\citet{Satyapal2021}}, as well as in Section\,\ref{sec:future}. The systematic issues, however, are harder to resolve. Below, we discuss some lessons that have been learned from searches for lower luminosity AGN associated with lower mass black holes. Following the $M-\sigma$ relation such lower-mass BHs are expected to reside in low mass and therefore low metallicity galaxies.  Indeed, the goal of a better census of lower luminosity AGN is closely related to the broader issue of a census of lower mass black holes, a population crucial to our understanding of the origins of SMBHs and secular black hole growth~\citep{Greene2012,KormendyHo2013,Smethurst2021}. 

\begin{figure}[H]
\includegraphics[width=0.5\textwidth]{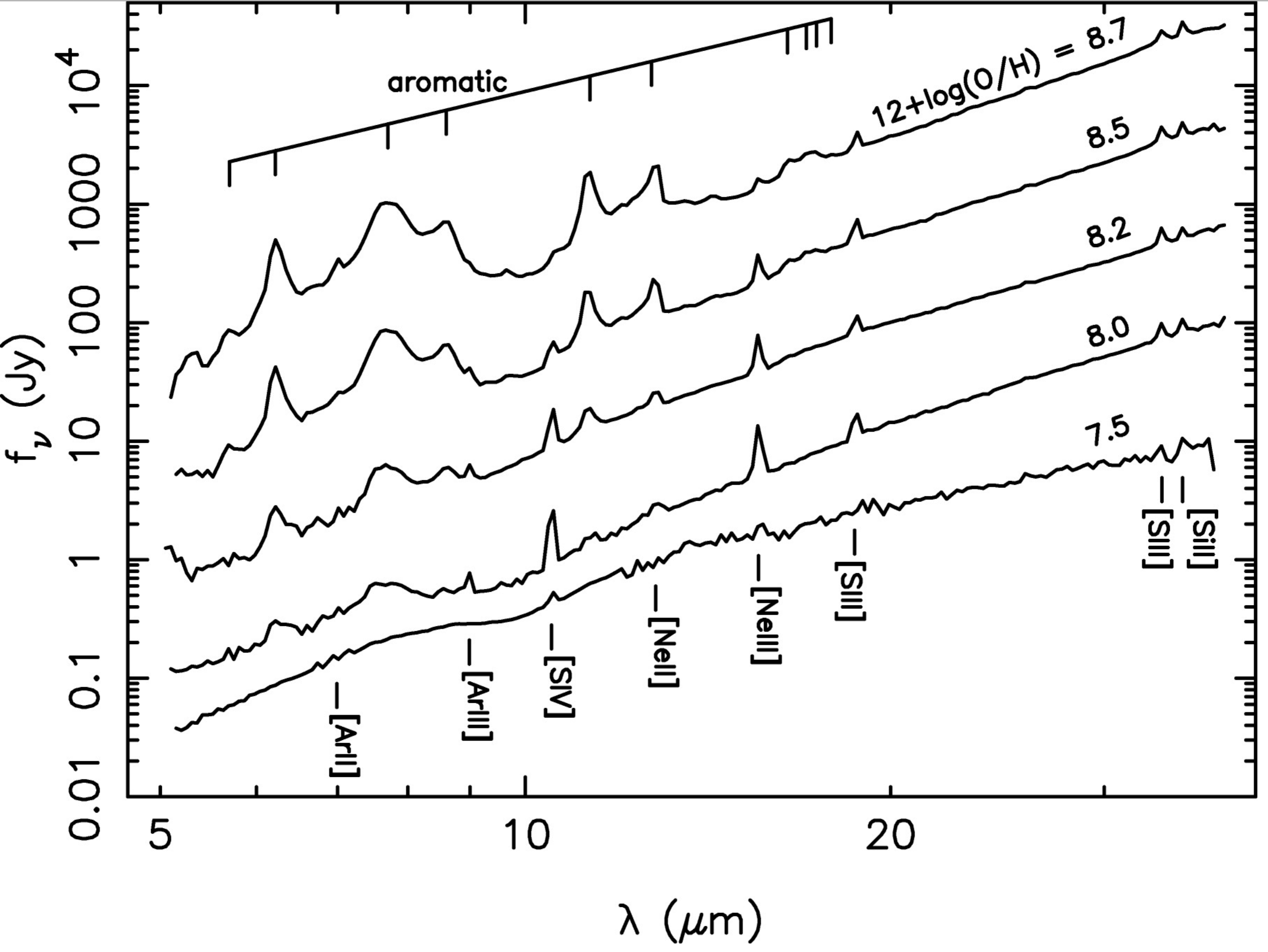}
\includegraphics[width=0.4\textwidth]{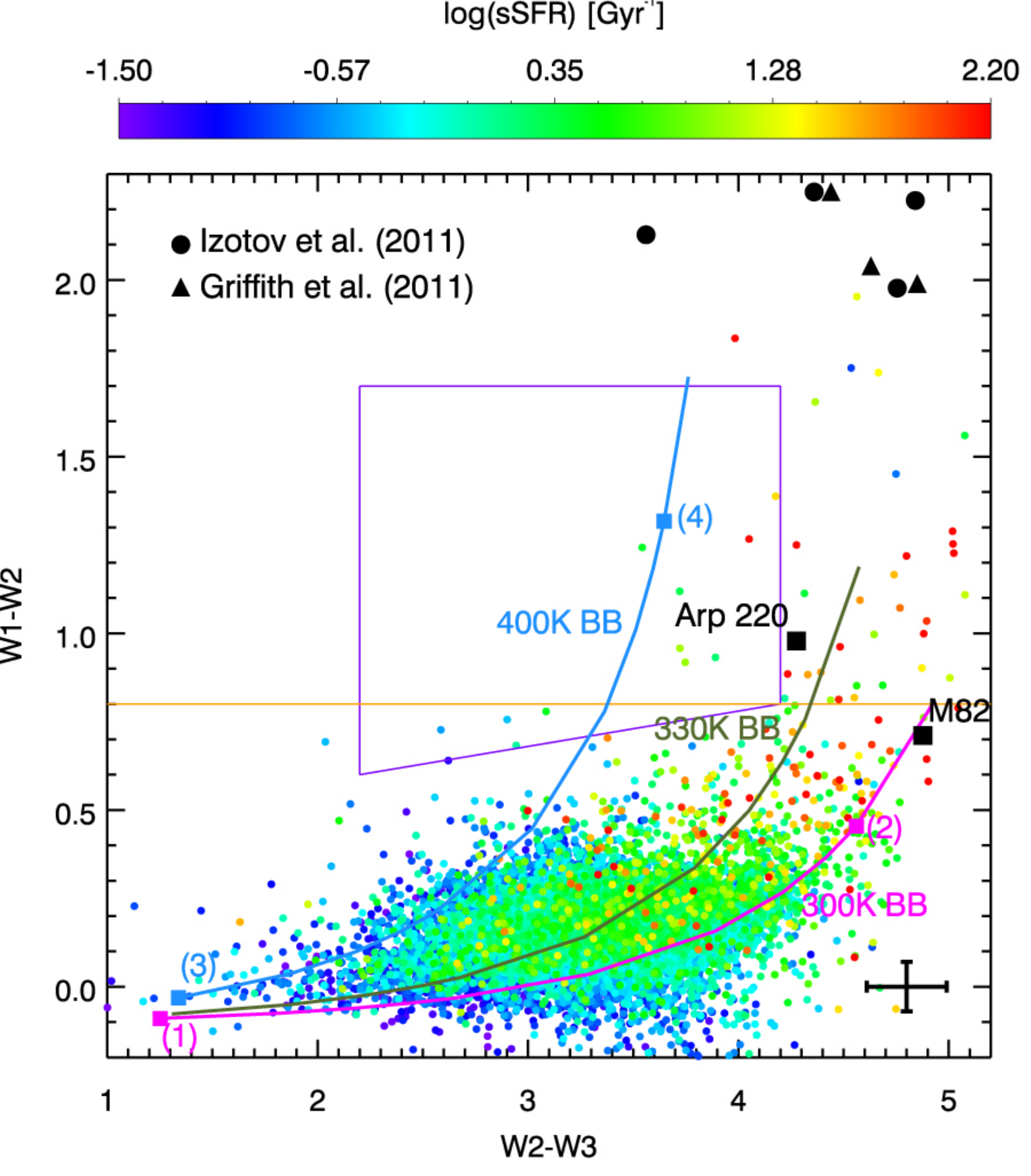}
\caption{Here,
 we highlight some issues with mid-IR AGN selection in dwarf galaxies. {\bf Left:} Average mid-IR spectra of starburst galaxies, without AGN, binned by metallicity as labeled, adapted from~\citet{Engelbracht2008}. {\bf Right:} The {\sl WISE} colors of dwarf galaxies color-coded by sSFR. The model tracks represent an elliptical template+a hot dust blackbody component of the indicated temperature. Note the bulk of the dwarf galaxies with red W1-W2 colors are consistent with intense starbursts. The few more robust candidates for AGN powered systems all lie within the purple polygon which represents the~\citet{Jarrett2011} AGN selection, adapted from~\citet{Hainline2016}, AAS\textsuperscript{©}. Reproduced with permission. 
\label{fig:missingAGN}}
\end{figure}

In Sections\,\ref{sec:observed_properties_lowres} and \ref{sec:miragn}, we discussed both PAH-based and color-based mid-IR selections. However, studies have shown that both approaches are complicated for galaxies dominated by their HII regions as might be expected in dwarf starburst galaxies, see, e.g.,~\citet{Roussel2006}. This is already alluded to in Section\,\ref{sec:observed_properties_lowres}, since HII regions emission can mimic AGN mid-IR colors and a lack of PAH features. In addition, there is a trend of weakening PAH features with lower metallicity. ~\citet{Engelbracht2008} examined the mid-IR spectra of a sample of dwarf galaxies and found a clear trend of lower-metallicity galaxies moving towards power law continuum, weak PAH mid-IR spectra without any AGN presence (see Figure\,\ref{fig:missingAGN}Left). 
 They also found that the PAH EW was correlated with line diagnostics (combining [NeIII]/[NeII] and [SIV]/[SIII]) that trace the hardness of the radiation field. These results suggest perhaps that young starbursts generate a sufficiently hard radiation field that PAH carriers are destroyed. Indeed, finding AGN in low metallicity systems is also complicated for traditional techniques such as the optical BPT diagram~\citep{Groves2006a}. Diagnostic mid-IR NLR line ratios such as [NeIII] 15.6\,$\upmu$m/[NeII]12.8\,$\upmu$m are also degenerate between AGN and low-metallicity galaxies, as seen in~\citet{Hao2005} and~\mbox{\citet{Spoon2022}}. Two studies,~\citet{Sartori2015} and~\citet{Hainline2016}, with slightly different parent sample selections both examine the consistency between optical spectroscopic and mid-IR ({\sl WISE}) color-color selection techniques for finding AGN in dwarf galaxies. Both find that there is very little overlap between the AGN selection through optical spectroscopic and mid-IR color diagnostics.~\citet{Sartori2015} finds that optical spectroscopic selection is biased toward somewhat redder and higher mass hosts whereas mid-IR color selection is biased toward blue, lower mass hosts. The latter is further confirmed by~\mbox{\citet{Hainline2016}} who find that the majority of dwarf galaxies with mid-IR AGN-like colors appear to be compact blue galaxies representing young starbursts (again the explanation is an intense enough radiation field to heat dust sufficiently to reproduce the red mid-IR colors) (see Figure\,\ref{fig:missingAGN} Right). They find that this makes the single W1-W2 color particularly unreliable whereas the two color technique of using W1--W2 vs. W2--W3~\citep{Jarrett2011} is somewhat more reliable though still suffering from significant starburst contamination. Both studies agree that this degeneracy with the observed spectra of young starbursts means that any mid-IR selected AGN in dwarf galaxies requires independent confirmation. 
 
\textls[-15]{ The mid-IR regime itself offers two potential avenues for breaking this degeneracy between AGN and young starbursts in dwarf galaxies. One is to look for the presence of high ionization potential mid-IR lines such as the [NeV]14.3\,$\upmu$m line (see Figure\,\ref{fig:fine_structure}), which are strong indicators of the presence of an AGN as these lines are largely obscuration independent and cannot be produced by any known stellar-associated phenomenon, (e.g.,~\mbox{\citep{Armus2007,Veilleux2009,Satyapal2021}}). The [NeV] line has been used, for example, in finding AGN in bulgeless spiral galaxies without an optical spectroscopic indication of an AGN~\citep{Satyapal2008}.~\mbox{\citet{Richardson2022}}} recently found that mid-IR emission line diagnostics are less sensitive to various degeneracies between physical parameters than the traditional BPT emission line diagnostics, and therefore can serve as a good means of finding AGN (really IMBHs) within dwarf galaxies. Specifically, they find that the [OIV]\,25.9\,$\upmu$m/[NeIII]\,15.6\,$\upmu$m is a good AGN diagnostics either alone or in combination with [SIV]\,10.5/[ArIII]\,6.99\,$\upmu$m. 

The other approach to finding AGN in dwarf galaxies is mid-IR variability which we discuss in more detail in Section\,\ref{sec:variability}. Recently,~\citet{Secrest2020} showed that, similar to optical variability studies, mid-IR variability identified AGN decrease significantly with stellar mass. It is worth keeping in mind though that this may be an indication of a genuine lack of such AGN or that the variability properties of said AGN differ sufficiently from those of their higher mass counterparts that existing studies fail to find them. 

Finally, we will digress slightly from the topic of AGN to discuss the role of mid-IR studies in finding and studying lower mass black holes through looking for tidal disruption events (TDEs). TDEs are biased towards intermediate mass black holes since, to be observable, they require the tidal disruption radius to be outside the event horizon which for typical stellar masses favors $M_{BH}<10^8$M$_{\odot}$ (for a review see~\citep{Komossa2015}).  Mid-IR flares have been observed associated with optically-detected TDE candidates. For example,~\mbox{\citet{Jiang2017}} report on a mid-IR flare associated with a dust echo from a TDE candidate in a Seyfert 1 dwarf galaxy with a known $10^6$\,M$_{\odot}$ black hole. Interestingly, this mid-IR flare happened 11 days {\it before} the optical flare detected by ASAS-SN which classified this event as a TDE. In other cases, the mid-IR flare has happened years {\it after} the optical flare~\citep{Dou2017}. The relative association between the optical and mid-IR variability would be a great way to better understand these events and through them the overall population of lower mass black holes.    

\section{Constraints on the Physical Properties of the AGN}
\label{sec:physical_properties}

This section summarizes key results in how one uses mid-IR spectroscopy, spatially resolved and time-resolved observations of AGN to gain further insight into the black hole mass, accretion rate and the obscuring dust structures. These results provide us a deeper understanding both the nature of mid-IR selected AGN and their demographics as discussed in the previous section, as well as the interaction of the AGN with its host galaxy as discussed in the following section.  

\subsection{Constraints on Black Hole Accretion Rate and Mass}
\label{sec:mass_accretionrate}

The AGN luminosity is powered by accretion onto a supermassive black hole, and therefore there is a direct relationship between the bolometric luminosity and the accretion rate, $\dot{M}$, as given by Equation\,\eqref{eq:accretionrate}. The two necessary steps in deriving the accretion rate from mid-IR observations are: (a) converting from a mid-IR continuum luminosity to the bolometric AGN luminosity, and (b) assuming an efficiency, $\eta$. The first requires adopted a bolometric correction, BC, which depends on our knowledge of the entire AGN SED (see~\citep{Lacy2015}). For a review of the infrared SEDs of AGN, see Lyu and Rieke in this Special Issue. This infrared regime is by far the dominant component for obscured AGN, but is also significant for unobscured, Type 1 AGN~\citep{Richards2006}.  Updated bolometric corrections, including from the mid-IR (e.g., the 15\,$\upmu$m continuum), are presented in~\citet{Shen2020}. On the issue of the adopted efficiency, this parameter is likely to decrease significantly from standard adopted values (around $\eta\approx0.1$) when we consider low accretion rate systems (see, e.g.,~\citet{Trump2011}).~\citet{Trump2011} also present a comprehensive discussion on how the specific accretion rate (parametrized by $L_{bol}/L_{Edd}$\endnote{This is also called the Eddington ratio and can be written as $\lambda_{Edd}$. The Eddington luminosity, $L_{Edd}$, is a function of the black hole mass only.}) governs the physical structures surrounding the black hole including the appearance of the BLR and dusty torus, neither of which structures are expected to build-up at the lowest accretion rates. This is consistent with observations that the frequency of hot-dust-deficient quasars increases with smaller $L_{bol}/L_{Edd}$~\citep{Lyu2017_dustdeficientquasars}. This is also consistent with the observations of \linebreak \citet{Ogle2006} that roughly half of powerful radio galaxies are mid-IR bright suggesting obscured quasar nuclei, whereas the other half are mid-IR weak consistent with low accretion rate, jet-dominated, systems.   

\begin{figure}[H]
\includegraphics[width=12.5 cm]{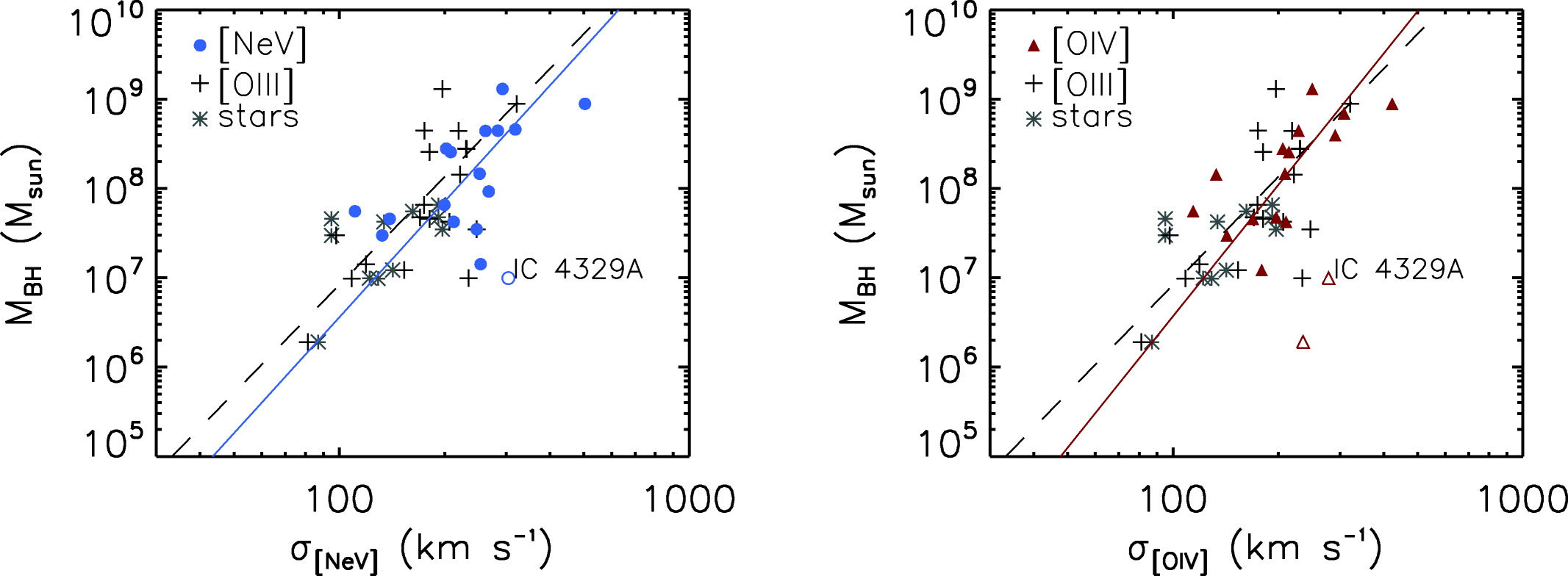}
\caption{The relationships between reverberation mapping determined black hole masses and mid-IR line velocity dispersions where we have [NeV]\,14.3\,$\upmu$m on the left and [OIV]\,26 $\upmu$m on the right.  Adapted from~\citet{Dasyra2008}, AAS\textsuperscript{©}. Reproduced with permission.  \label{fig:dasyra2008}}
\end{figure}

Deriving the Eddington ratio requires knowledge of $M_{BH}$. In principle one can do the reverse and adopt an Eddington ratio of say 1 and use that to derive a lower limit for $M_{BH}$. We can do better using the $M_{BH}-\sigma$ relation for AGN. The original relation is based on a relationship between the mass of the black hole and of the velocity dispersion of the stars based on stellar photospheric absorption lines~\citep{FerrareseMerritt2000}. A similar relationship, though with greater scatter, exists using the velocity dispersion of the [OIII]\,5007\AA\ line, associated with the narrow line region, NLR~\citep{GreeneHo2005}. The velocity dispersion of this NLR gas is a proxy for that of the bulge in galaxies so this relationship is to be expected as a consequence of the $M_{BH}-M_{bulge}$ relation.~\citet{Dasyra2008} expanded this work by using a sample of local AGN with reverberation mapping determined black hole masses and looked for relationships between these masses and both the velocity dispersion and luminosity of AGN-powered mid-IR fine structure lines which, as discussed in Section\,\ref{sec:observed_properties_highres}, also probe the NLR. Figure\,\ref{fig:dasyra2008}, shows the relationships between $M_{BH}$ and the two most prominent AGN mid-IR lines [NeV] and [OIV]. These are compared with similar relationships using stellar absorption lines and the optical [OIII]\,5007\AA\ line. The key conclusion is that the velocity dispersions of these lines do indeed correlate with the black hole masses and that the dispersion observed is comparable with that seen when using the optical NLR lines. This result is crucial in terms of our ability to use mid-IR spectroscopy to derive black hole masses, even in heavily obscured systems where the optical lines are harder to observe. 

\subsection{Spatially Resolved Mid-IR Imaging: Our Evolving View of the Obscuring `Torus'} \label{sec:highresimaging}

As discussed in~\citet{Hickox&Alexander2018_review}, obscuration in AGN can occur on multiple scales. They point to three particular sources of obscuration: (1) the nuclear molecular dusty torus (tens of pc scale, e.g.,~\citep{Garcia-Burillo2019}), (2) a circumnuclear starburst (100s of pc scales), and (3) the host galaxy itself (kpc scales).  We refer the reader to that review for more detailed discussion. In this section, we briefly summarize how spatially resolved mid-IR imaging (supported by simulations) has evolved our understanding of the obscuring nuclear torus. Obscuration on larger scales is addressed in Section\,\ref{sec:agn_host}.  

The AGN unification model suggests obscuration in AGN is due to an axisymmetric small scale gas and dust structure (the ``torus'') which obscures the accretion disk and BLR, for a review see~\citet{Netzer2015}. A pc-scale such structure has been directly observed using near-IR interferometry~\citep{Jaffe2004}. More recently, high spatial resolution studies of the molecular dusty tori of local AGN with the Atacama Large Millimeter Array (ALMA) have shown these to extend to 10s of pc~\citep{Combes2019,Garcia-Burillo2021}. The inner-edge of the torus corresponds to the dust sublimation temperature, and thus scales with luminosity as $r\propto L^{1/2}$ while the outer edge likely corresponds to the gravitational sphere of influence of the black hole or $\approx$10~pc for nearby Seyferts~\citep{Hickox&Alexander2018_review}. This nuclear torus is believed to be ``clumpy'', i.e., comprised of distinct optically-thick clouds as opposed to a smooth structure. The key evidence for this is that the mid-IR emission is not showing strong orientation dependence such as the tight correlation between the X-ray and mid-IR emission which is consistent between Seyfert 1s and Seyfert 2s~\citep{Lutz2004}. The covering factor of the tori shows a strong dependence on the Eddington ratio ($L_{bol}/L_{Edd}$) which shows that indeed most of the obscuration (at least in local AGN) takes place within the gravitational sphere of influence of the black hole (roughly 10pc).  

Recent observational and theoretical work has evolved the torus picture from a static structure to a much more dynamic one where a complex interplay of gas inflow and outflow produces an axisymmetric structure that mimics a lot of the torus characteristics but is certainly changeable in time. This leads, for example, to the existence of changing look AGN, likely driven by changes in accretion rate, as seen in recent mid-IR studies~\citep{Sheng2017,Lyu2022}. 

\begin{figure}[H]
\includegraphics[width=10.5cm]{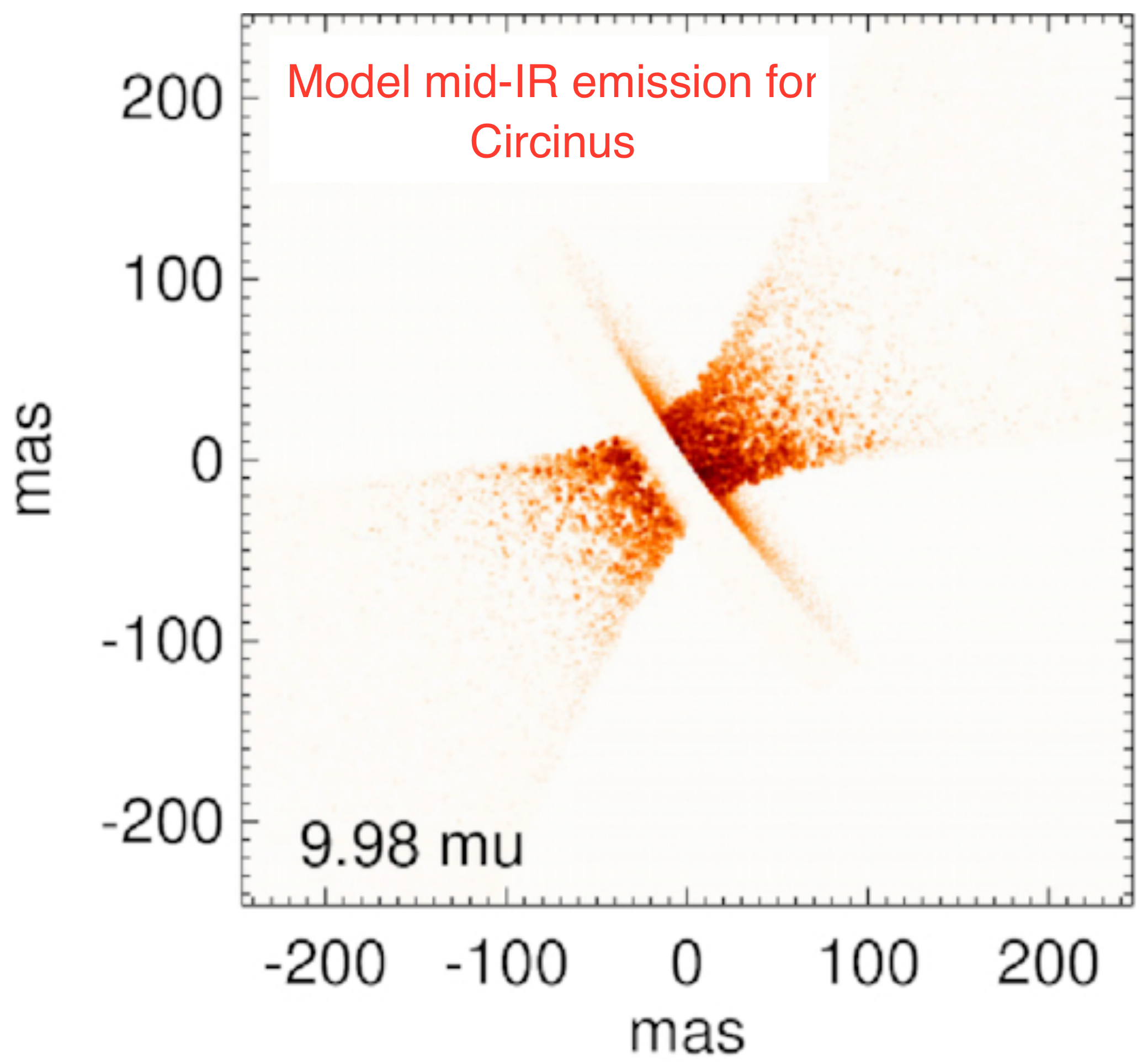}
\caption{A model for the mid-IR emission of the Circinus galaxy constrained by both mid-IR pc-scale interferometric observations with VLTI/MIDI as well as the global SED. This is the best-fit model which is a thin disk, consisting of a mixture of graphites and silicates (accounting for the absorption seen in this $\sim$10\,$\upmu$m model), plus a hollow hyperboloid, consisting of predominantly graphites.  Adapted from~\citet{Stalevski2019}. This model is consistent with the very recent VLTI/MATISSE high-resolution imaging of Circinus presented in~\citet{Isbell2022}. \label{fig:circinus_model}}
\end{figure}

Mid-IR interferometric observations have surprisingly revealed that significant mid-IR emission can arise from dust not in the ``torus'' structure but perpendicular to it \linebreak (e.g.,~\citep{Asmus2016,Asmus2019,Stalevski2019,Alonso-Herrero2021,Isbell2022}), as shown in Figure\,\ref{fig:circinus_model}. This polar structure, extending 10s to $\sim$100s of pc, is attributed to polar dusty winds. This
is consistent with the AGN-driven winds which include warm ionized gas
as well as cool neutral-atomic and molecular gas~\citep{Veilleux2013,Cicone2014}. This dusty wind is likely radiation pressure driven as argued in~\citet{Leftley2019}. AGN outflows are discussed in much more detail in the recent review of~\citet{Veilleux2020}.  

\begin{figure}[H]
\includegraphics[width=11.5cm]{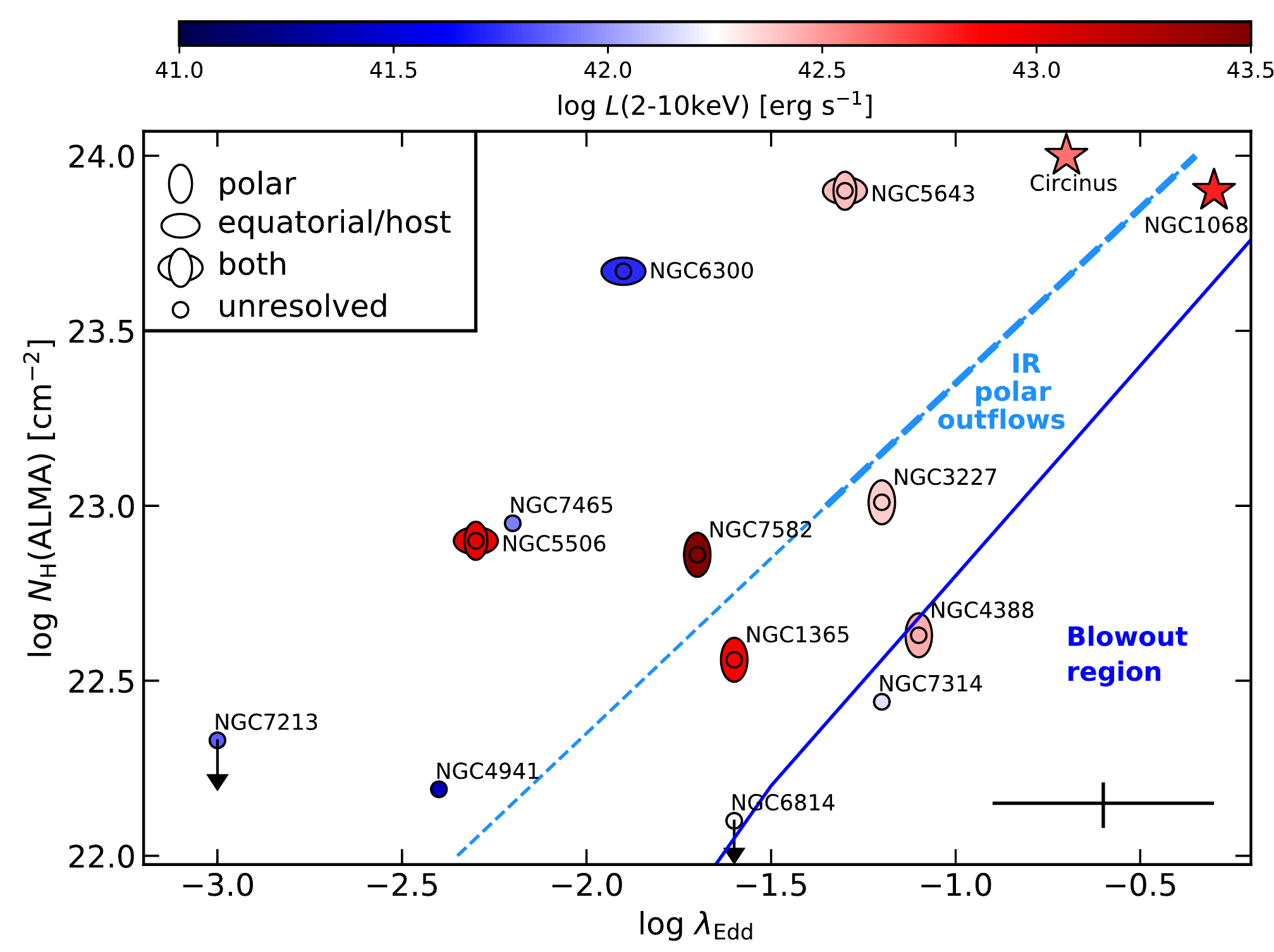}
\caption{An illustration on the dependence of mid-IR continuum morphology (as presented in the symbols) on the Eddington ratio and column density for a sample of local Seyferts. To the right of the solid line is the region where the AGN wind is able to blowout all surrounding material. To the right of the dashed line is where IR polar outflows are expected to develop, consistent with the observations. Both Circinus and NGC1068 also show polar dust emission, although not part of this particular study which is why they are given different symbols. Reproduced with permission from~\citet{Alonso-Herrero2021}, ESO\textsuperscript{©}.\label{fig:polar_dust}}
\end{figure}

Both observations and simulations suggest that such polar dust emission is likely most prevalent in systems with high Eddington ratios and/or high AGN luminosities~\citep{Alonso-Herrero2021}, see Figure\,\ref{fig:polar_dust}. Indeed, it appears that once selection effects are taken into account such polar dust is both ubiquitous in and potentially accounts for $>$50\% of the mid-IR emission of high Eddington ratio AGN~\citep{Asmus2016,Asmus2019}. This is leading us in a direction where we need to use torus+wind models to understand the nuclear dust distribution around AGN, as shown, for example, in Figure\,\ref{fig:circinus_model}. Recent observations by~\citet{Prieto2021} also suggest that, at least in some cases, the obscuring dust may actually take the form of pc-scale high opacity dusty filaments which in Type 2 AGN cross the line-of-sight to and obscure the central source and in Type 1 do not, or are of insufficient opacity to fully obscure it. Such filaments are reminiscent of the molecular gas filaments seen in the center of our own Galaxy. Such filaments are also consistent with the view of the torus as a dynamic structure arising from the interplay of gas inflow and outflow. This view is also supported by recent high resolution ALMA images of molecular gas in the nucleus of NGC1068
\citep{Imanishi2018,Garcia-Burillo2019}. 

\subsection{Mid-IR Variability as a Probe of the Obscuring Dust Structures} \label{sec:variability}

As discussed already in Section\,\ref{sec:agn_lowmass}, mid-IR variability is one of the means of selecting AGN (e.g.,~\citep{Garcia-Gonzalez2015,Polimera2018}). The primary advantage of selecting AGN using mid-IR variability as opposed to color-based selections, as discussed in Section\,\ref{sec:agn_color_selection}, is that this is much less biased against systems where the host galaxy's mid-IR emission dominates over the AGN emission.~\citet{Polimera2018} used mid-IR variability to select AGN, finding that roughly 1\% of their $z \sim$ 3 galaxies host an AGN. This selection method, however, is also subject to large fractions of false positives (at the level of 20--25\% as given in~\citet{Polimera2018}).  

\begin{figure}[H]
\includegraphics[width=6 cm]{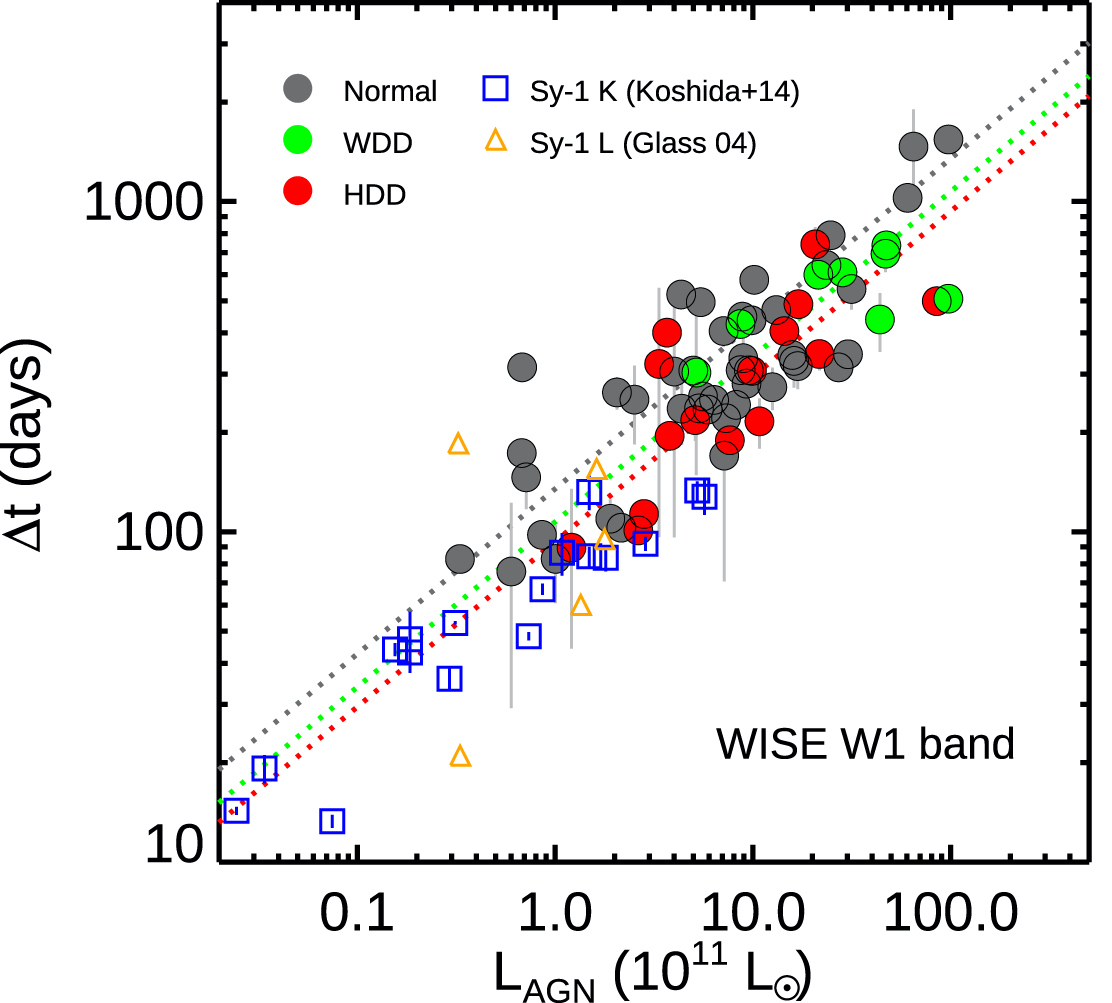}
\includegraphics[width=6.5 cm]{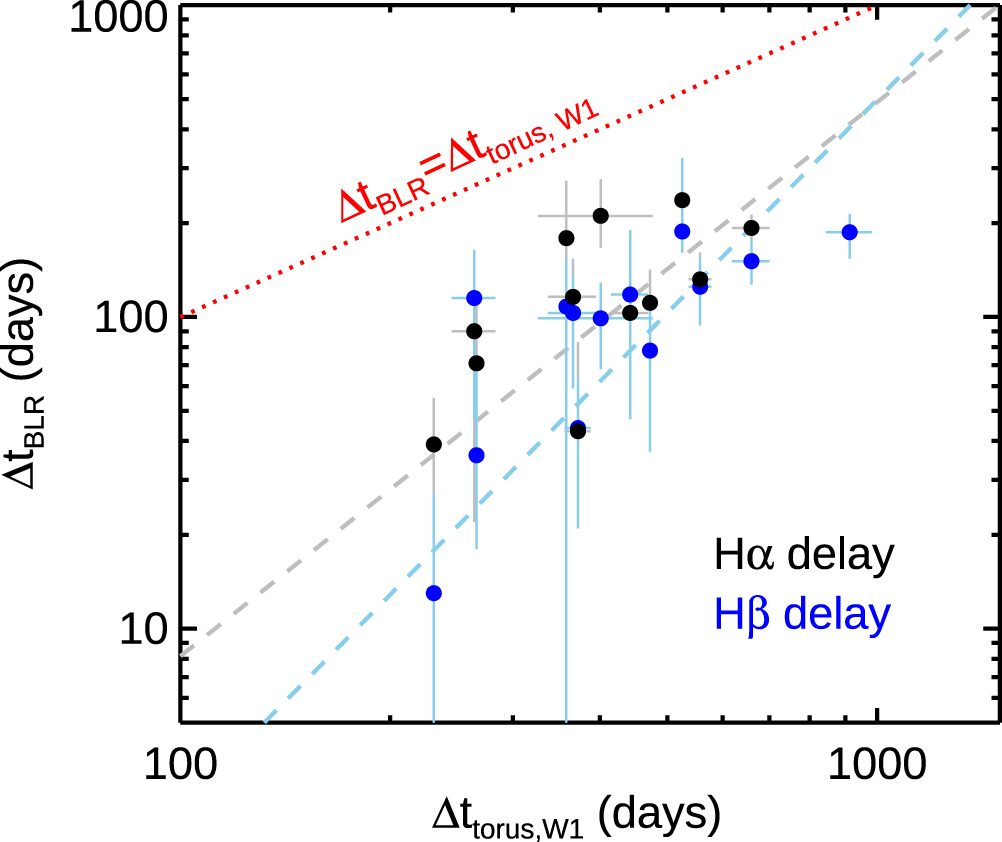}
\caption{{\bf Left}: The optical-to-mid-IR time-lag ($\Delta t$) scales with the AGN luminosity as $\Delta t\propto L_{AGN}^{0.5}$. Here, the filled symbols correspond to WISE 3.4$\upmu$m light curves for quasars, whereas the open symbols correspond to literature measurements in the K-band for Seyferts. The measured lag implies a size of the $\lambda_{rest}\approx$3--4 $\upmu$m dust emitting region to be $\approx0.3$\,pc for an $L_{AGN}\sim10^{12}$L$_{\odot}$ quasar. {\bf Right}: Comparison with the time lags to the BLR suggests that this mid-IR emitting dust is roughly 4$\times$ further away than the radius of the BLR. Reproduced with permission from~\citet{Lyu2019}, AAS\textsuperscript{©}. \label{fig:lyu2019}}
\end{figure}

Beyond its use as an AGN-selection tool, mid-IR variability can be a powerful tool in studying the structure of the torus. For example, just as the time-lags between the accretion disk continuum and the broad Balmer lines gives us a handle on the size of the BLR~\citep{Peterson1993,Peterson2004}, time lags between optical and mid-IR continuum light curves give us a handle on the size of the dust emitting region. The recent study of~\citet{Lyu2019} combines {\sl WISE} data with ground-based optical transient surveys (e.g., ASAS-SN) to do just that for 87 Palomar Green quasars (all at $z<0.5$). They find that the vast majority, 77\% of their sample, show clear dust reverberation signatures with dust light curves mimicking optical light curves but with a time lag corresponding to the light travel time to the dust emitting region. Figure\,\ref{fig:lyu2019} shows their results which have the expected scaling with the AGN luminosity ($\Delta t\propto r\propto L^{1/2}$) and allow them to infer relatively compact sizes to the 3--4~$\upmu$m dust emitting regions ($\approx$0.3 pc for a $L_{AGN}\sim10^{12}$L$_{\odot}$ AGN) with that region being just outside the BLR (Figure\,\ref{fig:lyu2019} {Right}). This dust emitting region is found to be smaller in dust deficient quasars (DD in their nomenclature) compared to the majority of quasars, see also~\citep{Lyu2017_dustdeficientquasars}. The minority of quasars that do not show such clear dust reverberation behavior correspond to radio loud objects seen relatively face-on (as in flat spectrum radio quasars) where much of the mid-IR emission comes from the jet and whose variability is therefore uncorrelated with the accretion disk variability.  

The mid-IR variability amplitude changes with both wavelength and rest timescale. \citet{Kozlowski2016} find that the level of mid-IR variability increases with rest-frame timelag. Typical amplitudes are small, $\approx$0.1 mag by about 1 year timescales. They also find that the mid-IR variability amplitude is smaller than the optical one on scales smaller than a few years, likely the result of processes that dampen shorter timescale variability more easily than longer timescale variability as the information re changing conditions such as changing accretion rates propagates out to the dust emitting regions.~\citet{Lyu2019} also find that the variability amplitude falls steeply with wavelength as expected due to the accretion disk variability signal (e.g., due to changes in accretion rate) attenuating as it travels to relatively cooler dust emitting regions that correspond to larger structures. Consistently,~\citet{Isbell2022} comment that their MATISSE high resolution in the N-band ($\approx$10\,$\upmu$m) images of Circinus do not show any significant change from the earlier MIDI images of the same galaxy, taken 6--7 years earlier. These observations together suggest that using mid-IR variability to select AGN works best when using multi-year data at relatively short wavelengths (i.e., 3--4\,$\upmu$m).

\section{Impact of AGN on the Host Galaxy and Galaxy-BH Co-Evolution}
\label{sec:agn_host}

In this section, we discuss how mid-IR spectroscopic observations give us some insights into what is the impact of AGN on their host galaxies, especially in the heavily dust obscured phase. This question also directly impacts the broader question of what processes are at play that explain the apparent co-evolution of supermassive black holes and their host galaxies. A review of AGN feeding and feedback is presented in the Vivian U review ``The Role of AGN in Infrared Galaxies from the Multiwavelength Perspective'' in this Special Issue. We also include a discussion on how we disentangle the role of AGN and star-formation in powering the IR emission of IR-luminous galaxies. 

\subsection{Gas Properties around AGN}
\label{sec:ism}

Characterizing the physical properties of the ISM in AGN host galaxies has two goals. One goal is to better understand what kinds of galaxies are more likely to host AGN. For example, multiple studies have examined the mass-metallicity relation for AGN-host galaxies
\citep{Groves2006a,Matsuoka2009,Matsuoka2018,Thomas2019} and generally find that AGN are found preferentially in high mass, high metallicity environments.~\citet{Thomas2019} in particular find a systematic offset (specifically, a 0.09\,dex increase in oxygen abundance) in the AGN mass-metallicity relation relative to the star-forming galaxy MZR.  This high metallicity appears to already be in place by $z \sim$4~\citep{Matsuoka2009} consistent with the idea of very rapid early enrichment. In principle, it is ideal to simultaneously model the black hole accretion history, star-formation history and related chemical enrichment accounting for both gas-phase metallicity and the metals locked in dust and including effects of gas inflow/outflow, e.g., as in~\citet{Valiante2011}. However, there is also an underlying issue, especially when considering cosmic noon AGN that reside in dust obscured galaxies. The above studies all use metallicities based on UV and optical emission lines that are subject to significant dust obscuration. Using mid-IR diagnostic lines would be preferable as these are less affected by dust obscuration and also much less dependent on the uncertain electron temperature than UV/optical lines~\citep{Nagao2011,Fernandez-Ontiveros2021}. Some key mid-IR diagnostics are the [NeIII]15\,$\upmu$m/[NeII]\,12\,$\upmu$m ratio which is a sensitive probe of both the ionization parameter and ISM pressure, provided the metallicity is known~\citep{kewley2019}. Through its sensitivity to the ionization parameter, it has also been used as an AGN indicator (e.g.,~\citep{Armus2007}). The same ratio divided by the hydrogen Pfund-$\alpha$ line at 7.46\,$\upmu$m is an ionization parameter independent metallicity diagnostic~\citep{kewley2019,Fernandez-Ontiveros2021}.

The second goal is to study the impact of an AGN on the molecular gas properties of their host galaxies, a key question in the quest to better understand BH-galaxy co-evolution. There are two potential impacts. One is through driving an outflow which removes gas, including molecular gas from the system (e.g.,~\citep{Cicone2014}). The other is through exciting the molecular gas, making it at least temporarily unavailable to star-formation. The recent literature is still somewhat divided on the topic of the impact of AGN on the molecular gas reservoir of their hosts. For example,~\citet{Valenetino2021} examine the ALMA CO properties of both mid-IR and X-ray AGN and find no significant effect in either amount of molecular gas or its excitation. On the other hand,~\citet{Circosta2021} find that, controlling for other parameters such as stellar mass and star-formation rate, AGN hosts have lower CO 3-2 luminosities suggestive of either gas removal or excitation to higher rotation states. Indeed,~\citet{Mashian2015} find that AGN are more likely than pure starbursts to show J > 14 CO transitions, suggesting the CO gas in AGN hosts is highly excited. This study follows on the earlier~\citet{vanderWerf2010} study of Mrk231 who find that while J < 8 transitions are dominated by starburst-associated excitation, the higher J transitions require an AGN, specifically CO excitation through the X-rays produced in said AGN. Indeed, X-ray dominated regions (XDR) can be far more efficient in heating/exciting the molecular gas, especially at high densities, due to their ability to penetrate much deeper into dense molecular clouds than UV photons, see~\citet{Maloney1996} and~\citet{vanderWerf2010}.  

\begin{figure}[H]
\includegraphics[width=10.5 cm]{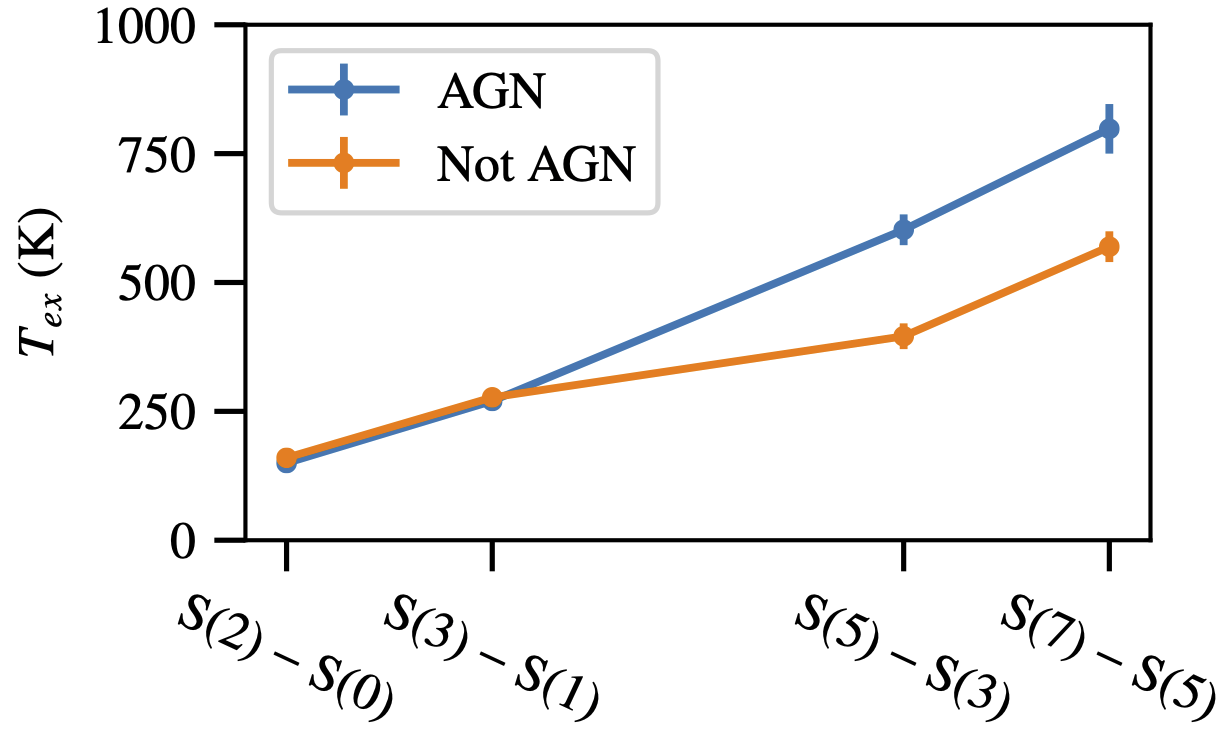}
\caption{The $H_2$ excitation temperature for mid-IR AGN vs. non-AGN calculated from each pair of mid-IR ro-vibrational transitions, as labeled. This result highlights that the presence of an AGN directly affects the ISM of its host galaxy. Adapted from~\citet{Lambrides2019}.\label{fig:effects_ism}}
\end{figure}

In the mid-IR, we observe ro-vibrational transitions of $H_2$, where the lower level transitions are labeled on the spectrum shown in Figure\,\ref{fig:intro}c. These transitions are excited radiatively or collisionally in PDRs (i.e., associated with star-formation), shocks or again XDRs~\citep{Maloney1996}. Shocks are expected as a result of supernovae, mergers, or radio jets. For~example,~\citet{Ogle2006} find significant $H_2$-PAH excess for radio galaxies over normal star-forming galaxies, consistent with jet-driven shocks. Fundamentally, the bulk of the $H_2$ that resides in cold molecular clouds is essentially invisible to us, but the above conditions can excite its mid-IR ro-vibrational transitions which in turn serve as a good temperature gauge for warm gas. 
\citet{Lambrides2019} analyze the mid-IR spectra of $\approx$2000 nearby galaxies (shown in Figure\,\ref{fig:mirseds}), separating out the AGN (where the mid-IR AGN fraction is $>$50\%), from the non-AGN. This allows them to study the dust and gas properties of AGN host galaxies vs. non-AGN. They explore diagnostic PAH ratios and find that in AGN host galaxies, the PAH have a wider range of sizes and fractional ionization than non-AGN galaxies. This suggests the AGN affects the properties of the PAH in their galaxies though not in a single direction.  Similar to the results of~\citet{Ogle2006} for radio galaxies, they find higher $H_2S(3)$/PAH11.3\,$\upmu$m ratios for the AGN-dominated vs. star-formation dominated systems. In addition, similar to the CO SLED results discussed above, they find stronger higher level $H_2$ transitions relative to lower level ones in AGN vs. non-AGN. This is illustrated in Figure\,\ref{fig:effects_ism} which shows the excitation temperature calculated based on different pairs of $H_2$ mid-IR transitions. The values calculated from the higher level transitions are higher for AGN than for non-AGN, potentially due to the influence of AGN-powered XDR.  

All of the above examples, point to the fact that the presence of an AGN directly affects the host galaxy's ISM, in particular molecular gas. Fundamentally, the degree to which AGN feedback takes the form of excitation vs. gas expulsion has important implications for the the nature of how an AGN can serve to regulate the star-formation within its host galaxy, as well as the ability of the same galaxy to re-trigger an AGN. Excitation might be more consistent with a picture of stochastic AGN activity providing long-term regulation on the level of star-formation in the host. Gas expulsion is more like the AGN-driven ``blowout'' following strong starburst and AGN peaks that arise in the coalescent stages of major mergers. Such blowouts result in permanently quenched ``red and dead'' ellipticals.  

\subsection{Dust Properties around AGN}
\label{sec:dust}

The mid-IR regime also provides strong evidence that an AGN affects its host galaxy's dust. As already brought up in Section\,\ref{sec:ism}, an AGN appears to affect both the size and ionization fraction of the PAH~\citep{Smith2007,Lambrides2019,Garcia-Bernete2022}. There may also be PAH carrier destruction as well but this appears to be localized to the nuclear regions around the AGN rather than the extended galaxies and affects shorter wavelength PAH features more than longer wavelength ones such as 11.3\,$\upmu$m (e.g.,~\citep{LeFloch2001,Alonso-Herrero2014}). 

\begin{figure}[H]
\includegraphics[width=13.5cm]{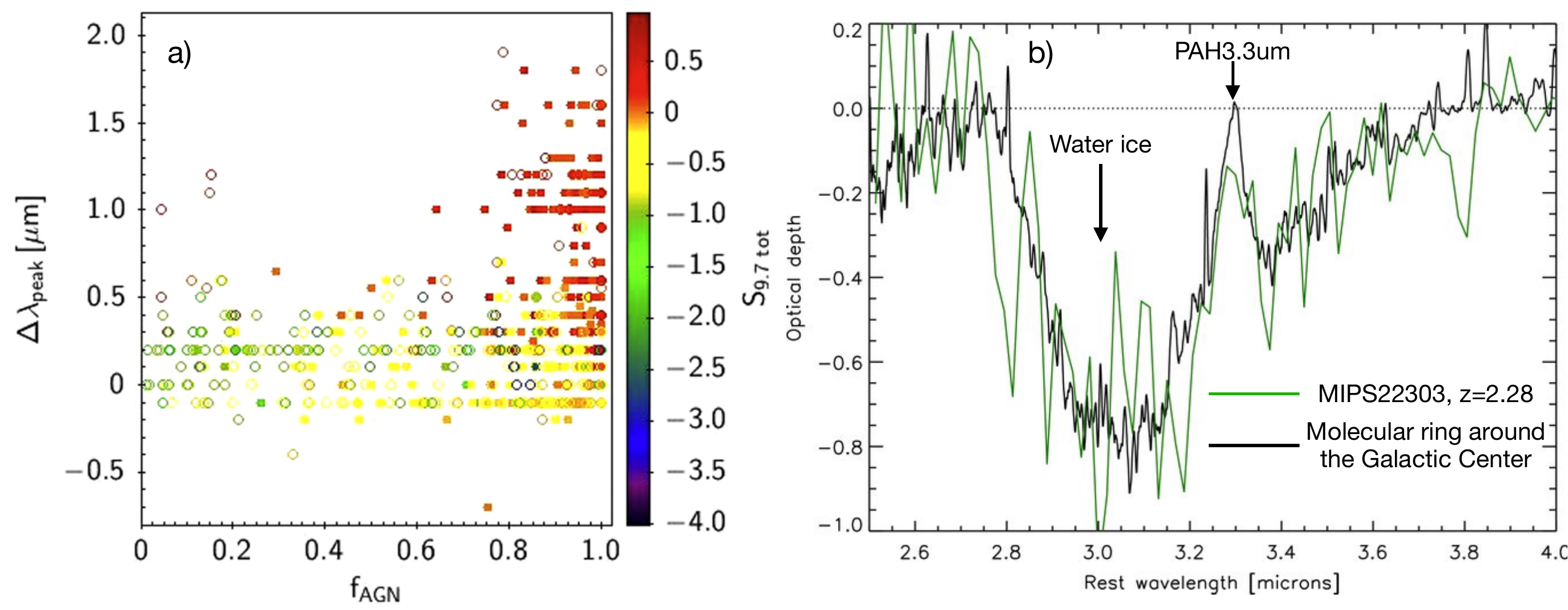}
\caption{({\bf a}) The 
 shift (relative to the canonical 9.7\,$\upmu$m) in the peak of the mid-IR silicate feature, see, e.g., Figure\,\ref{fig:intro}d, is only seen in objects with high mid-IR AGN fractions, here $f_{AGN}$, and silicates in emission, here red points. Adapted from~\citet{Hatziminaoglou2015}. ({\bf b}) Sources with deep silicate absorption also show other mid-IR absorption features such as the 3.0\,$\upmu$m water ice absorption with the 3.3\,$\upmu$m PAH feature often seen within that. Here, we show this behavior through the {\sl Spitzer}-IRS spectrum of a cosmic noon deep silicate absorption AGN compared with the {\sl ISO}-SWS spectrum of part of the molecular ring around the Milky Way Galactic Center. Adapted with permission from~\citet{Sajina2009} AAS\textsuperscript{©}. \label{fig:silicates_ices}}
\end{figure}

One of the most prominent features in the mid-IR is the silicate 9.7\,$\upmu$m absorption feature. The depth of this feature relative to the optical magnitude extinction ($A_V/\tau_{9.7}$) is a sensitive probe of the typical sizes of dust grains~\citep{Lyu2014,Shao2017}. In particular, while $A_V/\tau_{9.7}\approx18$ in the Galactic ISM, it is observed to be $A_V/\tau_{9.7}\approx5.5$ in AGN, as shown in~\citet{Lyu2014} and~\citet{Shao2017}. This decrease is an indication of typically larger dust grains consistent with the destruction of smaller grains in the harsher radiation fields around AGN, see also~\mbox{\citet{Xie2017}}. This is also supported by the centroid of the silicate feature in emission shifting to $>10$\,$\upmu$m in AGN, as seen for example in ~\citet{Hatziminaoglou2015}, see Figure\,\ref{fig:silicates_ices}a. Such a shift can be another indication of larger grain sizes. 
Spatially resolved studies reveal a more complex picture with significant differences in the $A_{9.7}/A_V$ between the toroidal or disky and polar components around AGN~\citep{Stalevski2019}. In heavily obscured AGN, mid-IR spectra allow us to study even the ices that are inferred to reside on the surfaces of these large silicate grains---these ices include water, carbon monoxide, carbon dioxide~\citep{Spoon2004,Sajina2009}. Ref.~\citep{Imanishi2008} examined the rest-frame 3--4\,$\upmu$m spectra of local ULIRGS, based on {\sl AKARI}~\citep{Matsuhara2006} data, and found surprisingly weak 3.3\,$\upmu$m PAH emission with very frequent absorption features due to water ice at 3.0\,$\upmu$m as well as carbonates dust at 3.4\,$\upmu$m. They argue that this is consistent with ubiquitous heavily obscured AGN among ULIRGS, with AGN fractions roughly twice the number inferred from Seyfert-like optical classifications.~\citet{Sajina2009} examined the {\sl Spitzer} rest-frame 2--4\,$\upmu$m spectra of a small sample of cosmic noon deep silicate absorption AGN. They detected absorption features due to water ice at 3.0\,$\upmu$m (see Figure\,\ref{fig:silicates_ices}b) as well as hydrocarbons at 3.4\,$\upmu$m and found that the ratios of $\tau_{3.0}/\tau_{9.7}$ and $\tau_{3.4}/\tau_{9.7}$ were both consistent with those of local ULIRGs studies by~\citet{Imanishi2008}. It is clear that in both cases, the obscuration is primarily by ice-covered large silicate grains. In addition, obscured AGN also show higher crystalline silicate fractions than the diffuse ISM~\citep{Spoon2004,Tsuchikawa2021}. This is believed to be due to the silicate dust being processed in the high temperature environments in the nuclear regions of the AGN, before being transported elsewhere in the galaxy. These studies cumulatively show that the presence of an AGN can modify the composition and size distribution of the dust within its host galaxy.         

The above discussion focuses on spectroscopic studies and the strengths or profiles of specific features. However, one common way to probe the composition and size distribution of dust grains is through modeling the extinction curve. For example,  
\citet{Wang2015_mir_extinction} show a model of the extinction curve extending to the mid-IR where 
data from both dense and diffuse Galactic environments both point to a flat
mid-IR extinction curve which requires the addition of a component of micron-sized dust grains. This component is essentially invisible in all other parts of the spectrum (creating grey extinction in the UV-through-optical regime), but~\citet{Wang2015_mir_extinction} suggests it accounts for 14\% of the dust mass and nearly 3\% of the IR dust emission of the Milky Way. Recent studies of the infrared (at least out to the near-IR) attenuation curves in IR-luminous galaxies also point toward flat/grey attenuation~\citep{LoFaro2017,Roebuck2019} which has significant implications on the inferred stellar population parameters such as stellar mass and SFR. In those works, the issue with the variable attenuation curve arises from the relative geometry between the power sources and the dust, see example discussion in~\citet{Roebuck2019}. However, if the intrinsic extinction curve in AGN hosts is also different from standard assumptions, due to variations in dust size distribution and/or composition (see above), this would need to be taken into account, in order to allow for the most accurate dust attenuation corrections.

\subsection{Disentangling Star-Formation from AGN-Powered Dust Emission: A Key to AGN-Galaxy Co-Evolution Studies}
\label{sec:coevolution}

A key piece of evidence of the co-evolution of black holes and their host galaxies is the similarity between the shapes of the cosmic SFRD and BHARD with redshift~\citep{MadauDickinson2014}, see also Figure\,\ref{fig:runburg_lum_density} for an updated version. However, neither is particularly well known past cosmic noon for obscured systems. Indeed, recent determinations of the SFRD accounting of high-z dusty galaxies, are pushing the peak to somewhat higher redshifts than previous estimates~\citep{Zavala2021}. A significant difficulty arises from the fact that frequently we find that obscured systems are composites where both star-formation and AGN activity take place (see Figure\,\ref{fig:role_composites}). Correctly accounting for the contribution of such systems to both the cosmic SFRD and BHARD therefore hinges on our ability to disentangle the degree to which the overall IR SED is powered by each process. One approach is by adopting an SED for the AGN as well as the star-forming component and assuming a model where the total emission is simply the sum of the two components, e.g., \linebreak \citet{Sajina2012,Kirkpatrick2015}. For systems with only moderate levels of obscuration (i.e., negligible obscuration in the mid-IR) this approach is reasonable, as the AGN SEDs~\citep{Mullaney2011,LyuRieke2017,Bernhard2021} (see review of Lyu and Rieke in this Special Issue) as well as the SEDs of pure star-forming galaxies are generally well characterized. 

\begin{figure}[H]
\includegraphics[width=11.5 cm]{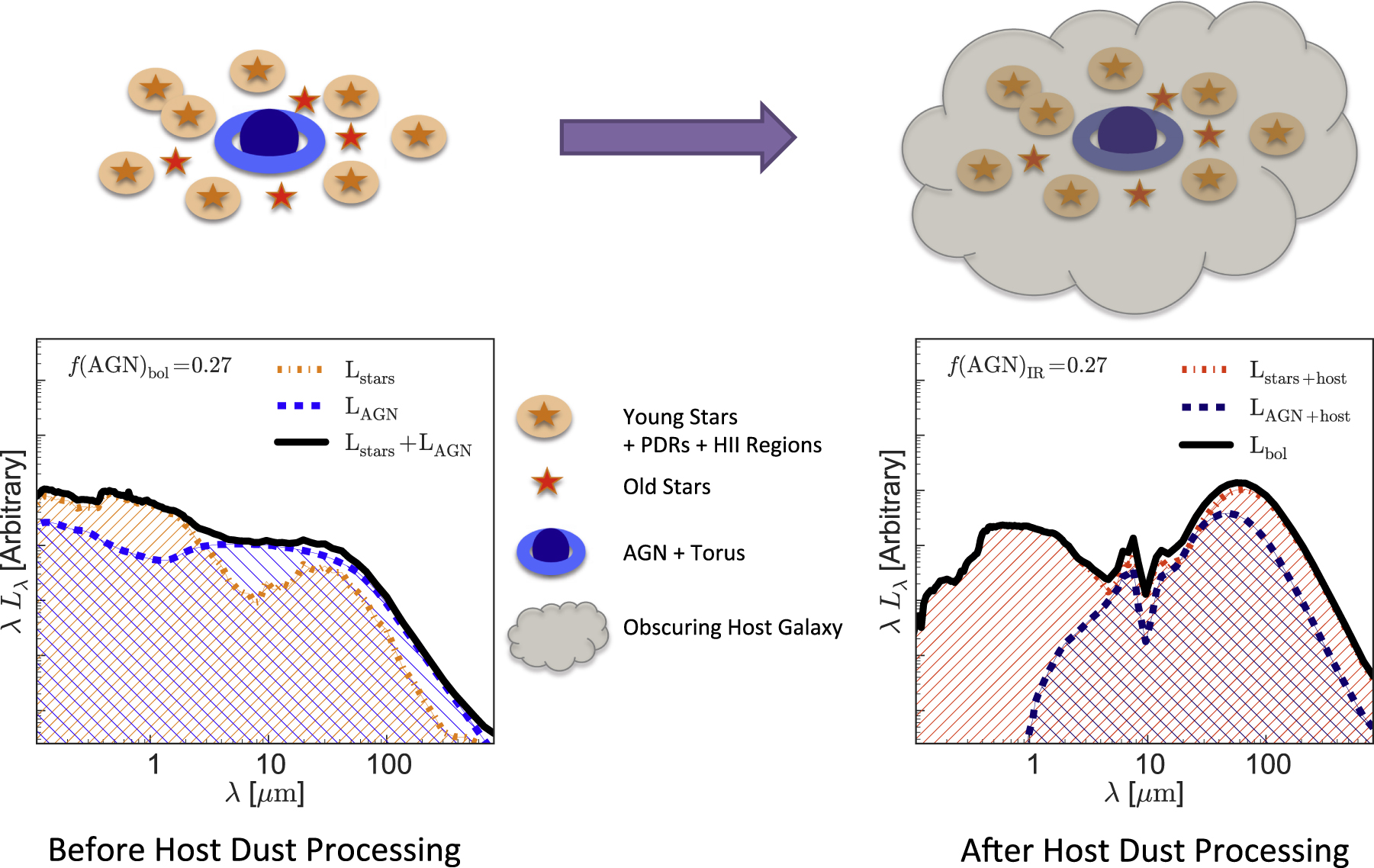}
\caption{An illustration, based on the coalescence stage of a gas-rich major merger simulation, of how the light from the AGN+dusty torus can be further modified by the dust of the host galaxy and lead to a much larger contribution of the AGN to the total IR emission of the galaxy than treatments not accounting for dust host galaxy obscured suggest. Reproduced with permission from~\mbox{\citet{Roebuck2016}}, AAS\textsuperscript{©}.\label{fig:roebuck2016_cartoon}}
\end{figure}

However, simulations suggest that for heavily obscured systems, we expect larger scale-host galaxy obscuration of the AGN, which further re-processes the AGN SED as shown in Figure\,\ref{fig:roebuck2016_cartoon}, see also~\citep{Snyder2013,Roebuck2016,McKinney2021}. The fact that host-scale AGN obscuration is indeed taking place is seen in the correlation between the depth of the silicate feature and the orientation of the host galaxy~\citep{Lacy2007,Goulding2012}. When such re-processing of the mid-IR emission of AGN is taking place, the AGN contribution to the far-IR increases, which can lead to overestimating the role of dusty star-formation while underestimating the role of dusty AGN in the particular galaxy.
The mid-IR regime has been used in multiple works to constrain the contribution of the AGN to the far-IR emission (e.g.,~\citep{Mullaney2011,Kirkpatrick2015,Symeonidis2016}). One common approach is to use the strength of the mid-IR PAH features to infer the SFR and hence the expected far-IR luminosity due to star-formation alone. This of course hinges on the accuracy of the PAH to SFR relation adopted. This is also complicated by the possibility of PAH carrier destruction in the vicinity of AGN,  see~\citet{Hernan-Caballero2011}, as well as modified PAH ratios~\citep{Smith2007,Lambrides2019} (as discussed in Section\,\ref{sec:ism}).  Another approach is a full modeling of the mid-IR spectra that does not rely on any one particular PAH feature~\citep{Sajina2007a,Sajina2012,Kirkpatrick2012,Kirkpatrick2015}.

The question of how much of the far-IR emission is powered by AGN in observational studies is discussed in detail in the review of Lyu and Rieke in this Special Issue. Here, for completeness, we summarize their main conclusions. The intrinsic SED of typical AGN is observed to ``turn over'' in the 20--40\,$\upmu$m range, such that beyond $\approx$30\,$\upmu$m, emission from dust heated by stars starts to dominate. Except in a small number of cases with very low levels of star-formation, the far-IR emission of galaxies does not have a significant AGN contribution. However, such contribution cannot be excluded in the case of heavily obscured systems. 
These conclusions are consistent with the observation of the steepening/reddening of the $f_{30}/f_{15}$ color in star-formation dominated systems vs. AGN-dominated systems discussed in Section\,\ref{sec:observed_properties_lowres}. However, AGN do appear to contribute to the heating of the warm dust responsible for the $\approx$30\,$\upmu$m emission ~\citep{Kirkpatrick2015,Lambrides2019} suggesting this regime is not a reliable SFR indicator, if an AGN is present. This affects the common practice of extrapolating from the 24\,$\upmu$m to the total IR luminosity and hence SFR.  On the theoretical side, hydrodynamic simulations including radiative transfer~\citep{Snyder2013,Roebuck2016,McKinney2021} show that for heavily obscured systems, see, e.g., Figure\,\ref{fig:roebuck2016_cartoon}, the intrinsic AGN SED can be heavily re-processed into the far-IR. \citet{Snyder2013} examined a variety of common observational mid-IR diagnostics of AGN and found that the $f_{30}/f_{15}$ ratio can miss such `buried AGN', but that they can be found by their characteristic strong silicate absorption features. Such systems have been found by {\sl Spitzer} as discussed in Section\,\ref{sec:observed_properties_lowres}, see example in Figure\,\ref{fig:intro}d. While representing a very interesting phase in AGN evolution, such systems are rare in number relative to the general AGN population. Therefore, in the absence of evidence to the contrary, it is a safe assumption that any particular galaxy or galaxy population's far-IR emission is powered by star-formation. However, while rare in number, such systems tend to have high $L_{bol}$ meaning their  contribution to the cosmic BHARD maybe non-negligible, although it has yet to be determined. 

\section{Future Prospects}
\label{sec:future}

As discussed in the context of the history of this field (Section\,\ref{sec:history}), the advances in the study of AGN in the mid-IR are very much driven by the advances in the available instrumentation. In this section, we begin by presenting a very broad overview of past, present and upcoming mid-IR spectrographs. We then focus in particular on the AGN science expected from the Mid-IR Instrument for {\sl JWST} (MIRI;~\citep{Rieke2015}), which after a more than 10-year hiatus brings back the ability to do mid-IR spectroscopy from space. We also briefly touch on how we anticipate these upcoming {\sl JWST}-MIRI studies to lay the ground work for future space-based IR telescopes. 

\subsection{Overview Mid-IR Spectrographs}

\begin{figure}[H]
\includegraphics[width=11.5 cm]{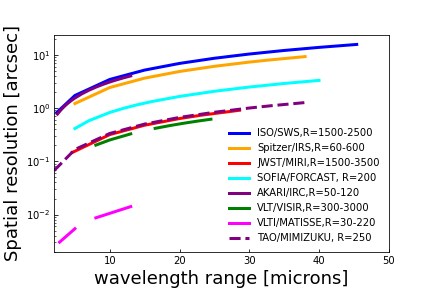}
\caption{A summary of the wavelength coverage, resolving power, and spatial resolution of past, present and upcoming mid-IR spectrographs. \label{fig:compare_instruments}}
\end{figure}

Figure\,\ref{fig:compare_instruments} shows a summary of the wavelength coverage, spatial resolution (assuming all diffraction limited) and representative spectral resolution of a variety of ground-based and space-borne mid-IR spectrographs. We see some obvious groupings---the top three curves are for relatively small diameter space-borne facilities. The middle group is for single dish ground-based facilities as well as the 6.5\,m {\sl JWST}. In between is the unique, 2.5\,m {\sl Stratospheric Observatory For Infrared Astronomy} (SOFIA) telescope, onboard an airplane flying at $\sim$37--45,000 feet. We show the values for the low-resolution mode of its FORCAST instrument (Faint Object Infrared Camera for the SOFIA Telescope)~\citep{Herter2012}. The highest spatial resolution shown is for the VLTI/MATISSE instrument~\citep{Lopez2022} which is an imaging spectro-interferometer achieving 3mas resolution at its shortest wavelength, see~\citep{Isbell2022} for recent results therewith.  The spectral resolution listed is for the N band (it operates in four resolution modes). 

Of course the wavelength coverage vs. spatial resolution parameter plane is only one aspect of how these instruments compare. The most obvious additional axis here would be the sensitivity. For example, on the new 6.5\, TAO telescope, the upcoming MIMIZUKU\citep{mimizuku2020} instrument has comparable spatial resolution to {\sl JWST}/MIRI, although with the advantage of a more extensive wavelength coverage out to 38\,$\upmu$m. Its sensitivity, however, is significantly worse than MIRI. For a rough comparison, the limiting flux in 10~ks observations at 10$\sigma$ with MIMIZUKU would be order of 10mJy, but is order of 0.1~mJy for MIRI. Ground-based instruments tend to be easier to use for time domain studies although there are notable exceptions such as the multi-epoch survey telescope {\sl WISE}, see~\citep{Kozlowski2016,Polimera2018,Lyu2019}. {\sl JWST}/MIRI does have support for time series observations for individual time variable objects. {\sl JWST} is also developing a time domain field (TDF) within its northern Continuous Viewing Zone~\citep{Jansen2018}. 

Figure\,\ref{fig:compare_instruments} does not show planned mid-IR instruments for the next generation ELTs. Such instruments would of course significantly improve on the sensitivity and spatial resolution of most existing ground-based instruments. We also do not include the planned Origins telescope (the next generation NASA great observatory in the mid to far-IR), as this will be covered in detailed in another paper in this Special Issue (by A.Cooray), although we do briefly discuss it in Section\,\ref{sec:diag_cosmic_time}. 

Of course, one of the most exciting instruments at present is the MIRI instrument on {\sl JWST} which not only brings back space-borne mid-IR spectroscopy, which was unavailable since the end of the cold mission of {\sl Spitzer} in 2009, but also does so with a much larger, 6.5\,m, telescope and is the first mid-IR IFU. MIRI \endnote{See \url{https://jwst-docs.stsci.edu/jwst-mid-infrared-instrument} for details.} has three observing modes: imaging in nine mid-IR bands; low resolution ($R \sim$100) slit/slitless spectroscopy; and medium resolution ($R \sim$1500--3500) IFU spectroscopy (MRS). In the following section, we go into more detail into AGN science with JWST/MIRI.

\subsection{AGN Studies with JWST}
\label{sec:jwst}

With renewed access from space to the mid-IR both in imaging and spectroscopy, JWST will build on the legacy of {\it Spitzer} and missions that came before it to further the study of AGN. 

\subsubsection{MIRI Spectroscopy}

\begin{figure}[H]
\includegraphics[width=12 cm]{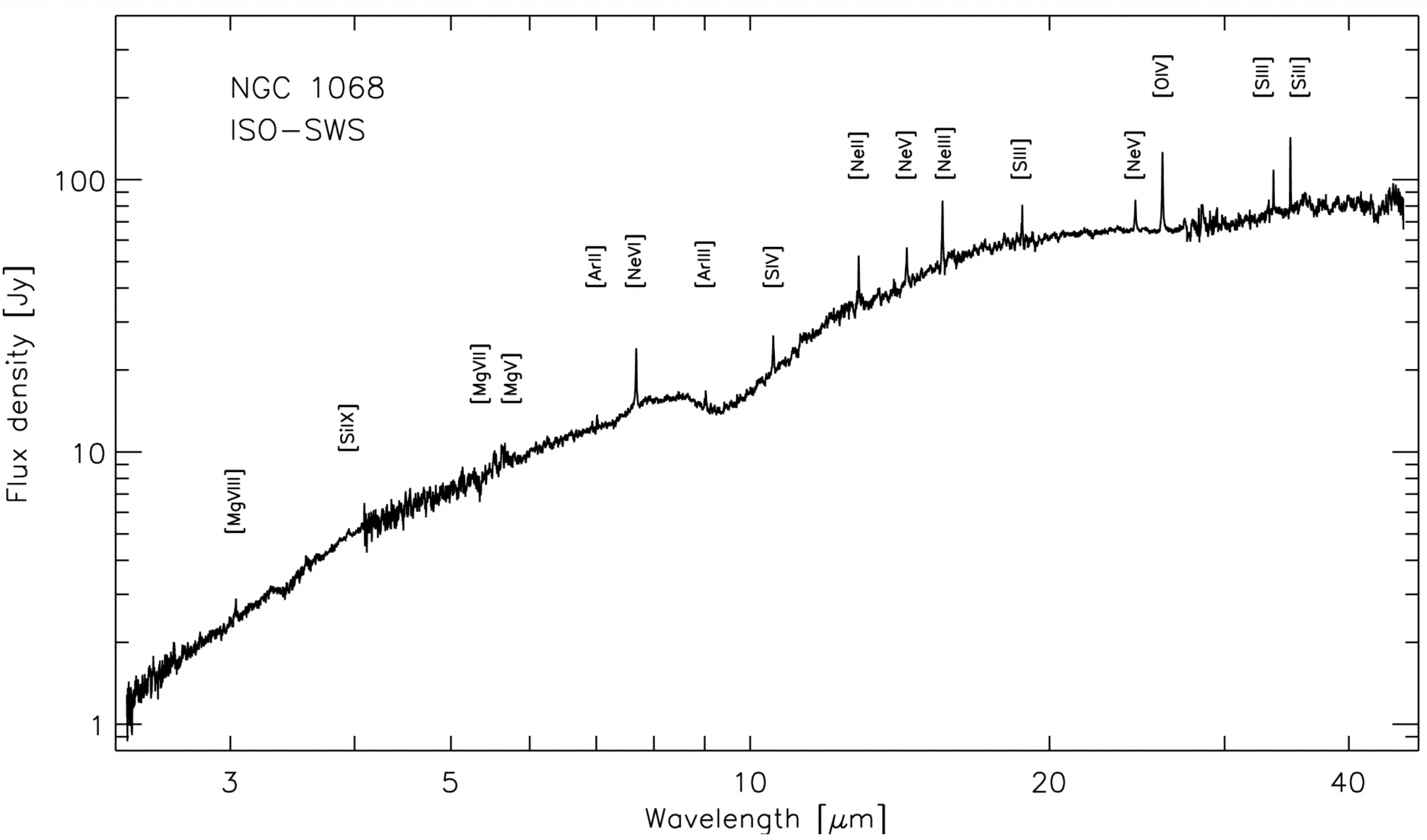}
\caption{ISO-SWS spectrum of the obscured AGN NGC 1068 showing various high ionization potential atomic lines including [NeVI] at 7.6$\,\upmu$m. This line was not studied with {\it Spitzer} since the low resolution of IRS at this wavelength was insufficient but shows considerable potential from the ISO spectra of NGC 1068 and Circinus 
~\citep{Lutz2000,Sturm2000}. Reproduced with permission from~\citet{Lutz2000}, AAS\textsuperscript{©}.   \label{fig:lutzfigure}}
\end{figure}

With the sensitivity, wavelength coverage and spectral resolution of {\it Spitzer}/IRS, observations of high redshift galaxies were limited to the broad PAH lines from 6--12$\,\upmu$m above the warm dust continuum. Similar studies will be possible with the LRS mode on MIRI, also extending to the rest-frame $\approx$3\,$\upmu$m to include the 3.3\,$\upmu$m PAH and key absorption features (see Figure\,\ref{fig:silicates_ices}b). However, {\sl Spitzer} did not have the sensitivity to detect the fine structure lines discussed in Section\,\ref{sec:observed_properties_highres} beyond the local Universe see, for example, Figure\,\ref{fig:intro}c. MIRI/MRS will be capable of  
detecting both narrow atomic lines that uniquely trace star formation and BHAR such as [NeII] and [NeV] (e.g., see Figure\,\ref{fig:fine_structure}), as well as powerful other diagnostic lines that will allow constraints of the physical characteristics of the ISM (see Section\,\ref{sec:observed_properties_highres}). 
{\it JWST}/MIRI will be able to detect all these powerful diagnostic lines out to higher redshifts. Specifically, the lines that have been well studied in the local Universe (see, e.g., Figure~\ref{fig:lutzfigure}) such as [NeV] 14.3$\,\mu$m will be visible out to $z\sim1$. The higher spectral resolution of {\it JWST}/MIRI, will allow lines such as [NeVI] 7.6$\,\mu$m, which sits in a crowded PAH region, to be better calibrated at low redshift, and subsequently applied out to cosmic noon, see Figure 
\,\ref{fig:FIRgap}. 

The power of MIRI is not just being able to detect fainter atomic lines in distant galaxies. The superior spectral and spatial resolution of MIRI/MRS can provide further information on the properties of AGN. With MIRI, the neon lines [NeV] and [NeVI] should be spectrally resolved and can be used to estimate the mass of the supermassive black holes, since they are expected to arise in the narrow line region around AGN, and their velocity dispersion correlates with the mass of the central black hole, as discussed in Section\,\ref{sec:mass_accretionrate}. Even when the lines remain unresolved, the [NeV] line luminosity is a reasonable estimator of the black hole mass~\citep{Dasyra2008}.

\begin{figure}[H]
\includegraphics[width=10.5 cm]{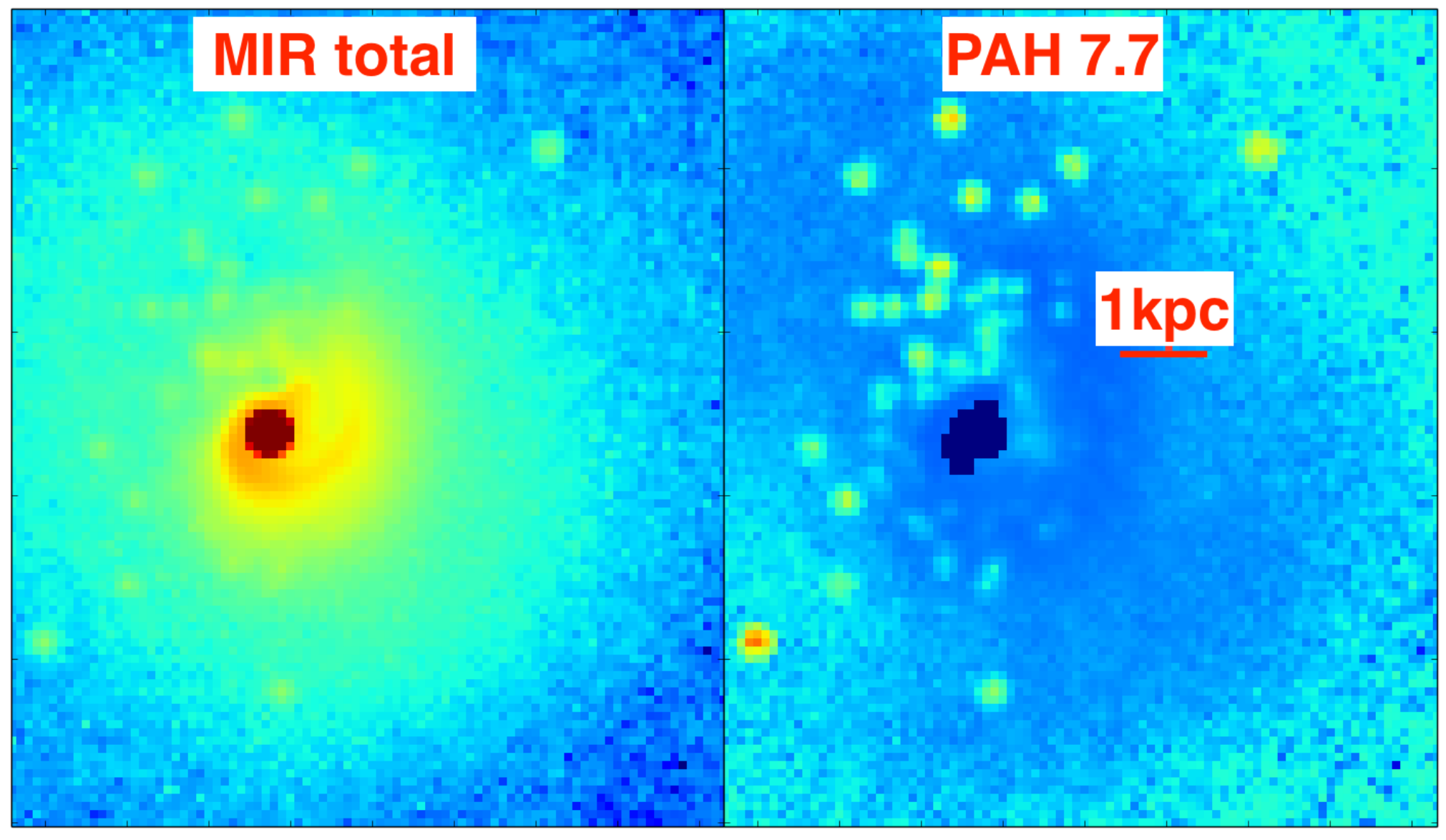}
\caption{Simulation of a major merger including both strong star-formation and an AGN (from GADGET+Sunrise) illustrating the spatial distribution of the total mid-IR emission as well as highlighting just the PAH emitting regions. One of the most exciting features of the upcoming JWST MIRI MRS IFU is the ability to provide high spectral resolution (with the sensitivity to probe sources out to high-$z$) but also to look at the spatial distribution of various lines/features. As a guide a 1 kpc scale is marked---this corresponds roughly to the spatial resolution of {\sl JWST}/MIRI in its shorter wavelengths at cosmic noon, $z \sim$1--3. 
\label{fig:simfigure}}
\end{figure}

Even in galaxies out to half the age of the Universe, the spatial resolution of MIRI will directly map the impact of an AGN heating its surroundings. Simulations predict that AGN in high redshift galaxies can provide significant heating throughout the host galaxy which contaminates the IR luminosity as a star formation rate indicator~\citep{Roebuck2016,McKinney2021}. Figure~\ref{fig:simfigure} shows a merger containing active star formation and AGN activity from the GADGET hydrodynamic simulations processed through the Sunrise radiative transfer code to interpret the mid-IR emission~\citep{Snyder2013,Roebuck2016}. By decomposing the mid-IR emission using the mid-IR spectral features, these simulations can be used to measure the spatial distribution of the AGN influence and the star-forming regions, where the latter appear extended and clumpy in simulations of high redshift, gas-rich merger systems (right panel of Figure~\ref{fig:simfigure}).

\subsubsection{MIRI Imaging}

While MIRI spectroscopy will be extremely powerful, it can only target one galaxy at a time and may require long integration times to detect the key diagnostic lines in high redshift galaxies. As mentioned earlier, one of the modes of MIRI is imaging which includes 9 bands spanning 5.6\,$\upmu$m to 25\,$\upmu$m. This can be thought of as extremely low resolution spectroscopy. It allows for simple mid-IR color-color diagnostics, similar to those developed for earlier instruments, see Section\,\ref{sec:agn_color_selection}.

\begin{figure}[H]
\includegraphics[width=10.5 cm]{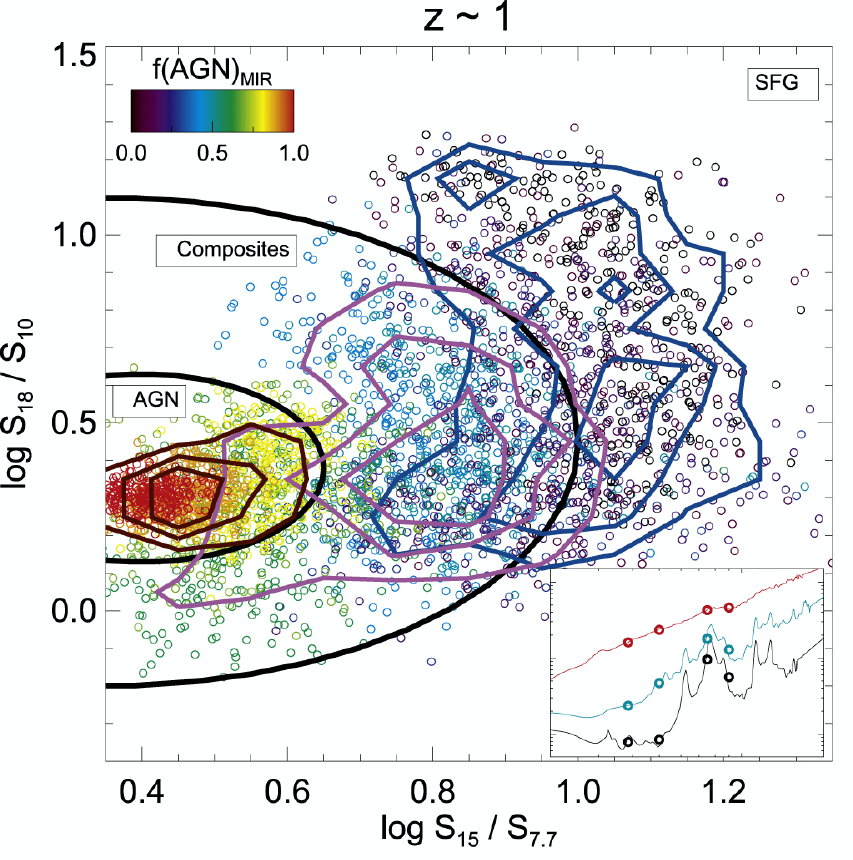}
\caption{AGN fraction determination based on JWST MIRI colors for $z \sim$1 galaxies. The color scheme represents the fraction of mid-IR emission that is AGN powered~\citep{Kirkpatrick2017}. The inset shows typical mid-IR spectra for an AGN, a composite or a star-forming galaxy (from top-to-bottom). Reproduced with permission from~\citet{Kirkpatrick2017}, AAS\textsuperscript{©}. \label{fig:kirkpatrick2017_z1}}
\end{figure}

Figure~\ref{fig:kirkpatrick2017_z1} shows an example of a combination of MIRI colors that effectively separates sources as a function of AGN fraction at $z \sim$1. This means that {\sl JWST}/MIRI will be the first to efficiently identify large numbers of composite sources which are key for studying the co-evolution of star formation and black hole growth in galaxies over cosmic time. 

These color diagnostics can be applied to large extragalactic surveys such as those planned by the MIRI GTO team (Program ID 1207, PI Rieke), the CEERS ERS (Program ID 1345, PI Finkelstein) and the PRIMER GO1 (Program ID1837, PI Dunlop) surveys to select large samples of AGN and composite systems. While color-color plots are limited to four imaging bands which restricts the AGN separation to certain redshift ranges, a more sophisticated study such as a principal component analysis could be applied to all nine MIRI imaging bands in order to identify AGN at any redshifts~\citep{Lin2018}. 

\subsection{Mid-IR Diagnostics of AGN over Cosmic Time}
\label{sec:diag_cosmic_time}

As discussed in Section\,\ref{sec:cosmic_bhard}, the volume-averaged history of star formation (SFRD) and black hole growth (BHAR density, BHARD) in the Universe both appear to increase up to a peak period at $z=1$--3. At higher redshifts, the BHARD appears to decline more steeply than star formation, indicating changes in their relative growth in the very early Universe~\citep{Vito2018} although more recent analysis does not see this, see Figure\,\ref{fig:runburg_lum_density}. However, these constraints on the average SFRD and BHARD have been based on studies of different galaxy populations at different wavelengths. In order to chart the relative growth of stars and supermassive black holes over cosmic time, we need measurements of the SFRs and the BHARs in the same galaxies.

\begin{figure}[H]
\includegraphics[width=13 cm]{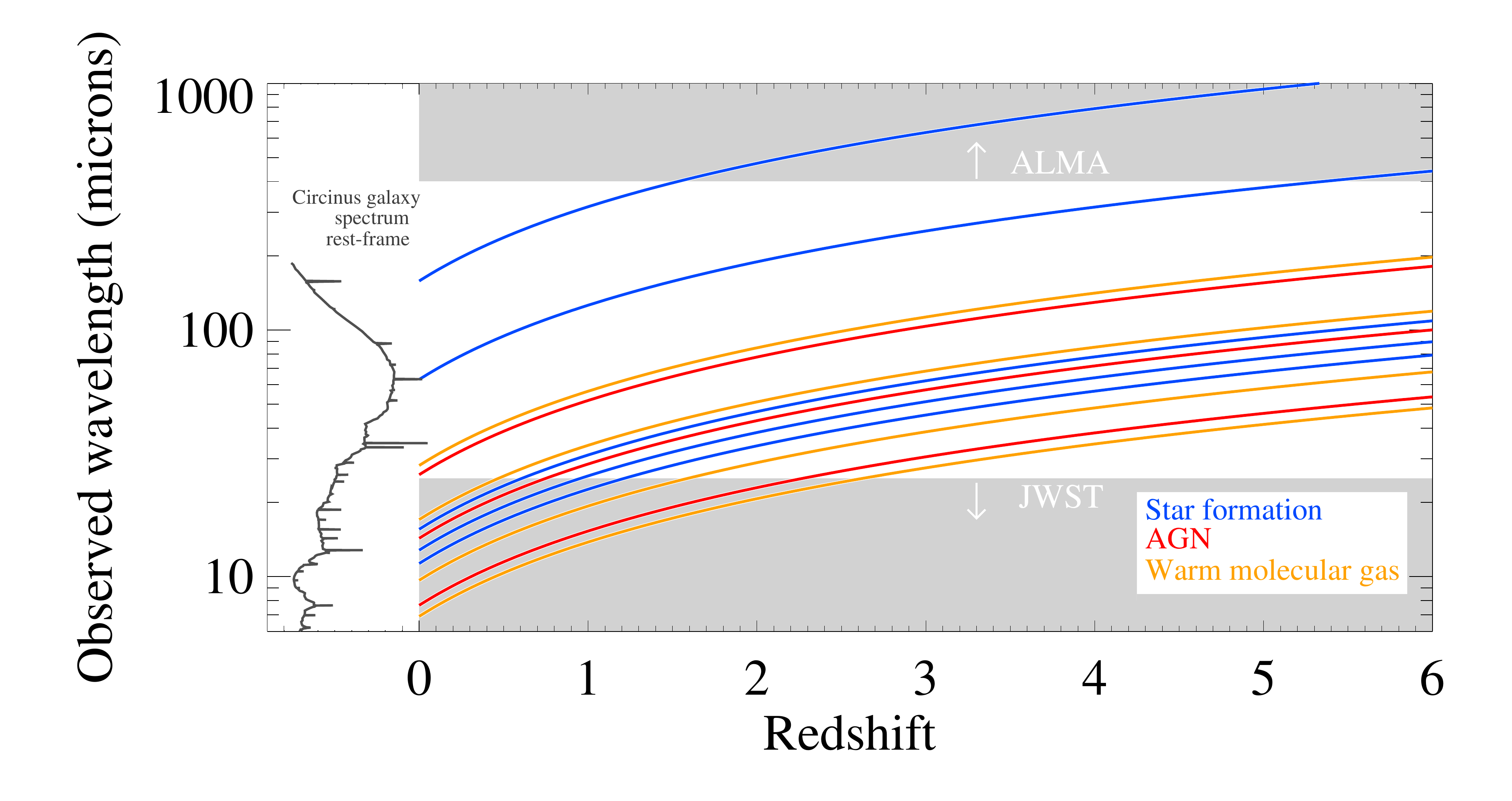}
\caption{Observed wavelengths of select infrared spectral lines that trace AGN, star formation and warm molecular gas as a function of redshift. The SED of the Circinus galaxy exhibiting these lines is shown tipped on its side at $z=0$ on the left panel of the plot. The powerful mid-infrared diagnostics of AGN are shifted into the far-IR for higher redshift galaxies. Currently, there is an order of magnitude gap in our wavelength coverage between {\it JWST} and ALMA which could be filled with a future cold far-IR space telescope. Figure is adapted from 
~\citet{Pope2019}. \label{fig:FIRgap}}
\end{figure}

While {\it JWST} will measure SFRs and BHARs in galaxies from mid-IR spectral lines out to $z \sim$1--2, it will not do so for statistical samples of galaxies, and these diagnostics shift to longer far-IR wavelengths at $z>2$, see Figure\,\ref{fig:FIRgap}. In order to simultaneously measure the BHARs and SFRs in galaxies at all redshifts will require a large, cold infrared space telescope to fill the order of magnitude gap in wavelength coverage between {\it JWST} and ALMA. Such missions have been studied both at the Flagship and Probe scales~\citep{Meixner2019,Glenn2021}. Given the recent Astro2020 decadal survey recommendations, a far-IR probe capable of identifying and studying AGN over all cosmic time might be possible in the coming decade. 

\section{Summary}
\label{sec:summary}

Mid-IR studies of AGN have an almost 60-year history of both ground-based and space-based observations. This review aimed to give a broad rather than comprehensive overview of some of the key results thereof. These include mid-IR techniques to find AGN which led to a more complete census of the AGN population and improved understanding of the AGN demographics; insights into how an AGN affects its host galaxy's ISM including multiphase gas and dust; and the realization that the geometry of the obscuring dust is dynamic and tightly coupled both to the accretion of gas that powers the AGN and the gas outflows powered by the AGN. 

The mid-IR provides diagnostics of AGN, star-formation, and the properties of both the multiphase gas and dust. This is therefore an ideal regime to study the interplay between them and upcoming facilities, such as JWST, will build on the decades of prior studies to give us an even clearer picture of how AGN and their host galaxies co-evolve. High spatial resolution imaging and further mid-IR studies in the time domain are needed, especially in light of our new model for a dynamic torus+wind structure regulated by the gas inflow and outflow of the AGN. 
A next-generation far-IR space telescope is needed to employ these powerful mid-IR diagnostics to quantify AGN activity in galaxies over all cosmic time.






\vspace{6pt} 



\authorcontributions{
The overall conceptualization of this review was done in discussion among all three authors. The bulk of writing was done by A.S., with M.L. writing Sections\,\ref{sec:agn_color_selection}--\ref{sec:cosmic_bhard} and A.P. writing Sections\,\ref{sec:jwst} and \ref{sec:diag_cosmic_time}.
All authors have read and agreed to the published version of the~manuscript.
}

\funding{This research received no external funding.}

\institutionalreview{Not applicable.}

\informedconsent{Not applicable.}

\dataavailability{Not applicable.}

\acknowledgments{The authors are very grateful to George Rieke, Jianwei Lyu, Almudena Alonso-Herrero, Evanthia Hatziminaoglou, Henrik Spoon, Sylvain Veilleux, Marko Stalevski, and Vivian U for useful discussions and feedback on a pre-publication draft. Their feedback has been invaluable in improving the content and presentation of this review. We are also grateful to the original authors of many of the figures shown here for giving us permission to reproduce their work for this review. Finally, we are grateful to the two anonymous referees for their reading of the submitted paper and feedback thereon which led to the final polish of this paper. }

\conflictsofinterest{The authors declare no conflict of interest.}



\abbreviations{Abbreviations}{
The following abbreviations are used in this manuscript:\\

\noindent 
\begin{tabular}{@{}ll}
AGN & Active Galactic Nuclei\\
BHAR & Black Hole Accretion Rate\\
BHARD & Black Hole Accretion Rate Density\\
EW & Equivalent Width\\
IFU & Integral Field Unit \\
ISM & Interstellar Medium\\
JWST & James Webb Space Telescope \\
MIRI & Mid-InfraRed Instrument (on JWST)\\
PAH & Polycyclic Aromatic Hydrocarbons\\
PDR & Photodissociation Region or Photon-Dominated Region\\
SED & Spectral Energy Distribution\\
SFR & Star Formation Rate\\
SFRD & Star Formation Rate Density\\
SLED & Spectral Line Energy Distribution \\
SMBH & Super Massive Black Hole \\
XDR & X-ray Dominated Region
\end{tabular}}





\begin{adjustwidth}{-\extralength}{0cm}
\printendnotes[custom] 
\reftitle{References}

\end{adjustwidth}

\begin{thebibliography}{999}

\bibitem[{Magorrian} {et~al.}(1998){Magorrian}, {Tremaine}, {Richstone},
  {Bender}, {Bower}, {Dressler}, {Faber}, {Gebhardt}, {Green}, {Grillmair},
  {Kormendy}, and {Lauer}]{Magorrian1998}
{Magorrian}, J.; {Tremaine}, S.; {Richstone}, D.; {Bender}, R.; {Bower}, G.;
  {Dressler}, A.; {Faber}, S.M.; {Gebhardt}, K.; {Green}, R.; {Grillmair}, C.;
  et~al.
\newblock {The Demography of Massive Dark Objects in Galaxy Centers}.
\newblock {\em Astron. J.} {\bf 1998}, {\em 115},~2285--2305. 
\newblock
  https://doi.org/{\changeurlcolor{black}\href{https://doi.org/10.1086/300353}{\detokenize{10.1086/300353}}}.

\bibitem[{G{\"u}ltekin} {et~al.}(2009){G{\"u}ltekin}, {Richstone},
  {Gebhardt}, {Lauer}, {Tremaine}, {Aller}, {Bender}, {Dressler}, {Faber},
  {Filippenko}, {Green}, {Ho}, {Kormendy}, {Magorrian}, {Pinkney}, and
  {Siopis}]{Gultekin2009}
{G{\"u}ltekin}, K.; {Richstone}, D.O.; {Gebhardt}, K.; {Lauer}, T.R.;
  {Tremaine}, S.; {Aller}, M.C.; {Bender}, R.; {Dressler}, A.; {Faber}, S.M.;
  {Filippenko}, A.V.;  et~al.
\newblock {The M-{\ensuremath{\sigma}} and M-L Relations in Galactic Bulges,
  and Determinations of Their Intrinsic Scatter}.
\newblock {\em Astrophys. J.} {\bf 2009}, {\em 698},~198--221.
\newblock
  https://doi.org/{\changeurlcolor{black}\href{https://doi.org/10.1088/0004-637X/698/1/198}{\detokenize{10.1088/0004-637X/698/1/198}}}.

\bibitem[{Madau} and {Dickinson}(2014)]{MadauDickinson2014}
{Madau}, P.; {Dickinson}, M.
\newblock {Cosmic Star-Formation History}.
\newblock {\em Annu. Rev. Aston. Astrophys.} {\bf 2014}, {\em
  52},~415--486. \linebreak
\newblock
  https://doi.org/{\changeurlcolor{black}\href{https://doi.org/10.1146/annurev-astro-081811-125615}{\detokenize{10.1146/annurev-astro-081811-125615}}}.

\bibitem[{Heckman} and {Best}(2014)]{HeckmanBest2014}
{Heckman}, T.M.; {Best}, P.N.
\newblock {The Coevolution of Galaxies and Supermassive Black Holes: Insights
  from Surveys of the Contemporary Universe}.
\newblock {\em Ann. Rev. Astron. Astrophys.} {\bf 2014}, {\em
  52},~589--660.
\newblock
  https://doi.org/{\changeurlcolor{black}\href{https://doi.org/10.1146/annurev-astro-081913-035722}{\detokenize{10.1146/annurev-astro-081913-035722}}}.

\bibitem[{Padovani} {et~al.}(2017){Padovani}, {Alexander}, {Assef}, {De
  Marco}, {Giommi}, {Hickox}, {Richards}, {Smol{\v{c}}i{\'c}},
  {Hatziminaoglou}, {Mainieri}, and {Salvato}]{Padovani2017}
{Padovani}, P.; {Alexander}, D.M.; {Assef}, R.J.; {De Marco}, B.; {Giommi}, P.;
  {Hickox}, R.C.; {Richards}, G.T.; {Smol{\v{c}}i{\'c}}, V.; {Hatziminaoglou},
  E.; {Mainieri}, V.;  et~al.
\newblock {Active galactic nuclei: What's in a name?}
\newblock {\em Astron. Astrophys. Rev.} {\bf 2017}, {\em 25},~2.
\newblock
  https://doi.org/{\changeurlcolor{black}\href{https://doi.org/10.1007/s00159-017-0102-9}{\detokenize{10.1007/s00159-017-0102-9}}}.

\bibitem[{Netzer}(2015)]{Netzer2015}
{Netzer}, H.
\newblock {Revisiting the Unified Model of Active Galactic Nuclei}.
\newblock {\em Ann. Rev. Astron. Astrophys.} {\bf 2015}, {\em
  53},~365--408.
\newblock
  https://doi.org/{\changeurlcolor{black}\href{https://doi.org/10.1146/annurev-astro-082214-122302}{\detokenize{10.1146/annurev-astro-082214-122302}}}.

\bibitem[{Hopkins} {et~al.}(2008){Hopkins}, {Hernquist}, {Cox}, and
  {Kere{\v{s}}}]{Hopkins2008}
{Hopkins}, P.F.; {Hernquist}, L.; {Cox}, T.J.; {Kere{\v{s}}}, D.
\newblock {A Cosmological Framework for the Co-Evolution of Quasars,
  Supermassive Black Holes, and Elliptical Galaxies. I. Galaxy Mergers and
  Quasar Activity}.
\newblock {\em Astrophys. J. Suppl.} {\bf 2008}, {\em
  175},~356--389.
\newblock
  https://doi.org/{\changeurlcolor{black}\href{https://doi.org/10.1086/524362}{\detokenize{10.1086/524362}}}.

\bibitem[{Sajina} {et~al.}(2012){Sajina}, {Yan}, {Fadda}, {Dasyra}, and
  {Huynh}]{Sajina2012}
{Sajina}, A.; {Yan}, L.; {Fadda}, D.; {Dasyra}, K.; {Huynh}, M.
\newblock {Spitzer- and Herschel-based Spectral Energy Distributions of 24
  {\ensuremath{\mu}}m Bright z \raisebox{-0.5ex}\textasciitilde 0.3-3.0
  Starbursts and Obscured Quasars}.
\newblock {\em Astrophys. J.} {\bf 2012}, {\em 757},~13.
\newblock
  https://doi.org/{\changeurlcolor{black}\href{https://doi.org/10.1088/0004-637X/757/1/13}{\detokenize{10.1088/0004-637X/757/1/13}}}.

\bibitem[{Kirkpatrick} {et~al.}(2012){Kirkpatrick}, {Pope}, {Alexander},
  {Charmandaris}, {Daddi}, {Dickinson}, {Elbaz}, {Gabor}, {Hwang}, {Ivison},
  {Mullaney}, {Pannella}, {Scott}, {Altieri}, {Aussel}, {Bournaud}, {Buat},
  {Coia}, {Dannerbauer}, {Dasyra}, {Kartaltepe}, {Leiton}, {Lin}, {Magdis},
  {Magnelli}, {Morrison}, {Popesso}, and {Valtchanov}]{Kirkpatrick2012}
{Kirkpatrick}, A.; {Pope}, A.; {Alexander}, D.M.; {Charmandaris}, V.; {Daddi},
  E.; {Dickinson}, M.; {Elbaz}, D.; {Gabor}, J.; {Hwang}, H.S.; {Ivison}, R.;
  et~al.
\newblock {GOODS-Herschel: Impact of Active Galactic Nuclei and Star Formation
  Activity on Infrared Spectral Energy Distributions at High Redshift}.
\newblock {\em Astrophys. J.} {\bf 2012}, {\em 759},~139.
\newblock
  https://doi.org/{\changeurlcolor{black}\href{https://doi.org/10.1088/0004-637X/759/2/139}{\detokenize{10.1088/0004-637X/759/2/139}}}.

\bibitem[{Kirkpatrick} {et~al.}(2015){Kirkpatrick}, {Pope}, {Sajina},
  {Roebuck}, {Yan}, {Armus}, {D{\'\i}az-Santos}, and
  {Stierwalt}]{Kirkpatrick2015}
{Kirkpatrick}, A.; {Pope}, A.; {Sajina}, A.; {Roebuck}, E.; {Yan}, L.; {Armus},
  L.; {D{\'\i}az-Santos}, T.; {Stierwalt}, S.
\newblock {The Role of Star Formation and an AGN in Dust Heating of z = 0.3-2.8
  Galaxies. I. Evolution with Redshift and Luminosity}.
\newblock {\em Astrophys. J.} {\bf 2015}, {\em 814},~9.
\newblock
  https://doi.org/{\changeurlcolor{black}\href{https://doi.org/10.1088/0004-637X/814/1/9}{\detokenize{10.1088/0004-637X/814/1/9}}}.

\bibitem[{Zamojski} {et~al.}(2011){Zamojski}, {Yan}, {Dasyra}, {Sajina},
  {Surace}, {Heckman}, and {Helou}]{Zamojski2011}
{Zamojski}, M.; {Yan}, L.; {Dasyra}, K.; {Sajina}, A.; {Surace}, J.; {Heckman},
  T.; {Helou}, G.
\newblock {HST/NICMOS Imaging of Bright High-redshift 24 {\ensuremath{\mu}}m
  Selected Galaxies: Merging Properties}.
\newblock {\em Astrophys. J.} {\bf 2011}, {\em 730},~125.
\newblock
  https://doi.org/{\changeurlcolor{black}\href{https://doi.org/10.1088/0004-637X/730/2/125}{\detokenize{10.1088/0004-637X/730/2/125}}}.

\bibitem[{Kocevski} {et~al.}(2015){Kocevski}, {Brightman}, {Nandra},
  {Koekemoer}, {Salvato}, {Aird}, {Bell}, {Hsu}, {Kartaltepe}, {Koo}, {Lotz},
  {McIntosh}, {Mozena}, {Rosario}, and {Trump}]{Kocevski2015}
{Kocevski}, D.D.; {Brightman}, M.; {Nandra}, K.; {Koekemoer}, A.M.; {Salvato},
  M.; {Aird}, J.; {Bell}, E.F.; {Hsu}, L.T.; {Kartaltepe}, J.S.; {Koo}, D.C.;
  et~al.
\newblock {Are Compton-thick AGNs the Missing Link between Mergers and Black
  Hole Growth?}
\newblock {\em Astrophys. J.} {\bf 2015}, {\em 814},~104.
\newblock
  https://doi.org/{\changeurlcolor{black}\href{https://doi.org/10.1088/0004-637X/814/2/104}{\detokenize{10.1088/0004-637X/814/2/104}}}.

\bibitem[{Treister} {et~al.}(2012){Treister}, {Schawinski}, {Urry}, and
  {Simmons}]{Treister2012}
{Treister}, E.; {Schawinski}, K.; {Urry}, C.M.; {Simmons}, B.D.
\newblock {Major Galaxy Mergers Only Trigger the Most Luminous Active Galactic
  Nuclei}.
\newblock {\em Astrophys. J. Lett.} {\bf 2012}, {\em 758},~L39.
\newblock
  https://doi.org/{\changeurlcolor{black}\href{https://doi.org/10.1088/2041-8205/758/2/L39}{\detokenize{10.1088/2041-8205/758/2/L39}}}.

\bibitem[{Hickox} and {Alexander}(2018)]{Hickox&Alexander2018_review}
{Hickox}, R.C.; {Alexander}, D.M.
\newblock {Obscured Active Galactic Nuclei}.
\newblock {\em Ann. Rev. Astron. Astrophys.} {\bf 2018}, {\em
  56},~625--671.
\newblock
  https://doi.org/{\changeurlcolor{black}\href{https://doi.org/10.1146/annurev-astro-081817-051803}{\detokenize{10.1146/annurev-astro-081817-051803}}}.

\bibitem[{Kormendy} and {Ho}(2013)]{KormendyHo2013}
{Kormendy}, J.; {Ho}, L.C.
\newblock {Coevolution (Or Not) of Supermassive Black Holes and Host Galaxies}.
\newblock {\em Ann. Rev. Astron. Astrophys.} {\bf 2013}, {\em
  51},~511--653.
\newblock
  https://doi.org/{\changeurlcolor{black}\href{https://doi.org/10.1146/annurev-astro-082708-101811}{\detokenize{10.1146/annurev-astro-082708-101811}}}.

\bibitem[{Lambrides} {et~al.}(2021){Lambrides}, {Chiaberge}, {Heckman},
  {Kirkpatrick}, {Meyer}, {Petric}, {Hall}, {Long}, {Watts}, {Gilli}, {Simons},
  {Tchernyshyov}, {Rodriguez-Gomez}, {Vito}, {de la Vega}, {Davis}, {Kocevski},
  and {Norman}]{Lambrides2021}
{Lambrides}, E.L.; {Chiaberge}, M.; {Heckman}, T.; {Kirkpatrick}, A.; {Meyer},
  E.T.; {Petric}, A.; {Hall}, K.; {Long}, A.; {Watts}, D.J.; {Gilli}, R.;
  et~al.
\newblock {Lower-luminosity Obscured AGN Host Galaxies Are Not Predominantly in
  Major-merging Systems at Cosmic Noon}.
\newblock {\em Astrophys. J.} {\bf 2021}, {\em 919},~129.
\newblock
  https://doi.org/{\changeurlcolor{black}\href{https://doi.org/10.3847/1538-4357/ac12c8}{\detokenize{10.3847/1538-4357/ac12c8}}}.

\bibitem[{Gilli} {et~al.}(2007){Gilli}, {Comastri}, and
  {Hasinger}]{Gilli2007}
{Gilli}, R.; {Comastri}, A.; {Hasinger}, G.
\newblock {The synthesis of the cosmic X-ray background in the Chandra and
  XMM-Newton era}.
\newblock {\em Astron. Astrophys.} {\bf 2007}, {\em 463},~79--96.
\newblock
  https://doi.org/{\changeurlcolor{black}\href{https://doi.org/10.1051/0004-6361:20066334}{\detokenize{10.1051/0004-6361:20066334}}}.

\bibitem[{Lacy} and {Sajina}(2020)]{lacy&sajina2020}
{Lacy}, M.; {Sajina}, A.
\newblock {Active galactic nuclei as seen by the Spitzer Space Telescope}.
\newblock {\em Nat. Astron.} {\bf 2020}, {\em 4},~352--363.
\newblock
  https://doi.org/{\changeurlcolor{black}\href{https://doi.org/10.1038/s41550-020-1071-x}{\detokenize{10.1038/s41550-020-1071-x}}}.

\bibitem[{Runburg} {et~al.}(2022){Runburg}, {Farrah}, {Sajina}, {Lacy},
  {Lidua}, {Hatziminaoglou}, {Brandt}, {Chen}, {Nyland}, {Shirley}, {Clements},
  and {Pitchford}]{Runburg2022}
{Runburg}, J.; {Farrah}, D.; {Sajina}, A.; {Lacy}, M.; {Lidua}, J.;
  {Hatziminaoglou}, E.; {Brandt}, W.N.; {Chen}, C.T.J.; {Nyland}, K.;
  {Shirley}, R.;  et~al.
\newblock {Consistent Analysis of the AGN LF in X-Ray and MIR in the XMM-LSS
  Field}.
\newblock {\em Astrophys. J.} {\bf 2022}, {\em 924},~133.
\newblock
  https://doi.org/{\changeurlcolor{black}\href{https://doi.org/10.3847/1538-4357/ac37b8}{\detokenize{10.3847/1538-4357/ac37b8}}}.

\bibitem[{Lacy} {et~al.}(2004){Lacy}, {Storrie-Lombardi}, {Sajina},
  {Appleton}, {Armus}, {Chapman}, {Choi}, {Fadda}, {Fang}, {Frayer},
  {Heinrichsen}, {Helou}, {Im}, {Marleau}, {Masci}, {Shupe}, {Soifer},
  {Surace}, {Teplitz}, {Wilson}, and {Yan}]{lacy2004}
{Lacy}, M.; {Storrie-Lombardi}, L.J.; {Sajina}, A.; {Appleton}, P.N.; {Armus},
  L.; {Chapman}, S.C.; {Choi}, P.I.; {Fadda}, D.; {Fang}, F.; {Frayer}, D.T.;
  et~al.
\newblock {Obscured and Unobscured Active Galactic Nuclei in the Spitzer Space
  Telescope First Look Survey}.
\newblock {\em Astrophys. J. Suppl.} {\bf 2004}, {\em
  154},~166--169.
\newblock
  https://doi.org/{\changeurlcolor{black}\href{https://doi.org/10.1086/422816}{\detokenize{10.1086/422816}}}.

\bibitem[{Stern} {et~al.}(2005){Stern}, {Eisenhardt}, {Gorjian}, {Kochanek},
  {Caldwell}, {Eisenstein}, {Brodwin}, {Brown}, {Cool}, {Dey}, {Green},
  {Jannuzi}, {Murray}, {Pahre}, and {Willner}]{Stern2005}
{Stern}, D.; {Eisenhardt}, P.; {Gorjian}, V.; {Kochanek}, C.S.; {Caldwell}, N.;
  {Eisenstein}, D.; {Brodwin}, M.; {Brown}, M.J.I.; {Cool}, R.; {Dey}, A.;
  et~al.
\newblock {Mid-Infrared Selection of Active Galaxies}.
\newblock {\em Astrophys. J.} {\bf 2005}, {\em 631},~163--168.
\newblock
  https://doi.org/{\changeurlcolor{black}\href{https://doi.org/10.1086/432523}{\detokenize{10.1086/432523}}}.

\bibitem[{Donley} {et~al.}(2007){Donley}, {Rieke}, {P{\'e}rez-Gonz{\'a}lez},
  {Rigby}, and {Alonso-Herrero}]{Donley2007}
{Donley}, J.L.; {Rieke}, G.H.; {P{\'e}rez-Gonz{\'a}lez}, P.G.; {Rigby}, J.R.;
  {Alonso-Herrero}, A.
\newblock {Spitzer Power-Law Active Galactic Nucleus Candidates in the Chandra
  Deep Field-North}.
\newblock {\em Astrophys. J.} {\bf 2007}, {\em 660},~167--190.
\newblock
  https://doi.org/{\changeurlcolor{black}\href{https://doi.org/10.1086/512798}{\detokenize{10.1086/512798}}}.

\bibitem[{Dey} {et~al.}(2008){Dey}, {Soifer}, {Desai}, {Brand}, {Le Floc'h},
  {Brown}, {Jannuzi}, {Armus}, {Bussmann}, {Brodwin}, {Bian}, {Eisenhardt},
  {Higdon}, {Weedman}, and {Willner}]{Dey2008}
{Dey}, A.; {Soifer}, B.T.; {Desai}, V.; {Brand}, K.; {Le Floc'h}, E.; {Brown},
  M.J.I.; {Jannuzi}, B.T.; {Armus}, L.; {Bussmann}, S.; {Brodwin}, M.;  et~al.
\newblock {A Significant Population of Very Luminous Dust-Obscured Galaxies at
  Redshift z $\sim$ 2}.
\newblock {\em Astrophys. J.} {\bf 2008}, {\em 677},~943--956.
\newblock
  https://doi.org/{\changeurlcolor{black}\href{https://doi.org/10.1086/529516}{\detokenize{10.1086/529516}}}.

\bibitem[{Assef} {et~al.}(2015){Assef}, {Eisenhardt}, {Stern}, {Tsai}, {Wu},
  {Wylezalek}, {Blain}, {Bridge}, {Donoso}, {Gonzales}, {Griffith}, and
  {Jarrett}]{Assef2015}
Assef, R.J.; {Eisenhardt}, P.R.M.; {Stern}, D.; {Tsai}, C.W.; {Wu}, J.;
  {Wylezalek}, D.; {Blain}, A.W.; {Bridge}, C.R.; {Donoso}, E.; {Gonzales}, A.;
   et~al.
\newblock {Half of the Most Luminous Quasars May Be Obscured: Investigating the
  Nature of WISE-Selected Hot Dust-Obscured Galaxies}.
\newblock {\em Astrophys. J.} {\bf 2015}, {\em 804},~27.
\newblock
  https://doi.org/{\changeurlcolor{black}\href{https://doi.org/10.1088/0004-637X/804/1/27}{\detokenize{10.1088/0004-637X/804/1/27}}}.

\bibitem[{Packham} {et~al.}(2005){Packham}, {Radomski}, {Roche}, {Aitken},
  {Perlman}, {Alonso-Herrero}, {Colina}, and {Telesco}]{Packham2005}
{Packham}, C.; {Radomski}, J.T.; {Roche}, P.F.; {Aitken}, D.K.; {Perlman}, E.;
  {Alonso-Herrero}, A.; {Colina}, L.; {Telesco}, C.M.
\newblock {The Extended Mid-Infrared Structure of the Circinus Galaxy}.
\newblock {\em Astrophys. J. Lett.} {\bf 2005}, {\em 618},~L17--L20.
\newblock
  https://doi.org/{\changeurlcolor{black}\href{https://doi.org/10.1086/427691}{\detokenize{10.1086/427691}}}.

\bibitem[{Roche} {et~al.}(2006){Roche}, {Packham}, {Telesco}, {Radomski},
  {Alonso-Herrero}, {Aitken}, {Colina}, and {Perlman}]{Roche2006}
{Roche}, P.F.; {Packham}, C.; {Telesco}, C.M.; {Radomski}, J.T.;
  {Alonso-Herrero}, A.; {Aitken}, D.K.; {Colina}, L.; {Perlman}, E.
\newblock {Mid-infrared, spatially resolved spectroscopy of the nucleus of the
  Circinus galaxy}.
\newblock {\em Mon. Not. R. Astron. Soc.} {\bf 2006},
  {\em 367},~1689--1698.
\newblock
  https://doi.org/{\changeurlcolor{black}\href{https://doi.org/10.1111/j.1365-2966.2006.10072.x}{\detokenize{10.1111/j.1365-2966.2006.10072.x}}}.

\bibitem[{H{\"o}nig} {et~al.}(2010){H{\"o}nig}, {Kishimoto}, {Gandhi},
  {Smette}, {Asmus}, {Duschl}, {Polletta}, and {Weigelt}]{Honig2010}
{H{\"o}nig}, S.F.; {Kishimoto}, M.; {Gandhi}, P.; {Smette}, A.; {Asmus}, D.;
  {Duschl}, W.; {Polletta}, M.; {Weigelt}, G.
\newblock {The dusty heart of nearby active galaxies. I. High-spatial
  resolution mid-IR spectro-photometry of Seyfert galaxies}.
\newblock {\em Astron. Astrophys.} {\bf 2010}, {\em 515},~A23.
\newblock
  https://doi.org/{\changeurlcolor{black}\href{https://doi.org/10.1051/0004-6361/200913742}{\detokenize{10.1051/0004-6361/200913742}}}.

\bibitem[{Alonso-Herrero} {et~al.}(2016){Alonso-Herrero}, {Poulton},
  {Roche}, {Hern{\'a}n-Caballero}, {Aretxaga}, {Mart{\'\i}nez-Paredes}, {Ramos
  Almeida}, {Pereira-Santaella}, {D{\'\i}az-Santos}, {Levenson}, {Packham},
  {Colina}, {Esquej}, {Gonz{\'a}lez-Mart{\'\i}n}, {Ichikawa}, {Imanishi},
  {Rodr{\'\i}guez Espinosa}, and {Telesco}]{Alonso-Herrero2016}
{Alonso-Herrero}, A.; {Poulton}, R.; {Roche}, P.F.; {Hern{\'a}n-Caballero}, A.;
  {Aretxaga}, I.; {Mart{\'\i}nez-Paredes}, M.; {Ramos Almeida}, C.;
  {Pereira-Santaella}, M.; {D{\'\i}az-Santos}, T.; {Levenson}, N.A.;  et~al.
\newblock {The complex evolutionary paths of local infrared bright galaxies: A
  high-angular resolution mid-infrared view}.
\newblock {\em Mon. Not. R. Astron. Soc.} {\bf 2016},
  {\em 463},~2405--2424.
\newblock
  https://doi.org/{\changeurlcolor{black}\href{https://doi.org/10.1093/mnras/stw2031}{\detokenize{10.1093/mnras/stw2031}}}.

\bibitem[{Lyu} {et~al.}(2019){Lyu}, {Rieke}, and {Smith}]{Lyu2019}
{Lyu}, J.; {Rieke}, G.H.; {Smith}, P.S.
\newblock {Mid-IR Variability and Dust Reverberation Mapping of Low-z Quasars.
  I. Data, Methods, and Basic Results}.
\newblock {\em Astrophys. J.} {\bf 2019}, {\em 886},~33.
\newblock
  https://doi.org/{\changeurlcolor{black}\href{https://doi.org/10.3847/1538-4357/ab481d}{\detokenize{10.3847/1538-4357/ab481d}}}.

\bibitem[{Juneau} {et~al.}(2022){Juneau}, {Goulding}, {Banfield}, {Bianchi},
  {Duc}, {Ho}, {Dopita}, {Scharw{\"a}chter}, {Bauer}, {Groves}, {Alexander},
  {Davies}, {Elbaz}, {Freeland}, {Hampton}, {Kewley}, {Nikutta}, {Shastri},
  {Shu}, {Vogt}, {Wang}, {Wong}, and {Woo}]{Juneau2022}
{Juneau}, S.; {Goulding}, A.D.; {Banfield}, J.; {Bianchi}, S.; {Duc}, P.A.;
  {Ho}, I.T.; {Dopita}, M.A.; {Scharw{\"a}chter}, J.; {Bauer}, F.E.; {Groves},
  B.;  et~al.
\newblock {The Black Hole-Galaxy Connection: Interplay between Feedback,
  Obscuration, and Host Galaxy Substructure}.
\newblock {\em Astrophys. J.} {\bf 2022}, {\em 925},~203.
\newblock
  https://doi.org/{\changeurlcolor{black}\href{https://doi.org/10.3847/1538-4357/ac425f}{\detokenize{10.3847/1538-4357/ac425f}}}.

\bibitem[{Kessler} {et~al.}(1996){Kessler}, {Steinz}, {Anderegg}, {Clavel},
  {Drechsel}, {Estaria}, {Faelker}, {Riedinger}, {Robson}, {Taylor}, and
  {Xim{\'e}nez de Ferr{\'a}n}]{Kessler1996}
{Kessler}, M.F.; {Steinz}, J.A.; {Anderegg}, M.E.; {Clavel}, J.; {Drechsel},
  G.; {Estaria}, P.; {Faelker}, J.; {Riedinger}, J.R.; {Robson}, A.; {Taylor},
  B.G.;  et~al.
\newblock {The Infrared Space Observatory (ISO) mission.}
\newblock {\em Astron. Astrophys.} {\bf 1996}, {\em 315},~L27--L31.

\bibitem[{Werner} {et~al.}(2004){Werner}, {Roellig}, {Low}, {Rieke},
  {Rieke}, {Hoffmann}, {Young}, {Houck}, {Brandl}, {Fazio}, {Hora}, {Gehrz},
  {Helou}, {Soifer}, {Stauffer}, {Keene}, {Eisenhardt}, {Gallagher}, {Gautier},
  {Irace}, {Lawrence}, {Simmons}, {Van Cleve}, {Jura}, {Wright}, and
  {Cruikshank}]{Werner2004}
{Werner}, M.W.; {Roellig}, T.L.; {Low}, F.J.; {Rieke}, G.H.; {Rieke}, M.;
  {Hoffmann}, W.F.; {Young}, E.; {Houck}, J.R.; {Brandl}, B.; {Fazio}, G.G.;
  et~al.
\newblock {The Spitzer Space Telescope Mission}.
\newblock {\em Astrophys. J. Suppl.} {\bf 2004}, {\em 154},~1--9.
\newblock
  https://doi.org/{\changeurlcolor{black}\href{https://doi.org/10.1086/422992}{\detokenize{10.1086/422992}}}.

\bibitem[{Wright} {et~al.}(2010){Wright}, {Eisenhardt}, {Mainzer},
  {Ressler}, {Cutri}, {Jarrett}, {Kirkpatrick}, {Padgett}, {McMillan},
  {Skrutskie}, {Stanford}, {Cohen}, {Walker}, {Mather}, {Leisawitz}, {Gautier},
  {McLean}, {Benford}, {Lonsdale}, {Blain}, {Mendez}, {Irace}, {Duval}, {Liu},
  {Royer}, {Heinrichsen}, {Howard}, {Shannon}, {Kendall}, {Walsh}, {Larsen},
  {Cardon}, {Schick}, {Schwalm}, {Abid}, {Fabinsky}, {Naes}, and
  {Tsai}]{Wright2010}
{Wright}, E.L.; {Eisenhardt}, P.R.M.; {Mainzer}, A.K.; {Ressler}, M.E.;
  {Cutri}, R.M.; {Jarrett}, T.; {Kirkpatrick}, J.D.; {Padgett}, D.; {McMillan},
  R.S.; {Skrutskie}, M.;  et~al.
\newblock {The Wide-field Infrared Survey Explorer (WISE): Mission Description
  and Initial On-orbit Performance}.
\newblock {\em Astron. J.} {\bf 2010}, {\em 140},~1868--1881.
\newblock
  https://doi.org/{\changeurlcolor{black}\href{https://doi.org/10.1088/0004-6256/140/6/1868}{\detokenize{10.1088/0004-6256/140/6/1868}}}.

\bibitem[{Leinert} {et~al.}(2003){Leinert}, {Graser}, {Przygodda}, {Waters},
  {Perrin}, {Jaffe}, {Lopez}, {Bakker}, {B{\"o}hm}, {Chesneau}, {Cotton},
  {Damstra}, {de Jong}, {Glazenborg-Kluttig}, {Grimm}, {Hanenburg}, {Laun},
  {Lenzen}, {Ligori}, {Mathar}, {Meisner}, {Morel}, {Morr}, {Neumann}, {Pel},
  {Schuller}, {Rohloff}, {Stecklum}, {Storz}, {von der L{\"u}he}, and
  {Wagner}]{Leinert2003}
{Leinert}, C.; {Graser}, U.; {Przygodda}, F.; {Waters}, L.B.F.M.; {Perrin}, G.;
  {Jaffe}, W.; {Lopez}, B.; {Bakker}, E.J.; {B{\"o}hm}, A.; {Chesneau}, O.;
  et~al.
\newblock {MIDI---The 10 $\upmu$m instrument on the VLTI}.
\newblock {\em Astrophys. Space Sci.} {\bf 2003}, {\em 286},~73--83.
\newblock
  https://doi.org/{\changeurlcolor{black}\href{https://doi.org/10.1023/A:1026158127732}{\detokenize{10.1023/A:1026158127732}}}.

\bibitem[{Lopez} {et~al.}(2022){Lopez}, {Lagarde}, {Petrov}, {Jaffe},
  {Antonelli}, {Allouche}, {Berio}, {Matter}, {Meilland}, {Millour},
  {Robbe-Dubois}, {Henning}, {Weigelt}, {Glindemann}, {Agocs}, {Bailet},
  {Beckmann}, {Bettonvil}, {van Boekel}, {Bourget}, {Bresson}, {Bristow},
  {Cruzal{\`e}bes}, {Eldswijk}, {Fante{\"\i} Caujolle}, {Gonz{\'a}lez Herrera},
  {Graser}, {Guajardo}, {Heininger}, {Hofmann}, {Kroes}, {Laun}, {Lehmitz},
  {Leinert}, {Meisenheimer}, {Morel}, {Neumann}, {Paladini}, {Percheron},
  {Riquelme}, {Schoeller}, {Stee}, {Venema}, {Woillez}, {Zins},
  {{\'A}brah{\'a}m}, {Abadie}, {Abuter}, {Accardo}, {Adler}, {Alonso},
  {Augereau}, {B{\"o}hm}, {Bazin}, {Beltran}, {Bensberg}, {Boland}, {Brast},
  {Burtscher}, {Castillo}, {Chelli}, {Cid}, {Clausse}, {Connot}, {Conzelmann},
  {Danchi}, {Delbo}, {Drevon}, {Dominik}, {van Duin}, {Ebert}, {Eisenhauer},
  {Flament}, {Frahm}, {G{\'a}mez Rosas}, {Gabasch}, {Gallenne}, {Garces},
  {Girard}, {Glazenborg}, {Gont{\'e}}, {Guitton}, {de Haan}, {Hanenburg},
  {Haubois}, {Hocd{\'e}}, {Hogerheijde}, {ter Horst}, {Hron}, {Hummel},
  {Hubin}, {Huerta}, {Idserda}, {Isbell}, {Ives}, {Jakob}, {Jask{\'o}},
  {Jochum}, {Klarmann}, {Klein}, {Kragt}, {Kuindersma}, {Kokoulina}, {Labadie},
  {Lacour}, {Leftley}, {Le Poole}, {Lizon}, {Lopez}, {Lykou}, {M{\'e}rand},
  {Marcotto}, {Mauclert}, {Maurer}, {Mehrgan}, {Meisner}, {Meixner}, {Mellein},
  {Menut}, {Mohr}, {Mosoni}, {Navarro}, {Nu{\ss}baum}, {Pallanca}, {Pantin},
  {Pasquini}, {Phan Duc}, {Pott}, {Pozna}, {Richichi}, {Ridinger}, {Rigal},
  {Rivinius}, {Roelfsema}, {Rohloff}, {Rousseau}, {Salabert}, {Schertl},
  {Schuhler}, {Schuil}, {Shabun}, {Soulain}, {Stephan}, {Toledo}, {Tristram},
  {Tromp}, {Vakili}, {Varga}, {Vinther}, {Waters}, {Wittkowski}, {Wolf},
  {Wrhel}, and {Yoffe}]{Lopez2022}
{Lopez}, B.; 
 {Lagarde}, S.; {Petrov}, R.G.; {Jaffe}, W.; {Antonelli}, P.;
  {Allouche}, F.; {Berio}, P.; {Matter}, A.; {Meilland}, A.; {Millour}, F.;
  et~al.
\newblock {MATISSE, the VLTI mid-infrared imaging spectro-interferometer}.
\newblock {\em Astron. Astrophys.} {\bf 2022}, {\em 659},~A192.
\newblock
  https://doi.org/{\changeurlcolor{black}\href{https://doi.org/10.1051/0004-6361/202141785}{\detokenize{10.1051/0004-6361/202141785}}}.

\bibitem[{Rieke} {et~al.}(2015){Rieke}, {Wright}, {B{\"o}ker}, {Bouwman},
  {Colina}, {Glasse}, {Gordon}, {Greene}, {G{\"u}del}, {Henning}, {Justtanont},
  {Lagage}, {Meixner}, {N{\o}rgaard-Nielsen}, {Ray}, {Ressler}, {van Dishoeck},
  and {Waelkens}]{Rieke2015}
{Rieke}, G.H.; {Wright}, G.S.; {B{\"o}ker}, T.; {Bouwman}, J.; {Colina}, L.;
  {Glasse}, A.; {Gordon}, K.D.; {Greene}, T.P.; {G{\"u}del}, M.; {Henning}, T.;
   et~al.
\newblock {The Mid-Infrared Instrument for the James Webb Space Telescope, I:
  Introduction}.
\newblock {\em Publ. Astron. Soc. Pac.} {\bf
  2015}, {\em 127},~584.
\newblock
  https://doi.org/{\changeurlcolor{black}\href{https://doi.org/10.1086/682252}{\detokenize{10.1086/682252}}}.

\bibitem[Lyu \& Rieke(2022)]{LyuRieke_review2022} Lyu, J. \& Rieke, G.\ 2022, Universe, 8, 304. doi:10.3390/universe8060304

\bibitem[U(2022)]{U_review2022} U, V.\ 2022, Universe, 8, 392. doi:10.3390/universe8080392
  
\bibitem[{Kleinmann} {et~al.}(1976){Kleinmann}, {Gillett}, and
  {Wright}]{Kleinmann1976}
{Kleinmann}, D.E.; {Gillett}, F.C.; {Wright}, E.L.
\newblock {8-13 Micron Spectrophotometry of NGC 1068}.
\newblock {\em Astrophys. J.} {\bf 1976}, {\em 208},~42--46.
\newblock
  https://doi.org/{\changeurlcolor{black}\href{https://doi.org/10.1086/154579}{\detokenize{10.1086/154579}}}.

\bibitem[{Le Floc'h} {et~al.}(2001){Le Floc'h}, {Mirabel}, {Laurent},
  {Charmandaris}, {Gallais}, {Sauvage}, {Vigroux}, and {Cesarsky}]{LeFloch2001}
{Le Floc'h}, E.; {Mirabel}, I.F.; {Laurent}, O.; {Charmandaris}, V.; {Gallais},
  P.; {Sauvage}, M.; {Vigroux}, L.; {Cesarsky}, C.
\newblock {Mid-Infrared observations of NGC 1068 with the Infrared Space
  Observatory}.
\newblock {\em Astron. Astrophys.} {\bf 2001}, {\em 367},~487--497.
\newblock
  https://doi.org/{\changeurlcolor{black}\href{https://doi.org/10.1051/0004-6361:20000569}{\detokenize{10.1051/0004-6361:20000569}}}.

\bibitem[{Tommasin} {et~al.}(2008){Tommasin}, {Spinoglio}, {Malkan},
  {Smith}, {Gonz{\'a}lez-Alfonso}, and {Charmandaris}]{Tommasin2008}
{Tommasin}, S.; {Spinoglio}, L.; {Malkan}, M.A.; {Smith}, H.;
  {Gonz{\'a}lez-Alfonso}, E.; {Charmandaris}, V.
\newblock {Spitzer IRS High-Resolution Spectroscopy of the 12
  {\ensuremath{\mu}}m Seyfert Galaxies. I. First Results}.
\newblock {\em Astrophys. J.} {\bf 2008}, {\em 676},~836--856.
\newblock
  https://doi.org/{\changeurlcolor{black}\href{https://doi.org/10.1086/527290}{\detokenize{10.1086/527290}}}.

\bibitem[{Rogalski}(2012)]{Rogalski2012}
{Rogalski}, A.
\newblock {History of infrared detectors}.
\newblock {\em Opto-Electron. Rev.} {\bf 2012}, {\em 20},~279--308.
\newblock
  https://doi.org/{\changeurlcolor{black}\href{https://doi.org/10.2478/s11772-012-0037-7}{\detokenize{10.2478/s11772-012-0037-7}}}.

\bibitem[{Low} and {Johnson}(1965)]{Low1965}
{Low}, F.J.; {Johnson}, H.L.
\newblock {The Spectrum of 3c 273.}
\newblock {\em Astrophys. J.} {\bf 1965}, {\em 141},~336.
\newblock
  https://doi.org/{\changeurlcolor{black}\href{https://doi.org/10.1086/148129}{\detokenize{10.1086/148129}}}.

\bibitem[{Low} and {Kleinmann}(1968)]{Low1968}
{Low}, J.; {Kleinmann}, D.E.
\newblock {Proceedings of the Conference on Seyfert Galaxies and Related
  Objects: 17. Infrared Observations of Seyfert Galaxies, Quasistellar Sources,
  and Planetary Nebulae}.
\newblock {\em Astron. J.} {\bf 1968}, {\em 73},~868.
\newblock
  https://doi.org/{\changeurlcolor{black}\href{https://doi.org/10.1086/110722}{\detokenize{10.1086/110722}}}.

\bibitem[{Rees} {et~al.}(1969){Rees}, {Silk}, {Werner}, and
  {Wickramasinghe}]{Rees1969}
{Rees}, M.J.; {Silk}, J.I.; {Werner}, M.W.; {Wickramasinghe}, N.C.
\newblock {Infrared Radiation from Dust in Seyfert Galaxies}.
\newblock {\em Nature} {\bf 1969}, {\em 223},~788--791.
\newblock
  https://doi.org/{\changeurlcolor{black}\href{https://doi.org/10.1038/223788a0}{\detokenize{10.1038/223788a0}}}.

\bibitem[{Rieke} and {Low}(1972{\natexlab{a}})]{RiekeLow1972}
\textls[-15]{{Rieke}, G.H.; {Low}, F.J.
\newblock {Infrared Photometry of Extragalactic Sources}.
\newblock {\em Astrophys. J. Lett.} {\bf 1972}, {\em 176},~L95.
\newblock
  https://doi.org/{\changeurlcolor{black}\href{https://doi.org/10.1086/181031}{\detokenize{10.1086/181031}}}.}

\bibitem[{Rieke} and {Low}(1972{\natexlab{b}})]{Rieke1972b}
\textls[-25]{{Rieke}, G.H.; {Low}, F.J.
\newblock {Variability of Extragalactic Sources at 10 Microns}.
\newblock {\em Astrophys. J. Lett.} {\bf 1972}, {\em 177},~L115.
\newblock
  https://doi.org/{\changeurlcolor{black}\href{https://doi.org/10.1086/181063}{\detokenize{10.1086/181063}}}.}

\bibitem[{Gillett} {et~al.}(1975){Gillett}, {Kleinmann}, {Wright}, and
  {Capps}]{Gillett1975}
{Gillett}, F.C.; {Kleinmann}, D.E.; {Wright}, E.L.; {Capps}, R.W.
\newblock {Observations of M82 and NGC 253 at 8 - 13 microns.}
\newblock {\em Astrophys. J. Lett.} {\bf 1975}, {\em 198},~L65--L68.
\newblock
  https://doi.org/{\changeurlcolor{black}\href{https://doi.org/10.1086/181813}{\detokenize{10.1086/181813}}}.

\bibitem[{Lebofsky} and {Rieke}(1979)]{LebofskyRieke1979}
\textls[-30]{{Lebofsky}, M.J.; {Rieke}, G.H.
\newblock {Extinction in infrared-emitting galactic nuclei.}
\newblock {\em Astrophys. J.} {\bf 1979}, {\em 229},~111--117.
\newblock
  https://doi.org/{\changeurlcolor{black}\href{https://doi.org/10.1086/156934}{\detokenize{10.1086/156934}}}.}

\bibitem[{Russell} {et~al.}(1977{\natexlab{a}}){Russell}, {Soifer}, and
  {Merrill}]{Russell1977a}
{Russell}, R.W.; {Soifer}, B.T.; {Merrill}, K.M.
\newblock {Observations of the unidentified 3.3 micrometer emission feature in
  nebulae.}
\newblock {\em Astrophys. J.} {\bf 1977}, {\em 213},~66--70.
\newblock
  https://doi.org/{\changeurlcolor{black}\href{https://doi.org/10.1086/155129}{\detokenize{10.1086/155129}}}.

\bibitem[{Russell} {et~al.}(1977{\natexlab{b}}){Russell}, {Soifer}, and
  {Willner}]{Russell1977b}
{Russell}, R.W.; {Soifer}, B.T.; {Willner}, S.P.
\newblock {The 4 to 8 micron spectrum of NGC 7027.}
\newblock {\em Astrophys. J. Lett.} {\bf 1977}, {\em 217},~L149.
\newblock
  https://doi.org/{\changeurlcolor{black}\href{https://doi.org/10.1086/182559}{\detokenize{10.1086/182559}}}.

\bibitem[{Neugebauer} {et~al.}(1984){Neugebauer}, {Habing}, {van Duinen},
  {Aumann}, {Baud}, {Beichman}, {Beintema}, {Boggess}, {Clegg}, {de Jong},
  {Emerson}, {Gautier}, {Gillett}, {Harris}, {Hauser}, {Houck}, {Jennings},
  {Low}, {Marsden}, {Miley}, {Olnon}, {Pottasch}, {Raimond}, {Rowan-Robinson},
  {Soifer}, {Walker}, {Wesselius}, and {Young}]{Neugebauer1984}
{Neugebauer}, G.; {Habing}, H.J.; {van Duinen}, R.; {Aumann}, H.H.; {Baud}, B.;
  {Beichman}, C.A.; {Beintema}, D.A.; {Boggess}, N.; {Clegg}, P.E.; {de Jong},
  T.;  et~al.
\newblock {The Infrared Astronomical Satellite (IRAS) mission.}
\newblock {\em Astrophys. J. Lett.} {\bf 1984}, {\em 278},~L1--L6.
\newblock
  https://doi.org/{\changeurlcolor{black}\href{https://doi.org/10.1086/184209}{\detokenize{10.1086/184209}}}.

\bibitem[{Gehrz} {et~al.}(2007){Gehrz}, {Roellig}, {Werner}, {Fazio},
  {Houck}, {Low}, {Rieke}, {Soifer}, {Levine}, and {Romana}]{Gehrz2007}
{Gehrz}, R.D.; {Roellig}, T.L.; {Werner}, M.W.; {Fazio}, G.G.; {Houck}, J.R.;
  {Low}, F.J.; {Rieke}, G.H.; {Soifer}, B.T.; {Levine}, D.A.; {Romana}, E.A.
\newblock {The NASA Spitzer Space Telescope}.
\newblock {\em Rev. Sci. Instrum.} {\bf 2007}, {\em
  78},~11302.
\newblock
  https://doi.org/{\changeurlcolor{black}\href{https://doi.org/10.1063/1.2431313}{\detokenize{10.1063/1.2431313}}}.

\bibitem[{Yan} {et~al.}(2007){Yan}, {Sajina}, {Fadda}, {Choi}, {Armus},
  {Helou}, {Teplitz}, {Frayer}, and {Surace}]{Yan2007}
{Yan}, L.; {Sajina}, A.; {Fadda}, D.; {Choi}, P.; {Armus}, L.; {Helou}, G.;
  {Teplitz}, H.; {Frayer}, D.; {Surace}, J.
\newblock {Spitzer Mid-Infrared Spectroscopy of Infrared Luminous Galaxies at z
  \raisebox{-0.5ex}\textasciitilde 2. I. The Spectra}.
\newblock {\em Astrophys. J.} {\bf 2007}, {\em 658},~778--793.
\newblock
  https://doi.org/{\changeurlcolor{black}\href{https://doi.org/10.1086/511516}{\detokenize{10.1086/511516}}}.

\bibitem[{Sajina} {et~al.}(2007){Sajina}, {Yan}, {Armus}, {Choi}, {Fadda},
  {Helou}, and {Spoon}]{Sajina2007a}
{Sajina}, A.; {Yan}, L.; {Armus}, L.; {Choi}, P.; {Fadda}, D.; {Helou}, G.;
  {Spoon}, H.
\newblock {Spitzer Mid-Infrared Spectroscopy of Infrared Luminous Galaxies at
  z\raisebox{-0.5ex}\textasciitilde2. II. Diagnostics}.
\newblock {\em Astrophys. J.} {\bf 2007}, {\em 664},~713--737.
\newblock
  https://doi.org/{\changeurlcolor{black}\href{https://doi.org/10.1086/519446}{\detokenize{10.1086/519446}}}.

\bibitem[{Pope} {et~al.}(2008){Pope}, {Bussmann}, {Dey}, {Meger},
  {Alexander}, {Brodwin}, {Chary}, {Dickinson}, {Frayer}, {Greve}, {Huynh},
  {Lin}, {Morrison}, {Scott}, and {Yan}]{Pope2008}
{Pope}, A.; {Bussmann}, R.S.; {Dey}, A.; {Meger}, N.; {Alexander}, D.M.;
  {Brodwin}, M.; {Chary}, R.R.; {Dickinson}, M.E.; {Frayer}, D.T.; {Greve},
  T.R.;  et~al.
\newblock {The Nature of Faint Spitzer-selected Dust-obscured Galaxies}.
\newblock {\em Astrophys. J.} {\bf 2008}, {\em 689},~127--133.
\newblock
  https://doi.org/{\changeurlcolor{black}\href{https://doi.org/10.1086/592739}{\detokenize{10.1086/592739}}}.

\bibitem[{Dasyra} {et~al.}(2008){Dasyra}, {Ho}, {Armus}, {Ogle}, {Helou},
  {Peterson}, {Lutz}, {Netzer}, and {Sturm}]{Dasyra2008}
{Dasyra}, K.M.; {Ho}, L.C.; {Armus}, L.; {Ogle}, P.; {Helou}, G.; {Peterson},
  B.M.; {Lutz}, D.; {Netzer}, H.; {Sturm}, E.
\newblock {High-Ionization Mid-Infrared Lines as Black Hole Mass and Bolometric
  Luminosity Indicators in Active Galactic Nuclei}.
\newblock {\em Astrophys. J. Lett.} {\bf 2008}, {\em 674},~L9.
\newblock
  https://doi.org/{\changeurlcolor{black}\href{https://doi.org/10.1086/528843}{\detokenize{10.1086/528843}}}.

\bibitem[{Mart{\'\i}nez-Sansigre} {et~al.}(2008){Mart{\'\i}nez-Sansigre},
  {Lacy}, {Sajina}, and {Rawlings}]{MartinezSansigre2008}
{Mart{\'\i}nez-Sansigre}, A.; {Lacy}, M.; {Sajina}, A.; {Rawlings}, S.
\newblock {Mid-Infrared Spectroscopy of High-Redshift Obscured Quasars}.
\newblock {\em Astrophys. J.} {\bf 2008}, {\em 674},~676--685.
\newblock
  https://doi.org/{\changeurlcolor{black}\href{https://doi.org/10.1086/525245}{\detokenize{10.1086/525245}}}.

\bibitem[{Seymour} {et~al.}(2008){Seymour}, {Ogle}, {De Breuck}, {Fazio},
  {Galametz}, {Haas}, {Lacy}, {Sajina}, {Stern}, {Willner}, and
  {Vernet}]{Seymour2008}
{Seymour}, N.; {Ogle}, P.; {De Breuck}, C.; {Fazio}, G.G.; {Galametz}, A.;
  {Haas}, M.; {Lacy}, M.; {Sajina}, A.; {Stern}, D.; {Willner}, S.P.;  et~al.
\newblock {Mid-Infrared Spectra of High-Redshift (z > 2) Radio Galaxies}.
\newblock {\em Astrophys. J. Lett.} {\bf 2008}, {\em 681},~L1.
\newblock
  https://doi.org/{\changeurlcolor{black}\href{https://doi.org/10.1086/590081}{\detokenize{10.1086/590081}}}.

\bibitem[{Fadda} {et~al.}(2010){Fadda}, {Yan}, {Lagache}, {Sajina}, {Lutz},
  {Wuyts}, {Frayer}, {Marcillac}, {Le Floc'h}, {Caputi}, {Spoon}, {Veilleux},
  {Blain}, and {Helou}]{Fadda2010}
{Fadda}, D.; {Yan}, L.; {Lagache}, G.; {Sajina}, A.; {Lutz}, D.; {Wuyts}, S.;
  {Frayer}, D.T.; {Marcillac}, D.; {Le Floc'h}, E.; {Caputi}, K.;  et~al.
\newblock {Ultra-deep Mid-infrared Spectroscopy of Luminous Infrared Galaxies
  at z \raisebox{-0.5ex}\textasciitilde 1 and z
  \raisebox{-0.5ex}\textasciitilde 2}.
\newblock {\em Astrophys. J.} {\bf 2010}, {\em 719},~425--450.
\newblock
  https://doi.org/{\changeurlcolor{black}\href{https://doi.org/10.1088/0004-637X/719/1/425}{\detokenize{10.1088/0004-637X/719/1/425}}}.

\bibitem[{Genzel} {et~al.}(1998){Genzel}, {Lutz}, {Sturm}, {Egami}, {Kunze},
  {Moorwood}, {Rigopoulou}, {Spoon}, {Sternberg}, {Tacconi-Garman}, {Tacconi},
  and {Thatte}]{Genzel1998}
{Genzel}, R.; {Lutz}, D.; {Sturm}, E.; {Egami}, E.; {Kunze}, D.; {Moorwood},
  A.F.M.; {Rigopoulou}, D.; {Spoon}, H.W.W.; {Sternberg}, A.; {Tacconi-Garman},
  L.E.;  et~al.
\newblock {What Powers Ultraluminous IRAS Galaxies?}
\newblock {\em Astrophys. J.} {\bf 1998}, {\em 498},~579--605.
\newblock
  https://doi.org/{\changeurlcolor{black}\href{https://doi.org/10.1086/305576}{\detokenize{10.1086/305576}}}.

\bibitem[{Sturm} {et~al.}(2000){Sturm}, {Lutz}, {Tran}, {Feuchtgruber},
  {Genzel}, {Kunze}, {Moorwood}, and {Thornley}]{Sturm2000}
{Sturm}, E.; {Lutz}, D.; {Tran}, D.; {Feuchtgruber}, H.; {Genzel}, R.; {Kunze},
  D.; {Moorwood}, A.F.M.; {Thornley}, M.D.
\newblock {ISO-SWS spectra of galaxies: Continuum and features}.
\newblock {\em Astron. Astrophys.} {\bf 2000}, {\em 358},~481--493.

\bibitem[{Brandl} {et~al.}(2006){Brandl}, {Bernard-Salas}, {Spoon},
  {Devost}, {Sloan}, {Guilles}, {Wu}, {Houck}, {Weedman}, {Armus}, {Appleton},
  {Soifer}, {Charmandaris}, {Hao}, {Higdon}, {Marshall}, and
  {Herter}]{Brandl2006}
{Brandl}, B.R.; {Bernard-Salas}, J.; {Spoon}, H.W.W.; {Devost}, D.; {Sloan},
  G.C.; {Guilles}, S.; {Wu}, Y.; {Houck}, J.R.; {Weedman}, D.W.; {Armus}, L.;
  et~al.
\newblock {The Mid-Infrared Properties of Starburst Galaxies from Spitzer-IRS
  Spectroscopy}.
\newblock {\em Astrophys. J.} {\bf 2006}, {\em 653},~1129--1144.
\newblock
  https://doi.org/{\changeurlcolor{black}\href{https://doi.org/10.1086/508849}{\detokenize{10.1086/508849}}}.

\bibitem[{Lambrides} {et~al.}(2019){Lambrides}, {Petric}, {Tchernyshyov},
  {Zakamska}, and {Watts}]{Lambrides2019}
{Lambrides}, E.L.; {Petric}, A.O.; {Tchernyshyov}, K.; {Zakamska}, N.L.;
  {Watts}, D.J.
\newblock {Mid-infrared spectroscopic evidence for AGN heating warm molecular
  gas}.
\newblock {\em Mon. Not. R. Astron. Soc.} {\bf 2019},
  {\em 487},~1823--1843.
\newblock
  https://doi.org/{\changeurlcolor{black}\href{https://doi.org/10.1093/mnras/stz1316}{\detokenize{10.1093/mnras/stz1316}}}.

\bibitem[{Shipley} {et~al.}(2016){Shipley}, {Papovich}, {Rieke}, {Brown},
  and {Moustakas}]{Shipley2016}
{Shipley}, H.V.; {Papovich}, C.; {Rieke}, G.H.; {Brown}, M.J.I.; {Moustakas},
  J.
\newblock {A New Star Formation Rate Calibration from Polycyclic Aromatic
  Hydrocarbon Emission Features and Application to High-redshift Galaxies}.
\newblock {\em Astrophys. J.} {\bf 2016}, {\em 818},~60.
\newblock
  https://doi.org/{\changeurlcolor{black}\href{https://doi.org/10.3847/0004-637X/818/1/60}{\detokenize{10.3847/0004-637X/818/1/60}}}.

\bibitem[{Hern{\'a}n-Caballero} {et~al.}(2009){Hern{\'a}n-Caballero},
  {P{\'e}rez-Fournon}, {Hatziminaoglou}, {Afonso-Luis}, {Rowan-Robinson},
  {Rigopoulou}, {Farrah}, {Lonsdale}, {Babbedge}, {Clements}, {Serjeant},
  {Pozzi}, {Vaccari}, {Montenegro-Montes}, {Valtchanov},
  {Gonz{\'a}lez-Solares}, {Oliver}, {Shupe}, {Gruppioni}, {Vila-Vilar{\'o}},
  {Lari}, and {La Franca}]{Hernan-Caballero2009}
{Hern{\'a}n-Caballero}, A.; {P{\'e}rez-Fournon}, I.; {Hatziminaoglou}, E.;
  {Afonso-Luis}, A.; {Rowan-Robinson}, M.; {Rigopoulou}, D.; {Farrah}, D.;
  {Lonsdale}, C.J.; {Babbedge}, T.; {Clements}, D.;  et~al.
\newblock {Mid-infrared spectroscopy of infrared-luminous galaxies at z
  \raisebox{-0.5ex}\textasciitilde 0.5-3}.
\newblock {\em Mon. Not. R. Astron. Soc.} {\bf 2009},
  {\em 395},~1695--1722.
\newblock
  https://doi.org/{\changeurlcolor{black}\href{https://doi.org/10.1111/j.1365-2966.2009.14660.x}{\detokenize{10.1111/j.1365-2966.2009.14660.x}}}.

\bibitem[{Hern{\'a}n-Caballero} and
  {Hatziminaoglou}(2011)]{Hernan-Caballero2011}
{Hern{\'a}n-Caballero}, A.; {Hatziminaoglou}, E.
\newblock {An atlas of mid-infrared spectra of star-forming and active
  galaxies}.
\newblock {\em Mon. Not. R. Astron. Soc.} {\bf 2011},
  {\em 414},~500--511.
\newblock
  https://doi.org/{\changeurlcolor{black}\href{https://doi.org/10.1111/j.1365-2966.2011.18413.x}{\detokenize{10.1111/j.1365-2966.2011.18413.x}}}.

\bibitem[{Esquej} {et~al.}(2014){Esquej}, {Alonso-Herrero},
  {Gonz{\'a}lez-Mart{\'\i}n}, {H{\"o}nig}, {Hern{\'a}n-Caballero}, {Roche},
  {Ramos Almeida}, {Mason}, {D{\'\i}az-Santos}, {Levenson}, {Aretxaga},
  {Rodr{\'\i}guez Espinosa}, and {Packham}]{Esquej2014}
{Esquej}, P.; {Alonso-Herrero}, A.; {Gonz{\'a}lez-Mart{\'\i}n}, O.;
  {H{\"o}nig}, S.F.; {Hern{\'a}n-Caballero}, A.; {Roche}, P.; {Ramos Almeida},
  C.; {Mason}, R.E.; {D{\'\i}az-Santos}, T.; {Levenson}, N.A.;  et~al.
\newblock {Nuclear Star Formation Activity and Black Hole Accretion in Nearby
  Seyfert Galaxies}.
\newblock {\em Astrophys. J.} {\bf 2014}, {\em 780},~86.
\newblock
  https://doi.org/{\changeurlcolor{black}\href{https://doi.org/10.1088/0004-637X/780/1/86}{\detokenize{10.1088/0004-637X/780/1/86}}}.

\bibitem[{Alonso-Herrero} {et~al.}(2014){Alonso-Herrero}, {Ramos Almeida},
  {Esquej}, {Roche}, {Hern{\'a}n-Caballero}, {H{\"o}nig},
  {Gonz{\'a}lez-Mart{\'\i}n}, {Aretxaga}, {Mason}, {Packham}, {Levenson},
  {Rodr{\'\i}guez Espinosa}, {Siebenmorgen}, {Pereira-Santaella},
  {D{\'\i}az-Santos}, {Colina}, {Alvarez}, and {Telesco}]{Alonso-Herrero2014}
{Alonso-Herrero}, A.; {Ramos Almeida}, C.; {Esquej}, P.; {Roche}, P.F.;
  {Hern{\'a}n-Caballero}, A.; {H{\"o}nig}, S.F.; {Gonz{\'a}lez-Mart{\'\i}n},
  O.; {Aretxaga}, I.; {Mason}, R.E.; {Packham}, C.;  et~al.
\newblock {Nuclear 11.3 {\ensuremath{\mu}}m PAH emission in local active
  galactic nuclei}.
\newblock {\em Mon. Not. R. Astron. Soc.} {\bf 2014},
  {\em 443},~2766--2782.
\newblock
  https://doi.org/{\changeurlcolor{black}\href{https://doi.org/10.1093/mnras/stu1293}{\detokenize{10.1093/mnras/stu1293}}}.

\bibitem[{Inami} {et~al.}(2013){Inami}, {Armus}, {Charmandaris}, {Groves},
  {Kewley}, {Petric}, {Stierwalt}, {D{\'\i}az-Santos}, {Surace}, {Rich},
  {Haan}, {Howell}, {Evans}, {Mazzarella}, {Marshall}, {Appleton}, {Lord},
  {Spoon}, {Frayer}, {Matsuhara}, and {Veilleux}]{Inami2013}
{Inami}, H.; {Armus}, L.; {Charmandaris}, V.; {Groves}, B.; {Kewley}, L.;
  {Petric}, A.; {Stierwalt}, S.; {D{\'\i}az-Santos}, T.; {Surace}, J.; {Rich},
  J.;  et~al.
\newblock {Mid-infrared Atomic Fine-structure Emission-line Spectra of Luminous
  Infrared Galaxies: Spitzer/IRS Spectra of the GOALS Sample}.
\newblock {\em Astrophys. J.} {\bf 2013}, {\em 777},~156.
\newblock
  https://doi.org/{\changeurlcolor{black}\href{https://doi.org/10.1088/0004-637X/777/2/156}{\detokenize{10.1088/0004-637X/777/2/156}}}.

\bibitem[{Armus} {et~al.}(2007){Armus}, {Charmandaris}, {Bernard-Salas},
  {Spoon}, {Marshall}, {Higdon}, {Desai}, {Teplitz}, {Hao}, {Devost}, {Brandl},
  {Wu}, {Sloan}, {Soifer}, {Houck}, and {Herter}]{Armus2007}
{Armus}, L.; {Charmandaris}, V.; {Bernard-Salas}, J.; {Spoon}, H.W.W.;
  {Marshall}, J.A.; {Higdon}, S.J.U.; {Desai}, V.; {Teplitz}, H.I.; {Hao}, L.;
  {Devost}, D.;  et~al.
\newblock {Observations of Ultraluminous Infrared Galaxies with the Infrared
  Spectrograph on the Spitzer Space Telescope. II. The IRAS Bright Galaxy
  Sample}.
\newblock {\em Astrophys. J.} {\bf 2007}, {\em 656},~148--167.
\newblock
  https://doi.org/{\changeurlcolor{black}\href{https://doi.org/10.1086/510107}{\detokenize{10.1086/510107}}}.

\bibitem[{Pier} and {Krolik}(1992)]{PierKrolik1992}
{Pier}, E.A.; {Krolik}, J.H.
\newblock {Infrared Spectra of Obscuring Dust Tori around Active Galactic
  Nuclei. I. Calculational Method and Basic Trends}.
\newblock {\em Astrophys. J.} {\bf 1992}, {\em 401},~99.
\newblock
  https://doi.org/{\changeurlcolor{black}\href{https://doi.org/10.1086/172042}{\detokenize{10.1086/172042}}}.

\bibitem[{Nenkova} {et~al.}(2000){Nenkova}, {Ivezi{\'c}}, and
  {Elitzur}]{Nenkova2000}
{Nenkova}, M.; {Ivezi{\'c}}, {\v{Z}}.; {Elitzur}, M.
\newblock {DUSTY: A Publicly Available Code for Modeling Dust Emission}.
\newblock In {\em Thermal Emission Spectroscopy and Analysis of Dust, Disks, and
  Regoliths}; Astronomical Society of the Pacific Conference Series; {Sitko}, M.L., {Sprague}, A.L., {Lynch}, D.K., Eds.; Astronomical Society of the Pacific: San Francisco, CA, USA, 2000; Volume 196, pp. 77--82.

\bibitem[{Siebenmorgen} {et~al.}(2005){Siebenmorgen}, {Haas}, {Kr{\"u}gel},
  and {Schulz}]{Siebenmorgen2005}
{Siebenmorgen}, R.; {Haas}, M.; {Kr{\"u}gel}, E.; {Schulz}, B.
\newblock {Discovery of 10 {\ensuremath{\mu}}m silicate emission in quasars.
  Evidence of the AGN unification scheme}.
\newblock {\em Astron. Astrophys.} {\bf 2005}, {\em 436},~L5--L8.
\newblock
  https://doi.org/{\changeurlcolor{black}\href{https://doi.org/10.1051/0004-6361:200500109}{\detokenize{10.1051/0004-6361:200500109}}}.

\bibitem[{Hao} {et~al.}(2005){Hao}, {Spoon}, {Sloan}, {Marshall}, {Armus},
  {Tielens}, {Sargent}, {van Bemmel}, {Charmandaris}, {Weedman}, and
  {Houck}]{Hao2005}
{Hao}, L.; {Spoon}, H.W.W.; {Sloan}, G.C.; {Marshall}, J.A.; {Armus}, L.;
  {Tielens}, A.G.G.M.; {Sargent}, B.; {van Bemmel}, I.M.; {Charmandaris}, V.;
  {Weedman}, D.W.;  et~al.
\newblock {The Detection of Silicate Emission from Quasars at 10 and 18
  Microns}.
\newblock {\em Astrophys. J. Lett.} {\bf 2005}, {\em 625},~L75--L78.
\newblock
  https://doi.org/{\changeurlcolor{black}\href{https://doi.org/10.1086/431227}{\detokenize{10.1086/431227}}}.

\bibitem[{Veilleux} {et~al.}(2009){Veilleux}, {Rupke}, {Kim}, {Genzel},
  {Sturm}, {Lutz}, {Contursi}, {Schweitzer}, {Tacconi}, {Netzer}, {Sternberg},
  {Mihos}, {Baker}, {Mazzarella}, {Lord}, {Sanders}, {Stockton}, {Joseph}, and
  {Barnes}]{Veilleux2009}
{Veilleux}, S.; {Rupke}, D.S.N.; {Kim}, D.C.; {Genzel}, R.; {Sturm}, E.;
  {Lutz}, D.; {Contursi}, A.; {Schweitzer}, M.; {Tacconi}, L.J.; {Netzer}, H.;
  et~al.
\newblock {Spitzer Quasar and Ulirg Evolution Study (QUEST). IV. Comparison of
  1 Jy Ultraluminous Infrared Galaxies with Palomar-Green Quasars}.
\newblock {\em Astrophys. J. Suppl.} {\bf 2009}, {\em
  182},~628--666.
\newblock
  https://doi.org/{\changeurlcolor{black}\href{https://doi.org/10.1088/0067-0049/182/2/628}{\detokenize{10.1088/0067-0049/182/2/628}}}.

\bibitem[{Spoon} {et~al.}(2022){Spoon}, {Hern{\'a}n-Caballero}, {Rupke},
  {Waters}, {Lebouteiller}, {Tielens}, {Loredo}, {Su}, and {Viola}]{Spoon2022}
{Spoon}, H.W.W.; {Hern{\'a}n-Caballero}, A.; {Rupke}, D.; {Waters}, L.B.F.M.;
  {Lebouteiller}, V.; {Tielens}, A.G.G.M.; {Loredo}, T.; {Su}, Y.; {Viola}, V.
\newblock {The Infrared Database of Extragalactic Observables from Spitzer. II.
  The Database and Diagnostic Power of Crystalline Silicate Features in Galaxy
  Spectra}.
\newblock {\em Astrophys. J. Suppl.} {\bf 2022}, {\em 259},~37.
\newblock
  https://doi.org/{\changeurlcolor{black}\href{https://doi.org/10.3847/1538-4365/ac4989}{\detokenize{10.3847/1538-4365/ac4989}}}.

\bibitem[{Sajina} {et~al.}(2007){Sajina}, {Yan}, {Lacy}, and
  {Huynh}]{Sajina2007b}
{Sajina}, A.; {Yan}, L.; {Lacy}, M.; {Huynh}, M.
\newblock {Discovery of Radio Jets in z \raisebox{-0.5ex}\textasciitilde 2
  Ultraluminous Infrared Galaxies with Deep 9.7 {\ensuremath{\mu}}m Silicate
  Absorption}.
\newblock {\em Astrophys. J. Lett.} {\bf 2007}, {\em 667},~L17--L20.
\newblock
  https://doi.org/{\changeurlcolor{black}\href{https://doi.org/10.1086/522050}{\detokenize{10.1086/522050}}}.

\bibitem[{Marshall} {et~al.}(2018){Marshall}, {Elitzur}, {Armus},
  {Diaz-Santos}, and {Charmandaris}]{Marshall2018}
{Marshall}, J.A.; {Elitzur}, M.; {Armus}, L.; {Diaz-Santos}, T.;
  {Charmandaris}, V.
\newblock {The Nature of Deeply Buried Ultraluminous Infrared Galaxies: A
  Unified Model for Highly Obscured Dusty Galaxy Emission}.
\newblock {\em Astrophys. J.} {\bf 2018}, {\em 858},~59.
\newblock
  https://doi.org/{\changeurlcolor{black}\href{https://doi.org/10.3847/1538-4357/aabcc0}{\detokenize{10.3847/1538-4357/aabcc0}}}.

\bibitem[{Snyder} {et~al.}(2013){Snyder}, {Hayward}, {Sajina}, {Jonsson},
  {Cox}, {Hernquist}, {Hopkins}, and {Yan}]{Snyder2013}
{Snyder}, G.F.; {Hayward}, C.C.; {Sajina}, A.; {Jonsson}, P.; {Cox}, T.J.;
  {Hernquist}, L.; {Hopkins}, P.F.; {Yan}, L.
\newblock {Modeling Mid-infrared Diagnostics of Obscured Quasars and
  Starbursts}.
\newblock {\em Astrophys. J.} {\bf 2013}, {\em 768},~168.
\newblock
  https://doi.org/{\changeurlcolor{black}\href{https://doi.org/10.1088/0004-637X/768/2/168}{\detokenize{10.1088/0004-637X/768/2/168}}}.

\bibitem[{Laurent} {et~al.}(2000){Laurent}, {Mirabel}, {Charmandaris},
  {Gallais}, {Madden}, {Sauvage}, {Vigroux}, and {Cesarsky}]{Laurent2000}
{Laurent}, O.; {Mirabel}, I.F.; {Charmandaris}, V.; {Gallais}, P.; {Madden},
  S.C.; {Sauvage}, M.; {Vigroux}, L.; {Cesarsky}, C.
\newblock {Mid-infrared diagnostics to distinguish AGNs from starbursts}.
\newblock {\em Astron. Astrophys.} {\bf 2000}, {\em 359},~887--899.

\bibitem[{Spoon} {et~al.}(2007){Spoon}, {Marshall}, {Houck}, {Elitzur},
  {Hao}, {Armus}, {Brandl}, and {Charmandaris}]{Spoon2007}
{Spoon}, H.W.W.; {Marshall}, J.A.; {Houck}, J.R.; {Elitzur}, M.; {Hao}, L.;
  {Armus}, L.; {Brandl}, B.R.; {Charmandaris}, V.
\newblock {Mid-Infrared Galaxy Classification Based on Silicate Obscuration and
  PAH Equivalent Width}.
\newblock {\em Astrophys. J. Lett.} {\bf 2007}, {\em 654},~L49--L52.
\newblock
  https://doi.org/{\changeurlcolor{black}\href{https://doi.org/10.1086/511268}{\detokenize{10.1086/511268}}}.

\bibitem[{Spinoglio} {et~al.}(2017){Spinoglio}, {Alonso-Herrero}, {Armus},
  {Baes}, {Bernard-Salas}, {Bianchi}, {Bocchio}, {Bolatto}, {Bradford},
  {Braine}, {Carrera}, {Ciesla}, {Clements}, {Dannerbauer}, {Doi},
  {Efstathiou}, {Egami}, {Fern{\'a}ndez-Ontiveros}, {Ferrara}, {Fischer},
  {Franceschini}, {Gallerani}, {Giard}, {Gonz{\'a}lez-Alfonso}, {Gruppioni},
  {Guillard}, {Hatziminaoglou}, {Imanishi}, {Ishihara}, {Isobe}, {Kaneda},
  {Kawada}, {Kohno}, {Kwon}, {Madden}, {Malkan}, {Marassi}, {Matsuhara},
  {Matsuura}, {Miniutti}, {Nagamine}, {Nagao}, {Najarro}, {Nakagawa}, {Onaka},
  {Oyabu}, {Pallottini}, {Piro}, {Pozzi}, {Rodighiero}, {Roelfsema}, {Sakon},
  {Santini}, {Schaerer}, {Schneider}, {Scott}, {Serjeant}, {Shibai}, {Smith},
  {Sobacchi}, {Sturm}, {Suzuki}, {Vallini}, {van der Tak}, {Vignali}, {Yamada},
  {Wada}, and {Wang}]{Spinoglio2017}
{Spinoglio}, L.; {Alonso-Herrero}, A.; {Armus}, L.; {Baes}, M.;
  {Bernard-Salas}, J.; {Bianchi}, S.; {Bocchio}, M.; {Bolatto}, A.; {Bradford},
  C.; {Braine}, J.;  et~al.
\newblock {Galaxy Evolution Studies with the SPace IR Telescope for Cosmology
  and Astrophysics (SPICA): The Power of IR Spectroscopy}.
\newblock {\em Publ. Astron. Soc. Aust.} {\bf
  2017}, {\em 34},~e057.
\newblock
  https://doi.org/{\changeurlcolor{black}\href{https://doi.org/10.1017/pasa.2017.48}{\detokenize{10.1017/pasa.2017.48}}}.

\bibitem[{Spinoglio} and {Malkan}(1992)]{SpinoglioMalkan1992}
{Spinoglio}, L.; {Malkan}, M.A.
\newblock {Infrared Line Diagnostics of Active Galactic Nuclei}.
\newblock {\em Astrophys. J.} {\bf 1992}, {\em 399},~504. \linebreak
\newblock
  https://doi.org/{\changeurlcolor{black}\href{https://doi.org/10.1086/171943}{\detokenize{10.1086/171943}}}.

\bibitem[{Gorjian} {et~al.}(2007){Gorjian}, {Cleary}, {Werner}, and
  {Lawrence}]{Gorjian2007}
{Gorjian}, V.; {Cleary}, K.; {Werner}, M.W.; {Lawrence}, C.R.
\newblock {A Relation between the Mid-Infrared [Ne V] 14.3 {\ensuremath{\mu}}m
  and [Ne III] 15.6 {\ensuremath{\mu}}m Lines in Active Galactic Nuclei}.
\newblock {\em Astrophys. J. Lett.} {\bf 2007}, {\em 655},~L73--L76.
\newblock
  https://doi.org/{\changeurlcolor{black}\href{https://doi.org/10.1086/511975}{\detokenize{10.1086/511975}}}.

\bibitem[{Fern{\'a}ndez-Ontiveros} {et~al.}(2021){Fern{\'a}ndez-Ontiveros},
  {P{\'e}rez-Montero}, {V{\'\i}lchez}, {Amor{\'\i}n}, and
  {Spinoglio}]{Fernandez-Ontiveros2021}
{Fern{\'a}ndez-Ontiveros}, J.A.; {P{\'e}rez-Montero}, E.; {V{\'\i}lchez}, J.M.;
  {Amor{\'\i}n}, R.; {Spinoglio}, L.
\newblock {Measuring chemical abundances with infrared nebular lines:
  HII-CHI-MISTRY-IR}.
\newblock {\em Astron. Astrophys.} {\bf 2021}, {\em 652},~A23.
\newblock
  https://doi.org/{\changeurlcolor{black}\href{https://doi.org/10.1051/0004-6361/202039716}{\detokenize{10.1051/0004-6361/202039716}}}.

\bibitem[{Kewley} {et~al.}(2019){Kewley}, {Nicholls}, and
  {Sutherland}]{kewley2019}
{Kewley}, L.J.; {Nicholls}, D.C.; {Sutherland}, R.S.
\newblock {Understanding Galaxy Evolution Through Emission Lines}.
\newblock {\em Ann. Rev. Astron. Astrophys.} {\bf 2019}, {\em
  57},~511--570.
\newblock
  https://doi.org/{\changeurlcolor{black}\href{https://doi.org/10.1146/annurev-astro-081817-051832}{\detokenize{10.1146/annurev-astro-081817-051832}}}.

\bibitem[{Rush} {et~al.}(1993){Rush}, {Malkan}, and
  {Spinoglio}]{1993ApJS...89....1R}
{Rush}, B.; {Malkan}, M.A.; {Spinoglio}, L.
\newblock {The Extended 12 Micron Galaxy Sample}.
\newblock {\em Astrophys. J. Suppl.} {\bf 1993}, {\em 89},~1.
\newblock
  https://doi.org/{\changeurlcolor{black}\href{https://doi.org/10.1086/191837}{\detokenize{10.1086/191837}}}.

\bibitem[{Haas} {et~al.}(2004){Haas}, {Siebenmorgen}, {Leipski}, {Ott},
  {Cunow}, {Meusinger}, {M{\"u}ller}, {Chini}, and
  {Schartel}]{2004A&A...419L..49H}
{Haas}, M.; {Siebenmorgen}, R.; {Leipski}, C.; {Ott}, S.; {Cunow}, B.;
  {Meusinger}, H.; {M{\"u}ller}, S.A.H.; {Chini}, R.; {Schartel}, N.
\newblock {Mid-infrared selection of AGN}.
\newblock {\em Astron. Astrophys.} {\bf 2004}, {\em 419},~L49--L53.
\newblock
  https://doi.org/{\changeurlcolor{black}\href{https://doi.org/10.1051/0004-6361:20040143}{\detokenize{10.1051/0004-6361:20040143}}}.

\bibitem[{Sajina} {et~al.}(2005){Sajina}, {Lacy}, and {Scott}]{Sajina2005}
{Sajina}, A.; {Lacy}, M.; {Scott}, D.
\newblock {Simulating the Spitzer Mid-Infrared Color-Color Diagrams}.
\newblock {\em Astrophys. J.} {\bf 2005}, {\em 621},~256--268.
\newblock
  https://doi.org/{\changeurlcolor{black}\href{https://doi.org/10.1086/426536}{\detokenize{10.1086/426536}}}.

\bibitem[{Alonso-Herrero} {et~al.}(2006){Alonso-Herrero},
  {P{\'e}rez-Gonz{\'a}lez}, {Alexander}, {Rieke}, {Rigopoulou}, {Le Floc'h},
  {Barmby}, {Papovich}, {Rigby}, {Bauer}, {Brandt}, {Egami}, {Willner}, {Dole},
  and {Huang}]{Alonso-Herrero2006}
{Alonso-Herrero}, A.; {P{\'e}rez-Gonz{\'a}lez}, P.G.; {Alexander}, D.M.;
  {Rieke}, G.H.; {Rigopoulou}, D.; {Le Floc'h}, E.; {Barmby}, P.; {Papovich},
  C.; {Rigby}, J.R.; {Bauer}, F.E.;  et~al.
\newblock {Infrared Power-Law Galaxies in the Chandra Deep Field-South: Active
  Galactic Nuclei and Ultraluminous Infrared Galaxies}.
\newblock {\em Astrophys. J.} {\bf 2006}, {\em 640},~167--184.
\newblock
  https://doi.org/{\changeurlcolor{black}\href{https://doi.org/10.1086/499800}{\detokenize{10.1086/499800}}}.

\bibitem[{Donley} {et~al.}(2012){Donley}, {Koekemoer}, {Brusa}, {Capak},
  {Cardamone}, {Civano}, {Ilbert}, {Impey}, {Kartaltepe}, {Miyaji}, {Salvato},
  {Sanders}, {Trump}, and {Zamorani}]{2012ApJ...748..142D}
{Donley}, J.L.; {Koekemoer}, A.M.; {Brusa}, M.; {Capak}, P.; {Cardamone}, C.N.;
  {Civano}, F.; {Ilbert}, O.; {Impey}, C.D.; {Kartaltepe}, J.S.; {Miyaji}, T.;
  et~al.
\newblock {Identifying Luminous Active Galactic Nuclei in Deep Surveys: Revised
  IRAC Selection Criteria}.
\newblock {\em Astrophys. J.} {\bf 2012}, {\em 748},~142.
\newblock
  https://doi.org/{\changeurlcolor{black}\href{https://doi.org/10.1088/0004-637X/748/2/142}{\detokenize{10.1088/0004-637X/748/2/142}}}.

\bibitem[{Mateos} {et~al.}(2012){Mateos}, {Alonso-Herrero}, {Carrera},
  {Blain}, {Watson}, {Barcons}, {Braito}, {Severgnini}, {Donley}, and
  {Stern}]{Mateos2012}
{Mateos}, S.; {Alonso-Herrero}, A.; {Carrera}, F.J.; {Blain}, A.; {Watson},
  M.G.; {Barcons}, X.; {Braito}, V.; {Severgnini}, P.; {Donley}, J.L.; {Stern},
  D.
\newblock {Using the Bright Ultrahard XMM-Newton survey to define an IR
  selection of luminous AGN based on WISE colours}.
\newblock {\em Mon. Not. R. Astron. Soc.} {\bf 2012},
  {\em 426},~3271--3281.
\newblock
  https://doi.org/{\changeurlcolor{black}\href{https://doi.org/10.1111/j.1365-2966.2012.21843.x}{\detokenize{10.1111/j.1365-2966.2012.21843.x}}}.

\bibitem[{Stern} {et~al.}(2012){Stern}, {Assef}, {Benford}, {Blain},
  {Cutri}, {Dey}, {Eisenhardt}, {Griffith}, {Jarrett}, {Lake}, {Masci},
  {Petty}, {Stanford}, {Tsai}, {Wright}, {Yan}, {Harrison}, and
  {Madsen}]{2012ApJ...753...30S}
{Stern}, D.; {Assef}, R.J.; {Benford}, D.J.; {Blain}, A.; {Cutri}, R.; {Dey},
  A.; {Eisenhardt}, P.; {Griffith}, R.L.; {Jarrett}, T.H.; {Lake}, S.;  et~al.
\newblock {Mid-infrared Selection of Active Galactic Nuclei with the Wide-Field
  Infrared Survey Explorer. I. Characterizing WISE-selected Active Galactic
  Nuclei in COSMOS}.
\newblock {\em Astrophys. J.} {\bf 2012}, {\em 753},~30.
\newblock
  https://doi.org/{\changeurlcolor{black}\href{https://doi.org/10.1088/0004-637X/753/1/30}{\detokenize{10.1088/0004-637X/753/1/30}}}.

\bibitem[{Assef} {et~al.}(2018){Assef}, {Stern}, {Noirot}, {Jun}, {Cutri},
  and {Eisenhardt}]{2018ApJS..234...23A}
{Assef}, R.J.; {Stern}, D.; {Noirot}, G.; {Jun}, H.D.; {Cutri}, R.M.;
  {Eisenhardt}, P.R.M.
\newblock {The WISE AGN Catalog}.
\newblock {\em Astrophys. J. Suppl.} {\bf 2018}, {\em 234},~23.
\newblock
  https://doi.org/{\changeurlcolor{black}\href{https://doi.org/10.3847/1538-4365/aaa00a}{\detokenize{10.3847/1538-4365/aaa00a}}}.

\bibitem[{Kirkpatrick} {et~al.}(2017){Kirkpatrick}, {Alberts}, {Pope},
  {Barro}, {Bonato}, {Kocevski}, {P{\'e}rez-Gonz{\'a}lez}, {Rieke},
  {Rodr{\'\i}guez-Mu{\~n}oz}, {Sajina}, {Grogin}, {Mantha}, {Pandya}, {Pforr},
  {Salvato}, and {Santini}]{Kirkpatrick2017}
{Kirkpatrick}, A.; {Alberts}, S.; {Pope}, A.; {Barro}, G.; {Bonato}, M.;
  {Kocevski}, D.D.; {P{\'e}rez-Gonz{\'a}lez}, P.; {Rieke}, G.H.;
  {Rodr{\'\i}guez-Mu{\~n}oz}, L.; {Sajina}, A.;  et~al.
\newblock {The AGN-Star Formation Connection: Future Prospects with JWST}.
\newblock {\em Astrophys. J.} {\bf 2017}, {\em 849},~111.
\newblock
  https://doi.org/{\changeurlcolor{black}\href{https://doi.org/10.3847/1538-4357/aa911d}{\detokenize{10.3847/1538-4357/aa911d}}}.

\bibitem[{Scoville} {et~al.}(2017){Scoville}, {Murchikova}, {Walter},
  {Vlahakis}, {Koda}, {Vanden Bout}, {Barnes}, {Hernquist}, {Sheth}, {Yun},
  {Sanders}, {Armus}, {Cox}, {Thompson}, {Robertson}, {Zschaechner}, {Tacconi},
  {Torrey}, {Hayward}, {Genzel}, {Hopkins}, {van der Werf}, and
  {Decarli}]{Scoville2017}
{Scoville}, N.; {Murchikova}, L.; {Walter}, F.; {Vlahakis}, C.; {Koda}, J.;
  {Vanden Bout}, P.; {Barnes}, J.; {Hernquist}, L.; {Sheth}, K.; {Yun}, M.;
  et~al.
\newblock {ALMA Resolves the Nuclear Disks of Arp 220}.
\newblock {\em Astrophys. J.} {\bf 2017}, {\em 836},~66.
\newblock
  https://doi.org/{\changeurlcolor{black}\href{https://doi.org/10.3847/1538-4357/836/1/66}{\detokenize{10.3847/1538-4357/836/1/66}}}.

\bibitem[{Alexander} {et~al.}(2008){Alexander}, {Chary}, {Pope}, {Bauer},
  {Brandt}, {Daddi}, {Dickinson}, {Elbaz}, and {Reddy}]{Alexander2008}
{Alexander}, D.M.; {Chary}, R.R.; {Pope}, A.; {Bauer}, F.E.; {Brandt}, W.N.;
  {Daddi}, E.; {Dickinson}, M.; {Elbaz}, D.; {Reddy}, N.A.
\newblock {Reliable Identification of Compton-thick Quasars at z
  {\ensuremath{\approx}} 2: Spitzer Mid-Infrared Spectroscopy of HDF-oMD49}.
\newblock {\em Astrophys. J.} {\bf 2008}, {\em 687},~835--847.
\newblock
  https://doi.org/{\changeurlcolor{black}\href{https://doi.org/10.1086/591928}{\detokenize{10.1086/591928}}}.

\bibitem[{Bauer} {et~al.}(2010){Bauer}, {Yan}, {Sajina}, and
  {Alexander}]{Bauer2010}
{Bauer}, F.E.; {Yan}, L.; {Sajina}, A.; {Alexander}, D.M.
\newblock {X-ray Constraints on the Active Galactic Nuclei Properties in
  Spitzer-Infrared Spectrograph Identified z \raisebox{-0.5ex}\textasciitilde 2
  Ultraluminous Infrared Galaxies}.
\newblock {\em Astrophys. J.} {\bf 2010}, {\em 710},~212--226.
\newblock
  https://doi.org/{\changeurlcolor{black}\href{https://doi.org/10.1088/0004-637X/710/1/212}{\detokenize{10.1088/0004-637X/710/1/212}}}.

\bibitem[{Hopkins} {et~al.}(2007){Hopkins}, {Richards}, and
  {Hernquist}]{Hopkins2007}
{Hopkins}, P.F.; {Richards}, G.T.; {Hernquist}, L.
\newblock {An Observational Determination of the Bolometric Quasar Luminosity
  Function}.
\newblock {\em Astrophys. J.} {\bf 2007}, {\em 654},~731--753.
\newblock
  https://doi.org/{\changeurlcolor{black}\href{https://doi.org/10.1086/509629}{\detokenize{10.1086/509629}}}.

\bibitem[{La Franca} {et~al.}(2005){La Franca}, {Fiore}, {Comastri},
  {Perola}, {Sacchi}, {Brusa}, {Cocchia}, {Feruglio}, {Matt}, {Vignali},
  {Carangelo}, {Ciliegi}, {Lamastra}, {Maiolino}, {Mignoli}, {Molendi}, and
  {Puccetti}]{LaFranca2005}
{La Franca}, F.; {Fiore}, F.; {Comastri}, A.; {Perola}, G.C.; {Sacchi}, N.;
  {Brusa}, M.; {Cocchia}, F.; {Feruglio}, C.; {Matt}, G.; {Vignali}, C.;
  et~al.
\newblock {The HELLAS2XMM Survey. VII. The Hard X-Ray Luminosity Function of
  AGNs up to z = 4: More Absorbed AGNs at Low Luminosities and High Redshifts}.
\newblock {\em Astrophys. J.} {\bf 2005}, {\em 635},~864--879.
\newblock
  https://doi.org/{\changeurlcolor{black}\href{https://doi.org/10.1086/497586}{\detokenize{10.1086/497586}}}.

\bibitem[{Aird} {et~al.}(2010){Aird}, {Nandra}, {Laird}, {Georgakakis},
  {Ashby}, {Barmby}, {Coil}, {Huang}, {Koekemoer}, {Steidel}, and
  {Willmer}]{Aird2010}
{Aird}, J.; {Nandra}, K.; {Laird}, E.S.; {Georgakakis}, A.; {Ashby}, M.L.N.;
  {Barmby}, P.; {Coil}, A.L.; {Huang}, J.S.; {Koekemoer}, A.M.; {Steidel},
  C.C.;  et~al.
\newblock {The evolution of the hard X-ray luminosity function of AGN}.
\newblock {\em Mon. Not. R. Astron. Soc.} {\bf 2010},
  {\em 401},~2531--2551.
\newblock
  https://doi.org/{\changeurlcolor{black}\href{https://doi.org/10.1111/j.1365-2966.2009.15829.x}{\detokenize{10.1111/j.1365-2966.2009.15829.x}}}.

\bibitem[{Lacy} {et~al.}(2015){Lacy}, {Ridgway}, {Sajina}, {Petric},
  {Gates}, {Urrutia}, and {Storrie-Lombardi}]{Lacy2015}
{Lacy}, M.; {Ridgway}, S.E.; {Sajina}, A.; {Petric}, A.O.; {Gates}, E.L.;
  {Urrutia}, T.; {Storrie-Lombardi}, L.J.
\newblock {The Spitzer Mid-infrared AGN Survey. II. The Demographics and Cosmic
  Evolution of the AGN Population}.
\newblock {\em Astrophys. J.} {\bf 2015}, {\em 802},~102.
\newblock
  https://doi.org/{\changeurlcolor{black}\href{https://doi.org/10.1088/0004-637X/802/2/102}{\detokenize{10.1088/0004-637X/802/2/102}}}.

\bibitem[{Cutri} {et~al.}(2002){Cutri}, {Nelson}, {Francis}, and
  {Smith}]{2002ASPC..284..127C}
{Cutri}, R.M.; {Nelson}, B.O.; {Francis}, P.J.; {Smith}, P.S.
\newblock {The 2MASS Red AGN Survey}.
\newblock In {\em IAU Colloq. 184: AGN Surveys}; Astronomical Society of the
  Pacific Conference Series;{Green}, R.F., {Khachikian}, E.Y.,
  {Sanders}, D.B., Eds.; Cambridge University Press: Cambridge, UK, 2002; Volume 284, p. 127.

\bibitem[{Glikman} {et~al.}(2004){Glikman}, {Gregg}, {Lacy}, {Helfand},
  {Becker}, and {White}]{2004ApJ...607...60G}
{Glikman}, E.; {Gregg}, M.D.; {Lacy}, M.; {Helfand}, D.J.; {Becker}, R.H.;
  {White}, R.L.
\newblock {FIRST-2Mass Sources below the APM Detection Threshold: A Population
  of Highly Reddened Quasars}.
\newblock {\em Astrophys. J.} {\bf 2004}, {\em 607},~60--7.
\newblock
  https://doi.org/{\changeurlcolor{black}\href{https://doi.org/10.1086/383305}{\detokenize{10.1086/383305}}}.

\bibitem[{Matute} {et~al.}(2006){Matute}, {La Franca}, {Pozzi}, {Gruppioni},
  {Lari}, and {Zamorani}]{2006A&A...451..443M}
{Matute}, I.; {La Franca}, F.; {Pozzi}, F.; {Gruppioni}, C.; {Lari}, C.;
  {Zamorani}, G.
\newblock {Active galactic nuclei in the mid-IR. Evolution and contribution to
  the cosmic infrared background}.
\newblock {\em Astron. Astrophys.} {\bf 2006}, {\em 451},~443--456.
\newblock
  https://doi.org/{\changeurlcolor{black}\href{https://doi.org/10.1051/0004-6361:20053710}{\detokenize{10.1051/0004-6361:20053710}}}.

\bibitem[{Glikman} {et~al.}(2018){Glikman}, {Lacy}, {LaMassa}, {Stern},
  {Djorgovski}, {Graham}, {Urrutia}, {Lovdal}, {Crnogorcevic}, {Daniels-Koch},
  {Hundal}, {Urry}, {Gates}, and {Murray}]{2018ApJ...861...37G}
{Glikman}, E.; {Lacy}, M.; {LaMassa}, S.; {Stern}, D.; {Djorgovski}, S.G.;
  {Graham}, M.J.; {Urrutia}, T.; {Lovdal}, L.; {Crnogorcevic}, M.;
  {Daniels-Koch}, H.;  et~al.
\newblock {Luminous WISE-selected Obscured, Unobscured, and Red Quasars in
  Stripe 82}.
\newblock {\em Astrophys. J.} {\bf 2018}, {\em 861},~37.
\newblock
  https://doi.org/{\changeurlcolor{black}\href{https://doi.org/10.3847/1538-4357/aac5d8}{\detokenize{10.3847/1538-4357/aac5d8}}}.

\bibitem[{Han} {et~al.}(2012){Han}, {Dai}, {Wang}, {Zhang}, and
  {Han}]{2012MNRAS.423..464H}
{Han}, Y.; {Dai}, B.; {Wang}, B.; {Zhang}, F.; {Han}, Z.
\newblock {Evolution of the luminosity function and obscuration of active
  galactic nuclei: Comparison between X-ray and infrared}.
\newblock {\em Mon. Not. R. Astron. Soc.} {\bf 2012}.
\newblock
  https://doi.org/{\changeurlcolor{black}\href{https://doi.org/10.1111/j.1365-2966.2012.20890.x}{\detokenize{10.1111/j.1365-2966.2012.20890.x}}}.


\bibitem[{Zavala} {et~al.}(2021){Zavala}, {Casey}, {Manning}, {Aravena},
  {Bethermin}, {Caputi}, {Clements}, {Cunha}, {Drew}, {Finkelstein},
  {Fujimoto}, {Hayward}, {Hodge}, {Kartaltepe}, {Knudsen}, {Koekemoer}, {Long},
  {Magdis}, {Man}, {Popping}, {Sanders}, {Scoville}, {Sheth}, {Staguhn},
  {Toft}, {Treister}, {Vieira}, and {Yun}]{Zavala2021}
{Zavala}, J.A.; {Casey}, C.M.; {Manning}, S.M.; {Aravena}, M.; {Bethermin}, M.;
  {Caputi}, K.I.; {Clements}, D.L.; {Cunha}, E.d.; {Drew}, P.; {Finkelstein},
  S.L.;  et~al.
\newblock {The Evolution of the IR Luminosity Function and Dust-obscured Star
  Formation over the Past 13 Billion Years}.
\newblock {\em Astrophys. J.} {\bf 2021}, {\em 909},~165.
\newblock
  https://doi.org/{\changeurlcolor{black}\href{https://doi.org/10.3847/1538-4357/abdb27}{\detokenize{10.3847/1538-4357/abdb27}}}.

\bibitem[{Engelbracht} {et~al.}(2008){Engelbracht}, {Rieke}, {Gordon},
  {Smith}, {Werner}, {Moustakas}, {Willmer}, and {Vanzi}]{Engelbracht2008}
{Engelbracht}, C.W.; {Rieke}, G.H.; {Gordon}, K.D.; {Smith}, J.D.T.; {Werner},
  M.W.; {Moustakas}, J.; {Willmer}, C.N.A.; {Vanzi}, L.
\newblock {Metallicity Effects on Dust Properties in Starbursting Galaxies}.
\newblock {\em Astrophys. J.} {\bf 2008}, {\em 678},~804--827.
\newblock
  https://doi.org/{\changeurlcolor{black}\href{https://doi.org/10.1086/529513}{\detokenize{10.1086/529513}}}.

\bibitem[{Jarrett} {et~al.}(2011){Jarrett}, {Cohen}, {Masci}, {Wright},
  {Stern}, {Benford}, {Blain}, {Carey}, {Cutri}, {Eisenhardt}, {Lonsdale},
  {Mainzer}, {Marsh}, {Padgett}, {Petty}, {Ressler}, {Skrutskie}, {Stanford},
  {Surace}, {Tsai}, {Wheelock}, and {Yan}]{Jarrett2011}
{Jarrett}, T.H.; {Cohen}, M.; {Masci}, F.; {Wright}, E.; {Stern}, D.;
  {Benford}, D.; {Blain}, A.; {Carey}, S.; {Cutri}, R.M.; {Eisenhardt}, P.;
  et~al.
\newblock {The Spitzer-WISE Survey of the Ecliptic Poles}.
\newblock {\em Astrophys. J.} {\bf 2011}, {\em 735},~112.
\newblock
  https://doi.org/{\changeurlcolor{black}\href{https://doi.org/10.1088/0004-637X/735/2/112}{\detokenize{10.1088/0004-637X/735/2/112}}}.

\bibitem[{Hainline} {et~al.}(2016){Hainline}, {Reines}, {Greene}, and
  {Stern}]{Hainline2016}
{Hainline}, K.N.; {Reines}, A.E.; {Greene}, J.E.; {Stern}, D.
\newblock {Mid-infrared Colors of Dwarf Galaxies: Young Starbursts Mimicking
  Active Galactic Nuclei}.
\newblock {\em Astrophys. J.} {\bf 2016}, {\em 832},~119.
\newblock
  https://doi.org/{\changeurlcolor{black}\href{https://doi.org/10.3847/0004-637X/832/2/119}{\detokenize{10.3847/0004-637X/832/2/119}}}.

\bibitem[{Satyapal} {et~al.}(2021){Satyapal}, {Kamal}, {Cann}, {Secrest},
  and {Abel}]{Satyapal2021}
{Satyapal}, S.; {Kamal}, L.; {Cann}, J.M.; {Secrest}, N.J.; {Abel}, N.P.
\newblock {The Diagnostic Potential of JWST in Characterizing Elusive AGNs}.
\newblock {\em Astrophys. J.} {\bf 2021}, {\em 906},~35.
\newblock
  https://doi.org/{\changeurlcolor{black}\href{https://doi.org/10.3847/1538-4357/abbfaf}{\detokenize{10.3847/1538-4357/abbfaf}}}.

\bibitem[{Greene}(2012)]{Greene2012}
{Greene}, J.E.
\newblock {Low-mass black holes as the remnants of primordial black hole
  formation}.
\newblock {\em Nat. Commun.} {\bf 2012}, {\em 3},~1304.
\newblock
  https://doi.org/{\changeurlcolor{black}\href{https://doi.org/10.1038/ncomms2314}{\detokenize{10.1038/ncomms2314}}}.

\bibitem[{Smethurst} {et~al.}(2021){Smethurst}, {Simmons}, {Coil},
  {Lintott}, {Keel}, {Masters}, {Glikman}, {Leung}, {Shanahan}, and
  {Garland}]{Smethurst2021}
{Smethurst}, R.J.; {Simmons}, B.D.; {Coil}, A.; {Lintott}, C.J.; {Keel}, W.;
  {Masters}, K.L.; {Glikman}, E.; {Leung}, G.C.K.; {Shanahan}, J.; {Garland},
  I.L.
\newblock {Kiloparsec-scale AGN outflows and feedback in merger-free galaxies}.
\newblock {\em Mon. Not. R. Astron. Soc.} {\bf 2021},
  {\em 507},~3985--3997.
\newblock
  https://doi.org/{\changeurlcolor{black}\href{https://doi.org/10.1093/mnras/stab2340}{\detokenize{10.1093/mnras/stab2340}}}.

\bibitem[{Roussel} {et~al.}(2006){Roussel}, {Helou}, {Smith}, {Draine},
  {Hollenbach}, {Moustakas}, {Spoon}, {Kennicutt}, {Rieke}, {Walter}, {Armus},
  {Dale}, {Sheth}, {Bendo}, {Engelbracht}, {Gordon}, {Meyer}, {Regan}, and
  {Murphy}]{Roussel2006}
{Roussel}, H.; {Helou}, G.; {Smith}, J.D.; {Draine}, B.T.; {Hollenbach}, D.J.;
  {Moustakas}, J.; {Spoon}, H.W.; {Kennicutt}, R.C.; {Rieke}, G.H.; {Walter},
  F.;  et~al.
\newblock {The Opaque Nascent Starburst in NGC 1377: Spitzer SINGS
  Observations}.
\newblock {\em Astrophys. J.} {\bf 2006}, {\em 646},~841--857.
\newblock
  https://doi.org/{\changeurlcolor{black}\href{https://doi.org/10.1086/505038}{\detokenize{10.1086/505038}}}.

\bibitem[{Groves} {et~al.}(2006){Groves}, {Heckman}, and
  {Kauffmann}]{Groves2006a}
{Groves}, B.A.; {Heckman}, T.M.; {Kauffmann}, G.
\newblock {Emission-line diagnostics of low-metallicity active galactic
  nuclei}.
\newblock {\em Mon. Not. R. Astron. Soc.} {\bf 2006},
  {\em 371},~1559--1569.
\newblock
  https://doi.org/{\changeurlcolor{black}\href{https://doi.org/10.1111/j.1365-2966.2006.10812.x}{\detokenize{10.1111/j.1365-2966.2006.10812.x}}}.

\bibitem[{Sartori} {et~al.}(2015){Sartori}, {Schawinski}, {Treister},
  {Trakhtenbrot}, {Koss}, {Shirazi}, and {Oh}]{Sartori2015}
{Sartori}, L.F.; {Schawinski}, K.; {Treister}, E.; {Trakhtenbrot}, B.; {Koss},
  M.; {Shirazi}, M.; {Oh}, K.
\newblock {The search for active black holes in nearby low-mass galaxies using
  optical and mid-IR data}.
\newblock {\em Mon. Not. R. Astron. Soc.} {\bf 2015},
  {\em 454},~3722--3742.
\newblock
  https://doi.org/{\changeurlcolor{black}\href{https://doi.org/10.1093/mnras/stv2238}{\detokenize{10.1093/mnras/stv2238}}}.

\bibitem[{Satyapal} {et~al.}(2008){Satyapal}, {Vega}, {Dudik}, {Abel}, and
  {Heckman}]{Satyapal2008}
{Satyapal}, S.; {Vega}, D.; {Dudik}, R.P.; {Abel}, N.P.; {Heckman}, T.
\newblock {Spitzer Uncovers Active Galactic Nuclei Missed by Optical Surveys in
  Seven Late-Type Galaxies}.
\newblock {\em Astrophys. J.} {\bf 2008}, {\em 677},~926--942.
\newblock
  https://doi.org/{\changeurlcolor{black}\href{https://doi.org/10.1086/529014}{\detokenize{10.1086/529014}}}.

\bibitem[{Richardson} {et~al.}(2022){Richardson}, {Simpson}, {Polimera},
  {Kannappan}, {Bellovary}, {Greene}, and {Jenkins}]{Richardson2022}
{Richardson}, C.T.; {Simpson}, C.; {Polimera}, M.S.; {Kannappan}, S.J.;
  {Bellovary}, J.M.; {Greene}, C.; {Jenkins}, S.
\newblock {Optical and JWST Mid-IR Emission Line Diagnostics for Simultaneous
  IMBH and Stellar Excitation in z 0 Dwarf Galaxies}.
\newblock {\em Astrophys. J.} {\bf 2022}, {\em 927},~165.
\newblock
  https://doi.org/{\changeurlcolor{black}\href{https://doi.org/10.3847/1538-4357/ac510c}{\detokenize{10.3847/1538-4357/ac510c}}}.

\bibitem[{Secrest} and {Satyapal}(2020)]{Secrest2020}
{Secrest}, N.J.; {Satyapal}, S.
\newblock {A Low Incidence of Mid-infrared Variability in Dwarf Galaxies}.
\newblock {\em Astrophys. J.} {\bf 2020}, {\em 900},~56.
\newblock
  https://doi.org/{\changeurlcolor{black}\href{https://doi.org/10.3847/1538-4357/ab9309}{\detokenize{10.3847/1538-4357/ab9309}}}.

\bibitem[{Komossa}(2015)]{Komossa2015}
{Komossa}, S.
\newblock {Tidal disruption of stars by supermassive black holes: Status of
  observations}.
\newblock {\em J. High Energy Astrophys.} {\bf 2015}, {\em
  7},~148--157.
\newblock
  https://doi.org/{\changeurlcolor{black}\href{https://doi.org/10.1016/j.jheap.2015.04.006}{\detokenize{10.1016/j.jheap.2015.04.006}}}.

\bibitem[{Jiang} {et~al.}(2017){Jiang}, {Wang}, {Yan}, {Xiao}, {Yang},
  {Dou}, {Wang}, {Cutri}, and {Mainzer}]{Jiang2017}
{Jiang}, N.; {Wang}, T.; {Yan}, L.; {Xiao}, T.; {Yang}, C.; {Dou}, L.; {Wang},
  H.; {Cutri}, R.; {Mainzer}, A.
\newblock {Mid-infrared Flare of TDE Candidate PS16dtm: Dust Echo and
  Implications for the Spectral Evolution}.
\newblock {\em Astrophys. J.} {\bf 2017}, {\em 850},~63.
\newblock
  https://doi.org/{\changeurlcolor{black}\href{https://doi.org/10.3847/1538-4357/aa93f5}{\detokenize{10.3847/1538-4357/aa93f5}}}.

\bibitem[{Dou} {et~al.}(2017){Dou}, {Wang}, {Yan}, {Jiang}, {Yang}, {Cutri},
  {Mainzer}, and {Peng}]{Dou2017}
{Dou}, L.; {Wang}, T.; {Yan}, L.; {Jiang}, N.; {Yang}, C.; {Cutri}, R.M.;
  {Mainzer}, A.; {Peng}, B.
\newblock {Discovery of a Mid-infrared Echo from the TDE Candidate in the
  Nucleus of ULIRG F01004-2237}.
\newblock {\em Astrophys. J. Lett.} {\bf 2017}, {\em 841},~L8.
\newblock
  https://doi.org/{\changeurlcolor{black}\href{https://doi.org/10.3847/2041-8213/aa7130}{\detokenize{10.3847/2041-8213/aa7130}}}.

\bibitem[{Richards} {et~al.}(2006){Richards}, {Lacy}, {Storrie-Lombardi},
  {Hall}, {Gallagher}, {Hines}, {Fan}, {Papovich}, {Vanden Berk}, {Trammell},
  {Schneider}, {Vestergaard}, {York}, {Jester}, {Anderson}, {Budav{\'a}ri}, and
  {Szalay}]{Richards2006}
{Richards}, G.T.; {Lacy}, M.; {Storrie-Lombardi}, L.J.; {Hall}, P.B.;
  {Gallagher}, S.C.; {Hines}, D.C.; {Fan}, X.; {Papovich}, C.; {Vanden Berk},
  D.E.; {Trammell}, G.B.;  et~al.
\newblock {Spectral Energy Distributions and Multiwavelength Selection of Type
  1 Quasars}.
\newblock {\em Astrophys. J. Suppl.} {\bf 2006}, {\em
  166},~470--497.
\newblock
  https://doi.org/{\changeurlcolor{black}\href{https://doi.org/10.1086/506525}{\detokenize{10.1086/506525}}}.

\bibitem[{Shen} {et~al.}(2020){Shen}, {Hopkins}, {Faucher-Gigu{\`e}re},
  {Alexander}, {Richards}, {Ross}, and {Hickox}]{Shen2020}
{Shen}, X.; {Hopkins}, P.F.; {Faucher-Gigu{\`e}re}, C.A.; {Alexander}, D.M.;
  {Richards}, G.T.; {Ross}, N.P.; {Hickox}, R.C.
\newblock {The bolometric quasar luminosity function at z = 0-7}.
\newblock {\em Mon. Not. R. Astron. Soc.} {\bf 2020},
  {\em 495},~3252--3275.
\newblock
  https://doi.org/{\changeurlcolor{black}\href{https://doi.org/10.1093/mnras/staa1381}{\detokenize{10.1093/mnras/staa1381}}}.

\bibitem[{Trump} {et~al.}(2011){Trump}, {Impey}, {Kelly}, {Civano}, {Gabor},
  {Diamond-Stanic}, {Merloni}, {Urry}, {Hao}, {Jahnke}, {Nagao}, {Taniguchi},
  {Koekemoer}, {Lanzuisi}, {Liu}, {Mainieri}, {Salvato}, and
  {Scoville}]{Trump2011}
{Trump}, J.R.; {Impey}, C.D.; {Kelly}, B.C.; {Civano}, F.; {Gabor}, J.M.;
  {Diamond-Stanic}, A.M.; {Merloni}, A.; {Urry}, C.M.; {Hao}, H.; {Jahnke}, K.;
   et~al.
\newblock {Accretion Rate and the Physical Nature of Unobscured Active
  Galaxies}.
\newblock {\em Astrophys. J.} {\bf 2011}, {\em 733},~60.
\newblock
  https://doi.org/{\changeurlcolor{black}\href{https://doi.org/10.1088/0004-637X/733/1/60}{\detokenize{10.1088/0004-637X/733/1/60}}}.

\bibitem[{Lyu} {et~al.}(2017){Lyu}, {Rieke}, and
  {Shi}]{Lyu2017_dustdeficientquasars}
{Lyu}, J.; {Rieke}, G.H.; {Shi}, Y.
\newblock {Dust-deficient Palomar-Green Quasars and the Diversity of AGN
  Intrinsic IR Emission}.
\newblock {\em Astrophys. J.} {\bf 2017}, {\em 835},~257.
\newblock
  https://doi.org/{\changeurlcolor{black}\href{https://doi.org/10.3847/1538-4357/835/2/257}{\detokenize{10.3847/1538-4357/835/2/257}}}.

\bibitem[{Ogle} {et~al.}(2006){Ogle}, {Whysong}, and {Antonucci}]{Ogle2006}
{Ogle}, P.; {Whysong}, D.; {Antonucci}, R.
\newblock {Spitzer Reveals Hidden Quasar Nuclei in Some Powerful FR II Radio
  Galaxies}.
\newblock {\em Astrophys. J.} {\bf 2006}, {\em 647},~161--171.
\newblock
  https://doi.org/{\changeurlcolor{black}\href{https://doi.org/10.1086/505337}{\detokenize{10.1086/505337}}}.

\bibitem[{Ferrarese} and {Merritt}(2000)]{FerrareseMerritt2000}
{Ferrarese}, L.; {Merritt}, D.
\newblock {A Fundamental Relation between Supermassive Black Holes and Their
  Host Galaxies}.
\newblock {\em Astrophys. J. Lett.} {\bf 2000}, {\em 539},~L9--L12.
\newblock
  https://doi.org/{\changeurlcolor{black}\href{https://doi.org/10.1086/312838}{\detokenize{10.1086/312838}}}.

\bibitem[{Greene} and {Ho}(2005)]{GreeneHo2005}
{Greene}, J.E.; {Ho}, L.C.
\newblock {A Comparison of Stellar and Gaseous Kinematics in the Nuclei of
  Active Galaxies}.
\newblock {\em Astrophys. J.} {\bf 2005}, {\em 627},~721--732.
\newblock
  https://doi.org/{\changeurlcolor{black}\href{https://doi.org/10.1086/430590}{\detokenize{10.1086/430590}}}.

\bibitem[{Garc{\'\i}a-Burillo} {et~al.}(2019){Garc{\'\i}a-Burillo},
  {Combes}, {Ramos Almeida}, {Usero}, {Alonso-Herrero}, {Hunt}, {Rouan},
  {Aalto}, {Querejeta}, {Viti}, {van der Werf}, {Vives-Arias}, {Fuente},
  {Colina}, {Mart{\'\i}n-Pintado}, {Henkel}, {Mart{\'\i}n}, {Krips},
  {Gratadour}, {Neri}, and {Tacconi}]{Garcia-Burillo2019}
{Garc{\'\i}a-Burillo}, S.; {Combes}, F.; {Ramos Almeida}, C.; {Usero}, A.;
  {Alonso-Herrero}, A.; {Hunt}, L.K.; {Rouan}, D.; {Aalto}, S.; {Querejeta},
  M.; {Viti}, S.;  et~al.
\newblock {ALMA images the many faces of the <ASTROBJ>NGC 1068</ASTROBJ> torus
  and its surroundings}.
\newblock {\em Astron. Astrophys.} {\bf 2019}, {\em 632},~A61.
\newblock
  https://doi.org/{\changeurlcolor{black}\href{https://doi.org/10.1051/0004-6361/201936606}{\detokenize{10.1051/0004-6361/201936606}}}.

\bibitem[{Jaffe} {et~al.}(2004){Jaffe}, {Meisenheimer}, {R{\"o}ttgering},
  {Leinert}, {Richichi}, {Chesneau}, {Fraix-Burnet}, {Glazenborg-Kluttig},
  {Granato}, {Graser}, {Heijligers}, {K{\"o}hler}, {Malbet}, {Miley},
  {Paresce}, {Pel}, {Perrin}, {Przygodda}, {Schoeller}, {Sol}, {Waters},
  {Weigelt}, {Woillez}, and {de Zeeuw}]{Jaffe2004}
{Jaffe}, W.; {Meisenheimer}, K.; {R{\"o}ttgering}, H.J.A.; {Leinert}, C.;
  {Richichi}, A.; {Chesneau}, O.; {Fraix-Burnet}, D.; {Glazenborg-Kluttig}, A.;
  {Granato}, G.L.; {Graser}, U.;  et~al.
\newblock {The central dusty torus in the active nucleus of NGC 1068}.
\newblock {\em Nature} {\bf 2004}, {\em 429},~47--49.
\newblock
  https://doi.org/{\changeurlcolor{black}\href{https://doi.org/10.1038/nature02531}{\detokenize{10.1038/nature02531}}}.

\bibitem[{Combes} {et~al.}(2019){Combes}, {Garc{\'\i}a-Burillo}, {Audibert},
  {Hunt}, {Eckart}, {Aalto}, {Casasola}, {Boone}, {Krips}, {Viti}, {Sakamoto},
  {Muller}, {Dasyra}, {van der Werf}, and {Martin}]{Combes2019}
{Combes}, F.; {Garc{\'\i}a-Burillo}, S.; {Audibert}, A.; {Hunt}, L.; {Eckart},
  A.; {Aalto}, S.; {Casasola}, V.; {Boone}, F.; {Krips}, M.; {Viti}, S.;
  et~al.
\newblock {ALMA observations of molecular tori around massive black holes}.
\newblock {\em Astron. Astrophys.} {\bf 2019}, {\em 623},~A79.
\newblock
  https://doi.org/{\changeurlcolor{black}\href{https://doi.org/10.1051/0004-6361/201834560}{\detokenize{10.1051/0004-6361/201834560}}}.

\bibitem[{Garc{\'\i}a-Burillo} {et~al.}(2021){Garc{\'\i}a-Burillo},
  {Alonso-Herrero}, {Ramos Almeida}, {Gonz{\'a}lez-Mart{\'\i}n}, {Combes},
  {Usero}, {H{\"o}nig}, {Querejeta}, {Hicks}, {Hunt}, {Rosario}, {Davies},
  {Boorman}, {Bunker}, {Burtscher}, {Colina}, {D{\'\i}az-Santos}, {Gandhi},
  {Garc{\'\i}a-Bernete}, {Garc{\'\i}a-Lorenzo}, {Ichikawa}, {Imanishi},
  {Izumi}, {Labiano}, {Levenson}, {L{\'o}pez-Rodr{\'\i}guez}, {Packham},
  {Pereira-Santaella}, {Ricci}, {Rigopoulou}, {Rouan}, {Shimizu}, {Stalevski},
  {Wada}, and {Williamson}]{Garcia-Burillo2021}
{Garc{\'\i}a-Burillo}, S.; {Alonso-Herrero}, A.; {Ramos Almeida}, C.;
  {Gonz{\'a}lez-Mart{\'\i}n}, O.; {Combes}, F.; {Usero}, A.; {H{\"o}nig}, S.;
  {Querejeta}, M.; {Hicks}, E.K.S.; {Hunt}, L.K.;  et~al.
\newblock {The Galaxy Activity, Torus and Outflow Survey (GATOS) I. ALMA images
  of dusty molecular tori in Seyfert galaxies}.
\newblock {\em arXiv e-Prints} {\bf 2021},  arXiv:2104.10227.

\bibitem[{Lutz} {et~al.}(2004){Lutz}, {Maiolino}, {Spoon}, and
  {Moorwood}]{Lutz2004}
{Lutz}, D.; {Maiolino}, R.; {Spoon}, H.W.W.; {Moorwood}, A.F.M.
\newblock {The relation between AGN hard X-ray emission and mid-infrared
  continuum from ISO spectra: Scatter and unification aspects}.
\newblock {\em Astron. Astrophys.} {\bf 2004}, {\em 418},~465--47.
\newblock
  https://doi.org/{\changeurlcolor{black}\href{https://doi.org/10.1051/0004-6361:20035838}{\detokenize{10.1051/0004-6361:20035838}}}.

\bibitem[{Sheng} {et~al.}(2017){Sheng}, {Wang}, {Jiang}, {Yang}, {Yan},
  {Dou}, and {Peng}]{Sheng2017}
{Sheng}, Z.; {Wang}, T.; {Jiang}, N.; {Yang}, C.; {Yan}, L.; {Dou}, L.; {Peng},
  B.
\newblock {Mid-infrared Variability of Changing-look AGNs}.
\newblock {\em Astrophys. J. Lett.} {\bf 2017}, {\em 846},~L7.
\newblock
  https://doi.org/{\changeurlcolor{black}\href{https://doi.org/10.3847/2041-8213/aa85de}{\detokenize{10.3847/2041-8213/aa85de}}}.

\bibitem[{Lyu} {et~al.}(2022){Lyu}, {Wu}, {Yan}, {Yu}, and {Liu}]{Lyu2022}
{Lyu}, B.; {Wu}, Q.; {Yan}, Z.; {Yu}, W.; {Liu}, H.
\newblock {WISE View of Changing-look Active Galactic Nuclei: Evidence for a
  Transitional Stage of AGNs}.
\newblock {\em Astrophys. J.} {\bf 2022}, {\em 927},~227.
\newblock
  https://doi.org/{\changeurlcolor{black}\href{https://doi.org/10.3847/1538-4357/ac5256}{\detokenize{10.3847/1538-4357/ac5256}}}.

\bibitem[{Stalevski} {et~al.}(2019){Stalevski}, {Tristram}, and
  {Asmus}]{Stalevski2019}
{Stalevski}, M.; {Tristram}, K.R.W.; {Asmus}, D.
\newblock {Dissecting the active galactic nucleus in Circinus - II. A thin
  dusty disc and a polar outflow on parsec scales}.
\newblock {\em Mon. Not. R. Astron. Soc.} {\bf 2019},
  {\em 484},~3334--3355.
\newblock
  https://doi.org/{\changeurlcolor{black}\href{https://doi.org/10.1093/mnras/stz220}{\detokenize{10.1093/mnras/stz220}}}.

\bibitem[{Isbell} {et~al.}(2022){Isbell}, {Meisenheimer}, {Pott},
  {Stalevski}, {Tristram}, {Sanchez-Bermudez}, {Hofmann}, {G{\'a}mez Rosas},
  {Jaffe}, {Burtscher}, {Leftley}, {Petrov}, {Lopez}, {Henning}, {Weigelt},
  {Allouche}, {Berio}, {Bettonvil}, {Cruzalebes}, {Dominik}, {Heininger},
  {Hogerheijde}, {Lagarde}, {Lehmitz}, {Matter}, {Millour}, {Robbe-Dubois},
  {Schertl}, {van Boekel}, {Varga}, and {Woillez}]{Isbell2022}
{Isbell}, J.W.; {Meisenheimer}, K.; {Pott}, J.U.; {Stalevski}, M.; {Tristram},
  K.R.W.; {Sanchez-Bermudez}, J.; {Hofmann}, K.H.; {G{\'a}mez Rosas}, V.;
  {Jaffe}, W.; {Burtscher}, L.;  et~al.
\newblock {The dusty heart of Circinus: I. Imaging the circumnuclear dust in
  N-band}.
\newblock {\em arXiv e-Prints} {\bf 2022}, arXiv:2205.01575.

\bibitem[{Asmus} {et~al.}(2016){Asmus}, {H{\"o}nig}, and
  {Gandhi}]{Asmus2016}
{Asmus}, D.; {H{\"o}nig}, S.F.; {Gandhi}, P.
\newblock {The Subarcsecond Mid-infrared View of Local Active Galactic Nuclei.
  III. Polar Dust Emission}.
\newblock {\em Astrophys. J.} {\bf 2016}, {\em 822},~109.
\newblock
  https://doi.org/{\changeurlcolor{black}\href{https://doi.org/10.3847/0004-637X/822/2/109}{\detokenize{10.3847/0004-637X/822/2/109}}}.

\bibitem[{Asmus}(2019)]{Asmus2019}
{Asmus}, D.
\newblock {New evidence for the ubiquity of prominent polar dust emission in
  AGN on tens of parsec scales}.
\newblock {\em Mon. Not. R. Astron. Soc.} {\bf 2019},
  {\em 489},~2177--2188.
\newblock
  https://doi.org/{\changeurlcolor{black}\href{https://doi.org/10.1093/mnras/stz2289}{\detokenize{10.1093/mnras/stz2289}}}.

\bibitem[{Alonso-Herrero} {et~al.}(2021){Alonso-Herrero},
  {Garc{\'\i}a-Burillo}, {H{\"o}nig}, {Garc{\'\i}a-Bernete}, {Ramos Almeida},
  {Gonz{\'a}lez-Mart{\'\i}n}, {L{\'o}pez-Rodr{\'\i}guez}, {Boorman}, {Bunker},
  {Burtscher}, {Combes}, {Davies}, {D{\'\i}az-Santos}, {Gandhi},
  {Garc{\'\i}a-Lorenzo}, {Hicks}, {Hunt}, {Ichikawa}, {Imanishi}, {Izumi},
  {Labiano}, {Levenson}, {Packham}, {Pereira-Santaella}, {Ricci}, {Rigopoulou},
  {Roche}, {Rosario}, {Rouan}, {Shimizu}, {Stalevski}, {Wada}, and
  {Williamson}]{Alonso-Herrero2021}
{Alonso-Herrero}, A.; {Garc{\'\i}a-Burillo}, S.; {H{\"o}nig}, S.F.;
  {Garc{\'\i}a-Bernete}, I.; {Ramos Almeida}, C.; {Gonz{\'a}lez-Mart{\'\i}n},
  O.; {L{\'o}pez-Rodr{\'\i}guez}, E.; {Boorman}, P.G.; {Bunker}, A.J.;
  {Burtscher}, L.;  et~al.
\newblock {The Galaxy Activity, Torus, and Outflow Survey (GATOS). II. Torus
  and polar dust emission in nearby Seyfert galaxies}.
\newblock {\em Astron. Astrophys.} {\bf 2021}, {\em 652},~A99.
\newblock
  https://doi.org/{\changeurlcolor{black}\href{https://doi.org/10.1051/0004-6361/202141219}{\detokenize{10.1051/0004-6361/202141219}}}.

\bibitem[{Veilleux} {et~al.}(2013){Veilleux}, {Mel{\'e}ndez}, {Sturm},
  {Gracia-Carpio}, {Fischer}, {Gonz{\'a}lez-Alfonso}, {Contursi}, {Lutz},
  {Poglitsch}, {Davies}, {Genzel}, {Tacconi}, {de Jong}, {Sternberg}, {Netzer},
  {Hailey-Dunsheath}, {Verma}, {Rupke}, {Maiolino}, {Teng}, and
  {Polisensky}]{Veilleux2013}
{Veilleux}, S.; {Mel{\'e}ndez}, M.; {Sturm}, E.; {Gracia-Carpio}, J.;
  {Fischer}, J.; {Gonz{\'a}lez-Alfonso}, E.; {Contursi}, A.; {Lutz}, D.;
  {Poglitsch}, A.; {Davies}, R.;  et~al.
\newblock {Fast Molecular Outflows in Luminous Galaxy Mergers: Evidence for
  Quasar Feedback from Herschel}.
\newblock {\em Astrophys. J.} {\bf 2013}, {\em 776},~27.
\newblock
  https://doi.org/{\changeurlcolor{black}\href{https://doi.org/10.1088/0004-637X/776/1/27}{\detokenize{10.1088/0004-637X/776/1/27}}}.

\bibitem[{Cicone} {et~al.}(2014){Cicone}, {Maiolino}, {Sturm},
  {Graci{\'a}-Carpio}, {Feruglio}, {Neri}, {Aalto}, {Davies}, {Fiore},
  {Fischer}, {Garc{\'\i}a-Burillo}, {Gonz{\'a}lez-Alfonso}, {Hailey-Dunsheath},
  {Piconcelli}, and {Veilleux}]{Cicone2014}
{Cicone}, C.; {Maiolino}, R.; {Sturm}, E.; {Graci{\'a}-Carpio}, J.; {Feruglio},
  C.; {Neri}, R.; {Aalto}, S.; {Davies}, R.; {Fiore}, F.; {Fischer}, J.;
  et~al.
\newblock {Massive molecular outflows and evidence for AGN feedback from CO
  observations}.
\newblock {\em Astron. Astrophys.} {\bf 2014}, {\em 562},~A21.
\newblock
  https://doi.org/{\changeurlcolor{black}\href{https://doi.org/10.1051/0004-6361/201322464}{\detokenize{10.1051/0004-6361/201322464}}}.

\bibitem[{Leftley} {et~al.}(2019){Leftley}, {H{\"o}nig}, {Asmus},
  {Tristram}, {Gandhi}, {Kishimoto}, {Venanzi}, and {Williamson}]{Leftley2019}
{Leftley}, J.H.; {H{\"o}nig}, S.F.; {Asmus}, D.; {Tristram}, K.R.W.; {Gandhi},
  P.; {Kishimoto}, M.; {Venanzi}, M.; {Williamson}, D.J.
\newblock {Parsec-scale Dusty Winds in Active Galactic Nuclei: Evidence for
  Radiation Pressure Driving}.
\newblock {\em Astrophys. J.} {\bf 2019}, {\em 886},~55.
\newblock
  https://doi.org/{\changeurlcolor{black}\href{https://doi.org/10.3847/1538-4357/ab4a0b}{\detokenize{10.3847/1538-4357/ab4a0b}}}.

\bibitem[{Veilleux} {et~al.}(2020){Veilleux}, {Maiolino}, {Bolatto}, and
  {Aalto}]{Veilleux2020}
{Veilleux}, S.; {Maiolino}, R.; {Bolatto}, A.D.; {Aalto}, S.
\newblock {Cool outflows in galaxies and their implications}.
\newblock {\em Astron. Astrophys. Rev.} {\bf 2020}, {\em 28},~2.
\newblock
  https://doi.org/{\changeurlcolor{black}\href{https://doi.org/10.1007/s00159-019-0121-9}{\detokenize{10.1007/s00159-019-0121-9}}}.

\bibitem[{Prieto} {et~al.}(2021){Prieto}, {Nadolny},
  {Fern{\'a}ndez-Ontiveros}, and {Mezcua}]{Prieto2021}
{Prieto}, A.; {Nadolny}, J.; {Fern{\'a}ndez-Ontiveros}, J.A.; {Mezcua}, M.
\newblock {Dust in the central parsecs of unobscured AGN: More challenges to
  the torus}.
\newblock {\em arXiv e-Prints} {\bf 2021}, arXiv:2106.07753.

\bibitem[{Imanishi} {et~al.}(2018){Imanishi}, {Nakanishi}, {Izumi}, and
  {Wada}]{Imanishi2018}
{Imanishi}, M.; {Nakanishi}, K.; {Izumi}, T.; {Wada}, K.
\newblock {ALMA Reveals an Inhomogeneous Compact Rotating Dense Molecular Torus
  at the NGC 1068 Nucleus}.
\newblock {\em Astrophys. J. Lett.} {\bf 2018}, {\em 853},~L25.
\newblock
  https://doi.org/{\changeurlcolor{black}\href{https://doi.org/10.3847/2041-8213/aaa8df}{\detokenize{10.3847/2041-8213/aaa8df}}}.

\bibitem[{Garc{\'\i}a-Gonz{\'a}lez}
  {et~al.}(2015){Garc{\'\i}a-Gonz{\'a}lez}, {Alonso-Herrero},
  {P{\'e}rez-Gonz{\'a}lez}, {Hern{\'a}n-Caballero}, {Sarajedini}, and
  {Villar}]{Garcia-Gonzalez2015}
{Garc{\'\i}a-Gonz{\'a}lez}, J.; {Alonso-Herrero}, A.; {P{\'e}rez-Gonz{\'a}lez},
  P.G.; {Hern{\'a}n-Caballero}, A.; {Sarajedini}, V.L.; {Villar}, V.
\newblock {Selection of AGN candidates in the GOODS-South Field through
  Spitzer/MIPS 24 {\ensuremath{\mu}}m variability}.
\newblock {\em Mon. Not. R. Astron. Soc.} {\bf 2015},
  {\em 446},~3199--3223.
\newblock
  https://doi.org/{\changeurlcolor{black}\href{https://doi.org/10.1093/mnras/stu2204}{\detokenize{10.1093/mnras/stu2204}}}.

\bibitem[{Polimera} {et~al.}(2018){Polimera}, {Sarajedini}, {Ashby},
  {Willner}, and {Fazio}]{Polimera2018}
{Polimera}, M.; {Sarajedini}, V.; {Ashby}, M.L.N.; {Willner}, S.P.; {Fazio},
  G.G.
\newblock {Morphologies of mid-IR variability-selected AGN host galaxies}.
\newblock {\em Mon. Not. R. Astron. Soc.} {\bf 2018},
  {\em 476},~1111--1119.
\newblock
  https://doi.org/{\changeurlcolor{black}\href{https://doi.org/10.1093/mnras/sty164}{\detokenize{10.1093/mnras/sty164}}}.

\bibitem[{Peterson}(1993)]{Peterson1993}
\textls[-15]{{Peterson}, B.M.
\newblock {Reverberation Mapping of Active Galactic Nuclei}.
\newblock {\em Publ. Astron. Soc. Pac.} {\bf
  1993}, {\em 105},~247.
\newblock
  https://doi.org/{\changeurlcolor{black}\href{https://doi.org/10.1086/133140}{\detokenize{10.1086/133140}}}.}

\bibitem[{Peterson} {et~al.}(2004){Peterson}, {Ferrarese}, {Gilbert},
  {Kaspi}, {Malkan}, {Maoz}, {Merritt}, {Netzer}, {Onken}, {Pogge},
  {Vestergaard}, and {Wandel}]{Peterson2004}
{Peterson}, B.M.; {Ferrarese}, L.; {Gilbert}, K.M.; {Kaspi}, S.; {Malkan},
  M.A.; {Maoz}, D.; {Merritt}, D.; {Netzer}, H.; {Onken}, C.A.; {Pogge}, R.W.;
  et~al.
\newblock {Central Masses and Broad-Line Region Sizes of Active Galactic
  Nuclei. II. A Homogeneous Analysis of a Large Reverberation-Mapping
  Database}.
\newblock {\em Astrophys. J.} {\bf 2004}, {\em 613},~682--699.
\newblock
  https://doi.org/{\changeurlcolor{black}\href{https://doi.org/10.1086/423269}{\detokenize{10.1086/423269}}}.

\bibitem[{Koz{\l}owski} {et~al.}(2016){Koz{\l}owski}, {Kochanek}, {Ashby},
  {Assef}, {Brodwin}, {Eisenhardt}, {Jannuzi}, and {Stern}]{Kozlowski2016}
{Koz{\l}owski}, S.; {Kochanek}, C.S.; {Ashby}, M.L.N.; {Assef}, R.J.;
  {Brodwin}, M.; {Eisenhardt}, P.R.; {Jannuzi}, B.T.; {Stern}, D.
\newblock {Quasar Variability in the Mid-Infrared}.
\newblock {\em Astrophys. J.} {\bf 2016}, {\em 817},~119.
\newblock
  https://doi.org/{\changeurlcolor{black}\href{https://doi.org/10.3847/0004-637X/817/2/119}{\detokenize{10.3847/0004-637X/817/2/119}}}.

\bibitem[{Matsuoka} {et~al.}(2009){Matsuoka}, {Nagao}, {Maiolino},
  {Marconi}, and {Taniguchi}]{Matsuoka2009}
{Matsuoka}, K.; {Nagao}, T.; {Maiolino}, R.; {Marconi}, A.; {Taniguchi}, Y.
\newblock {Chemical evolution of high-redshift radio galaxies}.
\newblock {\em Astron. Astrophys.} {\bf 2009}, {\em 503},~721--730.
\newblock
  https://doi.org/{\changeurlcolor{black}\href{https://doi.org/10.1051/0004-6361/200811478}{\detokenize{10.1051/0004-6361/200811478}}}.

\bibitem[{Matsuoka} {et~al.}(2018){Matsuoka}, {Nagao}, {Marconi},
  {Maiolino}, {Mannucci}, {Cresci}, {Terao}, and {Ikeda}]{Matsuoka2018}
{Matsuoka}, K.; {Nagao}, T.; {Marconi}, A.; {Maiolino}, R.; {Mannucci}, F.;
  {Cresci}, G.; {Terao}, K.; {Ikeda}, H.
\newblock {The mass-metallicity relation of high-z type-2 active galactic
  nuclei}.
\newblock {\em Astron. Astrophys.} {\bf 2018}, {\em 616},~L4.
\newblock
  https://doi.org/{\changeurlcolor{black}\href{https://doi.org/10.1051/0004-6361/201833418}{\detokenize{10.1051/0004-6361/201833418}}}.

\bibitem[{Thomas} {et~al.}(2019){Thomas}, {Kewley}, {Dopita}, {Groves},
  {Hopkins}, and {Sutherland}]{Thomas2019}
{Thomas}, A.D.; {Kewley}, L.J.; {Dopita}, M.A.; {Groves}, B.A.; {Hopkins},
  A.M.; {Sutherland}, R.S.
\newblock {The Mass-Metallicity Relation of Local Active Galaxies}.
\newblock {\em Astrophys. J.} {\bf 2019}, {\em 874},~100.
\newblock
  https://doi.org/{\changeurlcolor{black}\href{https://doi.org/10.3847/1538-4357/ab08a1}{\detokenize{10.3847/1538-4357/ab08a1}}}.

\bibitem[{Valiante} {et~al.}(2011){Valiante}, {Schneider}, {Salvadori}, and
  {Bianchi}]{Valiante2011}
{Valiante}, R.; {Schneider}, R.; {Salvadori}, S.; {Bianchi}, S.
\newblock {The origin of the dust in high-redshift quasars: The case of SDSS
  J1148+5251}.
\newblock {\em Mon. Not. R. Astron. Soc.} {\bf 2011},
  {\em 416},~1916--1935.
\newblock
  https://doi.org/{\changeurlcolor{black}\href{https://doi.org/10.1111/j.1365-2966.2011.19168.x}{\detokenize{10.1111/j.1365-2966.2011.19168.x}}}.

\bibitem[{Nagao} {et~al.}(2011){Nagao}, {Maiolino}, {Marconi}, and
  {Matsuhara}]{Nagao2011}
{Nagao}, T.; {Maiolino}, R.; {Marconi}, A.; {Matsuhara}, H.
\newblock {Metallicity diagnostics with infrared fine-structure lines}.
\newblock {\em Astron. Astrophys.} {\bf 2011}, {\em 526},~A149.
\newblock
  https://doi.org/{\changeurlcolor{black}\href{https://doi.org/10.1051/0004-6361/201015471}{\detokenize{10.1051/0004-6361/201015471}}}.

\bibitem[{Valentino} {et~al.}(2021){Valentino}, {Daddi}, {Puglisi},
  {Magdis}, {Kokorev}, {Liu}, {Madden}, {G{\'o}mez-Guijarro}, {Lee}, {Cortzen},
  {Circosta}, {Delvecchio}, {Mullaney}, {Gao}, {Gobat}, {Aravena}, {Jin},
  {Fujimoto}, {Silverman}, and {Dannerbauer}]{Valenetino2021}
{Valentino}, F.; {Daddi}, E.; {Puglisi}, A.; {Magdis}, G.E.; {Kokorev}, V.;
  {Liu}, D.; {Madden}, S.C.; {G{\'o}mez-Guijarro}, C.; {Lee}, M.Y.; {Cortzen},
  I.;  et~al.
\newblock {The effect of active galactic nuclei on the cold interstellar medium
  in distant star-forming galaxies}.
\newblock {\em Astron. Astrophys.} {\bf 2021}, {\em 654},~A165.
\newblock
  https://doi.org/{\changeurlcolor{black}\href{https://doi.org/10.1051/0004-6361/202141417}{\detokenize{10.1051/0004-6361/202141417}}}.

\bibitem[{Circosta} {et~al.}(2021){Circosta}, {Mainieri}, {Lamperti},
  {Padovani}, {Bischetti}, {Harrison}, {Kakkad}, {Zanella}, {Vietri},
  {Lanzuisi}, {Salvato}, {Brusa}, {Carniani}, {Cicone}, {Cresci}, {Feruglio},
  {Husemann}, {Mannucci}, {Marconi}, {Perna}, {Piconcelli}, {Puglisi},
  {Saintonge}, {Schramm}, {Vignali}, and {Zappacosta}]{Circosta2021}
{Circosta}, C.; {Mainieri}, V.; {Lamperti}, I.; {Padovani}, P.; {Bischetti},
  M.; {Harrison}, C.M.; {Kakkad}, D.; {Zanella}, A.; {Vietri}, G.; {Lanzuisi},
  G.;  et~al.
\newblock {SUPER. IV. CO(J = 3-2) properties of active galactic nucleus hosts
  at cosmic noon revealed by ALMA}.
\newblock {\em Astron. Astrophys.} {\bf 2021}, {\em 646},~A96.
\newblock
  https://doi.org/{\changeurlcolor{black}\href{https://doi.org/10.1051/0004-6361/202039270}{\detokenize{10.1051/0004-6361/202039270}}}.

\bibitem[{Mashian} {et~al.}(2015){Mashian}, {Sturm}, {Sternberg}, {Janssen},
  {Hailey-Dunsheath}, {Fischer}, {Contursi}, {Gonz{\'a}lez-Alfonso},
  {Graci{\'a}-Carpio}, {Poglitsch}, {Veilleux}, {Davies}, {Genzel}, {Lutz},
  {Tacconi}, {Verma}, {Wei{\ss}}, {Polisensky}, and {Nikola}]{Mashian2015}
{Mashian}, N.; {Sturm}, E.; {Sternberg}, A.; {Janssen}, A.; {Hailey-Dunsheath},
  S.; {Fischer}, J.; {Contursi}, A.; {Gonz{\'a}lez-Alfonso}, E.;
  {Graci{\'a}-Carpio}, J.; {Poglitsch}, A.;  et~al.
\newblock {High-J CO Sleds in Nearby Infrared Bright Galaxies Observed By
  Herschel/PACS}.
\newblock {\em Astrophys. J.} {\bf 2015}, {\em 802},~81.
\newblock
  https://doi.org/{\changeurlcolor{black}\href{https://doi.org/10.1088/0004-637X/802/2/81}{\detokenize{10.1088/0004-637X/802/2/81}}}.

\bibitem[{van der Werf} {et~al.}(2010){van der Werf}, {Isaak}, {Meijerink},
  {Spaans}, {Rykala}, {Fulton}, {Loenen}, {Walter}, {Wei{\ss}}, {Armus},
  {Fischer}, {Israel}, {Harris}, {Veilleux}, {Henkel}, {Savini}, {Lord},
  {Smith}, {Gonz{\'a}lez-Alfonso}, {Naylor}, {Aalto}, {Charmandaris}, {Dasyra},
  {Evans}, {Gao}, {Greve}, {G{\"u}sten}, {Kramer}, {Mart{\'\i}n-Pintado},
  {Mazzarella}, {Papadopoulos}, {Sanders}, {Spinoglio}, {Stacey}, {Vlahakis},
  {Wiedner}, and {Xilouris}]{vanderWerf2010}
{van der Werf}, P.P.; {Isaak}, K.G.; {Meijerink}, R.; {Spaans}, M.; {Rykala},
  A.; {Fulton}, T.; {Loenen}, A.F.; {Walter}, F.; {Wei{\ss}}, A.; {Armus}, L.;
  et~al.
\newblock {Black hole accretion and star formation as drivers of gas excitation
  and chemistry in Markarian 231}.
\newblock {\em Astron. Astrophys.} {\bf 2010}, {\em 518},~L42.
\newblock
  https://doi.org/{\changeurlcolor{black}\href{https://doi.org/10.1051/0004-6361/201014682}{\detokenize{10.1051/0004-6361/201014682}}}.

\bibitem[{Maloney} {et~al.}(1996){Maloney}, {Hollenbach}, and
  {Tielens}]{Maloney1996}
{Maloney}, P.R.; {Hollenbach}, D.J.; {Tielens}, A.G.G.M.
\newblock {X-Ray--irradiated Molecular Gas. I. Physical Processes and General
  Results}.
\newblock {\em Astrophys. J.} {\bf 1996}, {\em 466},~561.
\newblock
  https://doi.org/{\changeurlcolor{black}\href{https://doi.org/10.1086/177532}{\detokenize{10.1086/177532}}}.

\bibitem[{Hatziminaoglou} {et~al.}(2015){Hatziminaoglou},
  {Hern{\'a}n-Caballero}, {Feltre}, and {Pi{\~n}ol Ferrer}]{Hatziminaoglou2015}
{Hatziminaoglou}, E.; {Hern{\'a}n-Caballero}, A.; {Feltre}, A.; {Pi{\~n}ol
  Ferrer}, N.
\newblock {A Complete Census of Silicate Features in the Mid-infrared Spectra
  of Active Galaxies}.
\newblock {\em Astrophys. J.} {\bf 2015}, {\em 803},~110.
\newblock
  https://doi.org/{\changeurlcolor{black}\href{https://doi.org/10.1088/0004-637X/803/2/110}{\detokenize{10.1088/0004-637X/803/2/110}}}.

\bibitem[{Sajina} {et~al.}(2009){Sajina}, {Spoon}, {Yan}, {Imanishi},
  {Fadda}, and {Elitzur}]{Sajina2009}
{Sajina}, A.; {Spoon}, H.; {Yan}, L.; {Imanishi}, M.; {Fadda}, D.; {Elitzur},
  M.
\newblock {Detections of Water Ice, Hydrocarbons, and 3.3 {\ensuremath{\mu}}m
  PAH in z \raisebox{-0.5ex}\textasciitilde 2 ULIRGs}.
\newblock {\em Astrophys. J.} {\bf 2009}, {\em 703},~270--284.
\newblock
  https://doi.org/{\changeurlcolor{black}\href{https://doi.org/10.1088/0004-637X/703/1/270}{\detokenize{10.1088/0004-637X/703/1/270}}}.

\bibitem[{Smith} {et~al.}(2007){Smith}, {Draine}, {Dale}, {Moustakas},
  {Kennicutt}, {Helou}, {Armus}, {Roussel}, {Sheth}, {Bendo}, {Buckalew},
  {Calzetti}, {Engelbracht}, {Gordon}, {Hollenbach}, {Li}, {Malhotra},
  {Murphy}, and {Walter}]{Smith2007}
{Smith}, J.D.T.; {Draine}, B.T.; {Dale}, D.A.; {Moustakas}, J.; {Kennicutt},
  R.~C., J.; {Helou}, G.; {Armus}, L.; {Roussel}, H.; {Sheth}, K.; {Bendo},
  G.J.;  et~al.
\newblock {The Mid-Infrared Spectrum of Star-forming Galaxies: Global
  Properties of Polycyclic Aromatic Hydrocarbon Emission}.
\newblock {\em Astrophys. J.} {\bf 2007}, {\em 656},~770--791.
\newblock
  https://doi.org/{\changeurlcolor{black}\href{https://doi.org/10.1086/510549}{\detokenize{10.1086/510549}}}.

\bibitem[{Garc{\'\i}a-Bernete} {et~al.}(2022){Garc{\'\i}a-Bernete},
  {Rigopoulou}, {Alonso-Herrero}, {Pereira-Santaella}, {Roche}, and
  {Kerkeni}]{Garcia-Bernete2022}
\textls[-5]{{Garc{\'\i}a-Bernete}, I.; {Rigopoulou}, D.; {Alonso-Herrero}, A.;
  {Pereira-Santaella}, M.; {Roche}, P.F.; {Kerkeni}, B.
\newblock {Polycyclic aromatic hydrocarbons in Seyfert and star-forming
  galaxies}.
\newblock {\em Mon. Not. R. Astron. Soc.} {\bf 2022},
  {\em 509},~4256--4275.
\newblock
  https://doi.org/{\changeurlcolor{black}\href{https://doi.org/10.1093/mnras/stab3127}{\detokenize{10.1093/mnras/stab3127}}}.}

\bibitem[{Lyu} {et~al.}(2014){Lyu}, {Hao}, and {Li}]{Lyu2014}
{Lyu}, J.; {Hao}, L.; {Li}, A.
\newblock {Dust in Active Galactic Nuclei: Anomalous Silicate to Optical
  Extinction Ratios?}
\newblock {\em Astrophys. J. Lett.} {\bf 2014}, {\em 792},~L9.
\newblock
  https://doi.org/{\changeurlcolor{black}\href{https://doi.org/10.1088/2041-8205/792/1/L9}{\detokenize{10.1088/2041-8205/792/1/L9}}}.

\bibitem[{Shao} {et~al.}(2017){Shao}, {Jiang}, and {Li}]{Shao2017}
{Shao}, Z.; {Jiang}, B.W.; {Li}, A.
\newblock {On the Optical-to-silicate Extinction Ratio as a Probe of the Dust
  Size in Active Galactic Nuclei}.
\newblock {\em Astrophys. J.} {\bf 2017}, {\em 840},~27.
\newblock
  https://doi.org/{\changeurlcolor{black}\href{https://doi.org/10.3847/1538-4357/aa6ba4}{\detokenize{10.3847/1538-4357/aa6ba4}}}.

\bibitem[{Xie} {et~al.}(2017){Xie}, {Li}, and {Hao}]{Xie2017}
{Xie}, Y.; {Li}, A.; {Hao}, L.
\newblock {Silicate Dust in Active Galactic Nuclei}.
\newblock {\em Astrophys. J. Suppl.} {\bf 2017}, {\em 228},~6.
\newblock
  https://doi.org/{\changeurlcolor{black}\href{https://doi.org/10.3847/1538-4365/228/1/6}{\detokenize{10.3847/1538-4365/228/1/6}}}.

\bibitem[{Spoon} {et~al.}(2004){Spoon}, {Armus}, {Cami}, {Tielens}, {Chiar},
  {Peeters}, {Keane}, {Charmandaris}, {Appleton}, {Teplitz}, and
  {Burgdorf}]{Spoon2004}
{Spoon}, H.W.W.; {Armus}, L.; {Cami}, J.; {Tielens}, A.G.G.M.; {Chiar}, J.E.;
  {Peeters}, E.; {Keane}, J.V.; {Charmandaris}, V.; {Appleton}, P.N.;
  {Teplitz}, H.I.;  et~al.
\newblock {Fire and Ice: Spitzer Infrared Spectrograph (IRS) Mid-Infrared
  Spectroscopy of IRAS F00183-7111}.
\newblock {\em Astrophys. J. Suppl.} {\bf 2004}, {\em
  154},~184--187.
\newblock
  https://doi.org/{\changeurlcolor{black}\href{https://doi.org/10.1086/422813}{\detokenize{10.1086/422813}}}.

\bibitem[{Imanishi} {et~al.}(2008){Imanishi}, {Nakagawa}, {Ohyama},
{Shirahata}, {Wada}, {Onaka}, and {Oi}]{Imanishi2008}
\textls[-15]{{Imanishi}, M.; {Nakagawa}, T.; {Ohyama}, Y.; {Shirahata}, M.; {Wada}, T.;
  {Onaka}, T.; {Oi}, N.
\newblock {Systematic Infrared 2.5-5 {\ensuremath{\mu}}m Spectroscopy of Nearby
  Ultraluminous Infrared Galaxies with AKARI}.
\newblock {\em Publ. Astron. Soc. Jpn.} {\bf 2008},
  {\em 60},~S489.
\newblock
  https://doi.org/{\changeurlcolor{black}\href{https://doi.org/10.1093/pasj/60.sp2.S489}{\detokenize{10.1093/pasj/60.sp2.S489}}}.}

\bibitem[{Matsuhara} {et~al.}(2006){Matsuhara}, {Wada}, {Matsuura},
  {Nakagawa}, {Kawada}, {Ohyama}, {Pearson}, {Oyabu}, {Takagi}, {Serjeant},
  {White}, {Hanami}, {Watarai}, {Takeuchi}, {Kodama}, {Arimoto}, {Okamura},
  {Lee}, {Pak}, {Im}, {Lee}, {Kim}, {Jeong}, {Imai}, {Fujishiro}, {Shirahata},
  {Suzuki}, {Ihara}, and {Sakon}]{Matsuhara2006}
{Matsuhara}, H.; {Wada}, T.; {Matsuura}, S.; {Nakagawa}, T.; {Kawada}, M.;
  {Ohyama}, Y.; {Pearson}, C.P.; {Oyabu}, S.; {Takagi}, T.; {Serjeant}, S.;
  et~al.
\newblock {Deep Extragalactic Surveys around the Ecliptic Poles with AKARI
  (ASTRO-F)}.
\newblock {\em Publ. Astron. Soc. Jpn.} {\bf 2006},
  {\em 58},~673--694.
\newblock
  https://doi.org/{\changeurlcolor{black}\href{https://doi.org/10.1093/pasj/58.4.673}{\detokenize{10.1093/pasj/58.4.673}}}.

\bibitem[{Tsuchikawa} {et~al.}(2021){Tsuchikawa}, {Kaneda}, {Oyabu},
  {Kokusho}, {Kobayashi}, {Yamagishi}, and {Toba}]{Tsuchikawa2021}
{Tsuchikawa}, T.; {Kaneda}, H.; {Oyabu}, S.; {Kokusho}, T.; {Kobayashi}, H.;
  {Yamagishi}, M.; {Toba}, Y.
\newblock {A systematic study of silicate absorption features in heavily
  obscured AGNs observed by Spitzer/IRS}.
\newblock {\em Astron. Astrophys.} {\bf 2021}, {\em 651},~A117.
\newblock
  https://doi.org/{\changeurlcolor{black}\href{https://doi.org/10.1051/0004-6361/202140483}{\detokenize{10.1051/0004-6361/202140483}}}.

\bibitem[{Wang} {et~al.}(2015){Wang}, {Li}, and
  {Jiang}]{Wang2015_mir_extinction}
{Wang}, S.; {Li}, A.; {Jiang}, B.W.
\newblock {Very Large Interstellar Grains as Evidenced by the Mid-infrared
  Extinction}.
\newblock {\em Astrophys. J.} {\bf 2015}, {\em 811},~38.
\newblock
  https://doi.org/{\changeurlcolor{black}\href{https://doi.org/10.1088/0004-637X/811/1/38}{\detokenize{10.1088/0004-637X/811/1/38}}}.

\bibitem[{Lo Faro} {et~al.}(2017){Lo Faro}, {Buat}, {Roehlly},
  {Alvarez-Marquez}, {Burgarella}, {Silva}, and {Efstathiou}]{LoFaro2017}
{Lo Faro}, B.; {Buat}, V.; {Roehlly}, Y.; {Alvarez-Marquez}, J.; {Burgarella},
  D.; {Silva}, L.; {Efstathiou}, A.
\newblock {Characterizing the UV-to-NIR shape of the dust attenuation curve of
  IR luminous galaxies up to z {\ensuremath{\sim}} 2}.
\newblock {\em Mon. Not. R. Astron. Soc.} {\bf 2017},
  {\em 472},~1372--1391.
\newblock
  https://doi.org/{\changeurlcolor{black}\href{https://doi.org/10.1093/mnras/stx1901}{\detokenize{10.1093/mnras/stx1901}}}.

\bibitem[{Roebuck} {et~al.}(2019){Roebuck}, {Sajina}, {Hayward}, {Martis},
  {Marchesini}, {Krefting}, and {Pope}]{Roebuck2019}
{Roebuck}, E.; {Sajina}, A.; {Hayward}, C.C.; {Martis}, N.; {Marchesini}, D.;
  {Krefting}, N.; {Pope}, A.
\newblock {Simulations Find Our Accounting of Dust-obscured Star Formation May
  Be Incomplete}.
\newblock {\em Astrophys. J.} {\bf 2019}, {\em 881},~18.
\newblock
  https://doi.org/{\changeurlcolor{black}\href{https://doi.org/10.3847/1538-4357/ab2bf5}{\detokenize{10.3847/1538-4357/ab2bf5}}}.

\bibitem[{Roebuck} {et~al.}(2016){Roebuck}, {Sajina}, {Hayward}, {Pope},
  {Kirkpatrick}, {Hernquist}, and {Yan}]{Roebuck2016}
{Roebuck}, E.; {Sajina}, A.; {Hayward}, C.C.; {Pope}, A.; {Kirkpatrick}, A.;
  {Hernquist}, L.; {Yan}, L.
\newblock {The Role of Star Formation and AGN in Dust Heating of z=0.3-2.8
  Galaxies - II. Informing IR AGN Fraction Estimates through Simulations}.
\newblock {\em Astrophys. J.} {\bf 2016}, {\em 833},~60.
\newblock
  https://doi.org/{\changeurlcolor{black}\href{https://doi.org/10.3847/1538-4357/833/1/60}{\detokenize{10.3847/1538-4357/833/1/60}}}.

\bibitem[{Mullaney} {et~al.}(2011){Mullaney}, {Alexander}, {Goulding}, and
  {Hickox}]{Mullaney2011}
{Mullaney}, J.R.; {Alexander}, D.M.; {Goulding}, A.D.; {Hickox}, R.C.
\newblock {Defining the intrinsic AGN infrared spectral energy distribution and
  measuring its contribution to the infrared output of composite galaxies}.
\newblock {\em Mon. Not. R. Astron. Soc.} {\bf 2011},
  {\em 414},~1082--1110.
\newblock
  https://doi.org/{\changeurlcolor{black}\href{https://doi.org/10.1111/j.1365-2966.2011.18448.x}{\detokenize{10.1111/j.1365-2966.2011.18448.x}}}.

\bibitem[{Lyu} and {Rieke}(2017)]{LyuRieke2017}
\textls[-5]{{Lyu}, J.; {Rieke}, G.H.
\newblock {The Intrinsic Far-infrared Continua of Type-1 Quasars}.
\newblock {\em Astrophys. J.} {\bf 2017}, {\em 841},~76.
\newblock
  https://doi.org/{\changeurlcolor{black}\href{https://doi.org/10.3847/1538-4357/aa7051}{\detokenize{10.3847/1538-4357/aa7051}}}.}

\bibitem[{Bernhard} {et~al.}(2021){Bernhard}, {Tadhunter}, {Mullaney},
  {Grimmett}, {Rosario}, and {Alexander}]{Bernhard2021}
{Bernhard}, E.; {Tadhunter}, C.; {Mullaney}, J.R.; {Grimmett}, L.P.; {Rosario},
  D.J.; {Alexander}, D.M.
\newblock {The post-Herschel view of intrinsic AGN emission: Constructing
  templates for galaxy and AGN emission at IR wavelengths}.
\newblock {\em Mon. Not. R. Astron. Soc.} {\bf 2021},
  {\em 503},~2598--2621.
\newblock
  https://doi.org/{\changeurlcolor{black}\href{https://doi.org/10.1093/mnras/stab419}{\detokenize{10.1093/mnras/stab419}}}.

\bibitem[{McKinney} {et~al.}(2021){McKinney}, {Hayward}, {Rosenthal},
  {Mart{\'\i}nez-Galarza}, {Pope}, {Sajina}, and {Smith}]{McKinney2021}
{McKinney}, J.; {Hayward}, C.C.; {Rosenthal}, L.J.; {Mart{\'\i}nez-Galarza},
  J.R.; {Pope}, A.; {Sajina}, A.; {Smith}, H.A.
\newblock {Dust-enshrouded AGNs Can Dominate Host-galaxy-scale Cold Dust
  Emission}.
\newblock {\em Astrophys. J.} {\bf 2021}, {\em 921},~55.
\newblock
  https://doi.org/{\changeurlcolor{black}\href{https://doi.org/10.3847/1538-4357/ac185f}{\detokenize{10.3847/1538-4357/ac185f}}}.

\bibitem[{Lacy} {et~al.}(2007){Lacy}, {Sajina}, {Petric}, {Seymour},
  {Canalizo}, {Ridgway}, {Armus}, and {Storrie-Lombardi}]{Lacy2007}
{Lacy}, M.; {Sajina}, A.; {Petric}, A.O.; {Seymour}, N.; {Canalizo}, G.;
  {Ridgway}, S.E.; {Armus}, L.; {Storrie-Lombardi}, L.J.
\newblock {Large Amounts of Optically Obscured Star Formation in the Host
  Galaxies of Some Type 2 Quasars}.
\newblock {\em Astrophys. J. Lett.} {\bf 2007}, {\em 669},~L61--L64.
\newblock
  https://doi.org/{\changeurlcolor{black}\href{https://doi.org/10.1086/523851}{\detokenize{10.1086/523851}}}.

\bibitem[{Goulding} {et~al.}(2012){Goulding}, {Alexander}, {Bauer},
  {Forman}, {Hickox}, {Jones}, {Mullaney}, and {Trichas}]{Goulding2012}
{Goulding}, A.D.; {Alexander}, D.M.; {Bauer}, F.E.; {Forman}, W.R.; {Hickox},
  R.C.; {Jones}, C.; {Mullaney}, J.R.; {Trichas}, M.
\newblock {Deep Silicate Absorption Features in Compton-thick Active Galactic
  Nuclei Predominantly Arise due to Dust in the Host Galaxy}.
\newblock {\em Astrophys. J.} {\bf 2012}, {\em 755},~5.
\newblock
  https://doi.org/{\changeurlcolor{black}\href{https://doi.org/10.1088/0004-637X/755/1/5}{\detokenize{10.1088/0004-637X/755/1/5}}}.

\bibitem[{Symeonidis} {et~al.}(2016){Symeonidis}, {Giblin}, {Page},
  {Pearson}, {Bendo}, {Seymour}, and {Oliver}]{Symeonidis2016}
{Symeonidis}, M.; {Giblin}, B.M.; {Page}, M.J.; {Pearson}, C.; {Bendo}, G.;
  {Seymour}, N.; {Oliver}, S.J.
\newblock {AGN are cooler than you think: The intrinsic far-IR emission from
  QSOs}.
\newblock {\em Mon. Not. R. Astron. Soc.} {\bf 2016},
  {\em 459},~257--276.
\newblock
  https://doi.org/{\changeurlcolor{black}\href{https://doi.org/10.1093/mnras/stw667}{\detokenize{10.1093/mnras/stw667}}}.

\bibitem[{Herter} {et~al.}(2012){Herter}, {Adams}, {De Buizer}, {Gull},
  {Schoenwald}, {Henderson}, {Keller}, {Nikola}, {Stacey}, and
  {Vacca}]{Herter2012}
{Herter}, T.L.; {Adams}, J.D.; {De Buizer}, J.M.; {Gull}, G.E.; {Schoenwald},
  J.; {Henderson}, C.P.; {Keller}, L.D.; {Nikola}, T.; {Stacey}, G.; {Vacca},
  W.D.
\newblock {First Science Observations with SOFIA/FORCAST: The FORCAST
  Mid-infrared Camera}.
\newblock {\em Astrophys. J. Lett.} {\bf 2012}, {\em 749},~L18.
\newblock
  https://doi.org/{\changeurlcolor{black}\href{https://doi.org/10.1088/2041-8205/749/2/L18}{\detokenize{10.1088/2041-8205/749/2/L18}}}.


\bibitem[{Kamizuka} {et~al.}(2020){Kamizuka}, {Miyata}, {Sako}, {Ohsawa},
  {Asano}, {Uchiyama}, {Mori}, {Yoshida}, {Tachibana}, {Michifuji}, {Uchiyama},
  {Sakon}, {Onaka}, {Kataza}, {Aoki}, {Doi}, {Hatsukade}, {Kato}, {Kohno},
  {Konishi}, {Minezaki}, {Morokuma}, {Numata}, {Motohara}, {Sameshima},
  {Soyano}, {Takahashi}, {Tanab{\'e}}, {Tanaka}, {Tarusawa}, {Koshida},
  {Tamura}, {Terao}, {Kushibiki}, {Nakamura}, and {Yoshii}]{mimizuku2020}
{Kamizuka}, T.; {Miyata}, T.; {Sako}, S.; {Ohsawa}, R.; {Asano}, K.;
  {Uchiyama}, M.S.; {Mori}, T.; {Yoshida}, Y.; {Tachibana}, K.; {Michifuji},
  T.;  et~al.
\newblock {The University of Tokyo Atacama Observatory 6.5m telescope: On-sky
  performance evaluations of the mid-infrared instrument MIMIZUKU on the Subaru
  telescope}.
\newblock In Proceedings of the Ground-Based and Airborne Instrumentation for Astronomy VIII, Online, 13 December 2020; Volume 11447,  p. 114475X.
\newblock
  https://doi.org/{\changeurlcolor{black}\href{https://doi.org/10.1117/12.2560789}{\detokenize{10.1117/12.2560789}}}.

\bibitem[{Jansen} and {Windhorst}(2018)]{Jansen2018}
{Jansen}, R.A.; {Windhorst}, R.A.
\newblock {The James Webb Space Telescope North Ecliptic Pole Time-domain
  Field. I. Field Selection of a JWST Community Field for Time-domain Studies}.
\newblock {\em Publ. Astron. Soc. Pac.} {\bf
  2018}, {\em 130},~124001.
\newblock
  https://doi.org/{\changeurlcolor{black}\href{https://doi.org/10.1088/1538-3873/aae476}{\detokenize{10.1088/1538-3873/aae476}}}.

\bibitem[{Lutz} {et~al.}(2000){Lutz}, {Sturm}, {Genzel}, {Moorwood},
  {Alexander}, {Netzer}, and {Sternberg}]{Lutz2000}
{Lutz}, D.; {Sturm}, E.; {Genzel}, R.; {Moorwood}, A.F.M.; {Alexander}, T.;
  {Netzer}, H.; {Sternberg}, A.
\newblock {ISO-SWS Spectroscopy of NGC 1068}.
\newblock {\em Astrophys. J.} {\bf 2000}, {\em 536},~697--709.
\newblock
  https://doi.org/{\changeurlcolor{black}\href{https://doi.org/10.1086/308972}{\detokenize{10.1086/308972}}}.

\bibitem[{Lin} {et~al.}(2018){Lin}, {Pope}, and {Kirkpatrick}]{Lin2018}
{Lin}, K.W.; {Pope}, A.; {Kirkpatrick}, A.
\newblock {Hunting for Active Galactic Nuclei in JWST/MIRI Imaging}.
\newblock In {\em American Astronomical Society Meeting Abstracts \#231};  American Astronomical Society: Washington, DC, USA, 2018; Volume 231, p. 250.17.

\bibitem[{Vito} {et~al.}(2018){Vito}, {Brandt}, {Yang}, {Gilli}, {Luo},
  {Vignali}, {Xue}, {Comastri}, {Koekemoer}, {Lehmer}, {Liu}, {Paolillo},
  {Ranalli}, {Schneider}, {Shemmer}, {Volonteri}, and {Wang}]{Vito2018}
{Vito}, F.; {Brandt}, W.N.; {Yang}, G.; {Gilli}, R.; {Luo}, B.; {Vignali}, C.;
  {Xue}, Y.Q.; {Comastri}, A.; {Koekemoer}, A.M.; {Lehmer}, B.D.;  et~al.
\newblock {High-redshift AGN in the Chandra Deep Fields: The obscured fraction
  and space density of the sub-L$_{*}$ population}.
\newblock {\em Mon. Not. R. Astron. Soc.} {\bf 2018},
  {\em 473},~2378--2406.
\newblock
  https://doi.org/{\changeurlcolor{black}\href{https://doi.org/10.1093/mnras/stx2486}{\detokenize{10.1093/mnras/stx2486}}}.

\bibitem[{Meixner} {et~al.}(2019){Meixner}, {Cooray}, {Leisawitz},
  {Staguhn}, {Armus}, {Battersby}, {Bauer}, {Bergin}, {Bradford},
  {Ennico-Smith}, {Fortney}, {Kataria}, {Melnick}, {Milam}, {Narayanan},
  {Padgett}, {Pontoppidan}, {Pope}, {Roellig}, {Sandstrom}, {Stevenson}, {Su},
  {Vieira}, {Wright}, {Zmuidzinas}, {Sheth}, {Benford}, {Mamajek}, {Neff}, {De
  Beck}, {Gerin}, {Helmich}, {Sakon}, {Scott}, {Vavrek}, {Wiedner}, {Carey},
  {Burgarella}, {Moseley}, {Amatucci}, {Carter}, {DiPirro}, {Wu}, {Beaman},
  {Beltran}, {Bolognese}, {Bradley}, {Corsetti}, {D'Asto}, {Denis}, {Derkacz},
  {Earle}, {Fantano}, {Folta}, {Gavares}, {Generie}, {Hilliard}, {Howard},
  {Jamil}, {Jamison}, {Lynch}, {Martins}, {Petro}, {Ramspacher}, {Rao},
  {Sandin}, {Stoneking}, {Tompkins}, and {Webster}]{Meixner2019}
{Meixner}, M.; {Cooray}, A.; {Leisawitz}, D.; {Staguhn}, J.; {Armus}, L.;
  {Battersby}, C.; {Bauer}, J.; {Bergin}, E.; {Bradford}, C.M.; {Ennico-Smith},
  K.;  et~al.
\newblock {Origins Space Telescope Mission Concept Study Report}.
\newblock {\em arXiv e-Prints} {\bf 2019}, arXiv:1912.06213.

\bibitem[{Glenn} {et~al.}(2021){Glenn}, {Bradford}, {Rosolowsky}, {Amini},
  {Alatalo}, {Armus}, {Benson}, {Chang}, {Darling}, {Day}, {Domber}, {Farrah},
  {Hensley}, {Lipscy}, {Moore}, {Oliver}, {Perido}, {Redding}, {Rodgers},
  {Shirley}, {Smith}, {Steeves}, {Tucker}, and {Zmuidzinas}]{Glenn2021}
{Glenn}, J.; {Bradford}, C.M.; {Rosolowsky}, E.; {Amini}, R.; {Alatalo}, K.;
  {Armus}, L.; {Benson}, A.J.; {Chang}, T.C.; {Darling}, J.; {Day}, P.K.;
  et~al.
\newblock {Galaxy Evolution Probe}.
\newblock {\em J. Astron. Telesc. Instrum. Syst.}
  {\bf 2021}, {\em 7},~34004.
\newblock
  https://doi.org/{\changeurlcolor{black}\href{https://doi.org/10.1117/1.JATIS.7.3.034004}{\detokenize{10.1117/1.JATIS.7.3.034004}}}.

\bibitem[{Pope} {et~al.}(2019){Pope}, {Armus}, {Murphy}, {Aalto},
  {Alexander}, {Appleton}, {Barger}, {Bradford}, {Capak}, {Casey},
  {Charmandaris}, {Chary}, {Cooray}, {Condon}, {Diaz Santos}, {Dickinson},
  {Farrah}, {Ferkinhoff}, {Grogin}, {Hickox}, {Kirkpatrick}, {Kotaro},
  {Matthews}, {Narayanan}, {Riechers}, {Sajina}, {Sargent}, {Scott}, {Smith},
  {Stacey}, {Veilleux}, and {Vieira}]{Pope2019}
{Pope}, A.; {Armus}, L.; {Murphy}, E.; {Aalto}, S.; {Alexander}, D.;
  {Appleton}, P.; {Barger}, A.; {Bradford}, M.; {Capak}, P.; {Casey}, C.;
  et~al.
\newblock {Simultaneous Measurements of Star Formation and Supermassive Black
  Hole Growth in Galaxies}.
\newblock {\em Bull. Am. Astron. Soc.} {\bf 2019}, {\em
  51},~330.

\end{thebibliography}
\end{document}